\newif\ifincludefigs
\includefigstrue 

\documentclass[aps,prd,superscriptaddress,nofootinbib,floatfix,notitlepage,twocolumn]{revtex4-1}

\usepackage{amsmath,amssymb,epsfig}
\usepackage{hyperref}
\usepackage{natbib}
\usepackage{aas_macros}

\usepackage{mathrsfs}
\usepackage{color}
\usepackage [autostyle, english = american]{csquotes}
\MakeOuterQuote{"}
\usepackage{scalerel}

\begin{document}
\renewcommand{\vec}[1]{\mathbf{#1}}
\newcommand{\vl}{\vec{l}}
\newcommand{\VL}{\vec{L}}
\newcommand{\va}{\ensuremath{\boldsymbol\alpha}}
\newcommand{\vx}{\vec{x}}
\newcommand{\beq}{\begin{equation}}
\newcommand{\eeq}{\end{equation}}
\newcommand{\expt}{\mathrm{expt}}
\newcommand{\la}{\langle}
\newcommand{\ra}{\rangle}
\newcommand{\lla}{\left\langle}
\newcommand{\rra}{\right\rangle}
\renewcommand{\(}{\left(}
\renewcommand{\)}{\right)}
\renewcommand{\[}{\left[}
\renewcommand{\]}{\right]}
\newcommand{\drm}{\mathrm{d}}
\newcommand{\Mpc}{\mathrm{Mpc}}
\newcommand{\FWHM}{\mathrm{FWHM}}
\newcommand{\arcmin}{\mathrm{arcmin}}
\newcommand{\m}{\mathrm{m}}
\newcommand{\K}{\mathrm{K}}
\newcommand{\eV}{\mathrm{eV}}
\newcommand{\sims}{\mathrm{sims}}
\newcommand{\cross}{\mathrm{cross}}
\newcommand{\auto}{\mathrm{auto}}
\renewcommand{\NG}{\mathrm{NG}}
\newcommand{\CMB}{\mathrm{CMB}}
\newcommand{\source}{\mathrm{source}}
\newcommand{\chiCMB}
{\chi_{\scaleto{\mathrm{CMB}}{4pt}}}
\newcommand{\hkappa}{\hat{\kappa}}
\renewcommand{\max}{\mathrm{max}}


\thispagestyle{empty}

\title{On the effect of non-Gaussian lensing deflections on CMB lensing measurements}

\author{Vanessa B\"ohm}
\affiliation{Berkeley Center for Cosmological Physics, University of California, Berkeley, CA 94720, USA}
\affiliation{Lawrence Berkeley National Laboratory, 1 Cyclotron Road, Berkeley, CA 93720, USA}
\author{Blake D. Sherwin}
\affiliation{Department of Applied Mathematics and Theoretical Physics, University of Cambridge, Wilberforce Road, Cambridge CB3 0WA}
\author{Jia Liu}
\affiliation{Department of Astrophysical Sciences, Princeton University, Princeton, NJ 08544, USA}
\author{J. Colin Hill}
\affiliation{Center for Computational Astrophysics, Flatiron Institute,162 5th Avenue, 10010, New York, NY, USA}
\affiliation{Institute for Advanced Study, Einstein Drive, Princeton, NJ 08540, USA}
\author{Marcel Schmittfull}
\affiliation{Institute for Advanced Study, Einstein Drive, Princeton, NJ 08540, USA}
\author{Toshiya Namikawa}
\affiliation{Leung Center for Cosmology and Particle Astrophysics, National Taiwan University, Taipei, 10617, Taiwan}

\date{\today}

\begin{abstract}
We investigate the impact of non-Gaussian lensing deflections on measurements of the CMB lensing power spectrum. We find that the false assumption of their Gaussianity significantly biases these measurements in current and future experiments at the percent level. The bias is detected by comparing CMB lensing reconstructions from simulated CMB data lensed with Gaussian deflection fields to reconstructions from simulations lensed with fully non-Gaussian deflection fields. The non-Gaussian deflections are produced by ray-tracing through snapshots of an N-body simulation and capture both the non-Gaussianity induced by non-linear structure formation and by multiple correlated deflections. We find that the amplitude of the measured bias is in agreement with analytical predictions by B\"ohm et al. 2016. The bias is largest in temperature-based measurements and we do not find evidence for it in measurements from a combination of polarization fields ($EB,EB$). We argue that the non-Gaussian bias should be even more important for measurements of cross-correlations of CMB lensing with low-redshift tracers of large-scale structure.
\end{abstract}

\maketitle
\section{Introduction} 
Photons of the cosmic microwave background (CMB) get deflected by the the cosmic matter distribution between the surface of last scattering and the observer. This effect is known as CMB lensing (see e.g., Refs.~\cite{CMBLensRev1} and ~\cite{CMBLensRev2} for reviews). Coherent deflections distort the observed CMB fluctuations in both temperature and polarization in a characteristic way. The statistics of the deflections contain a vast amount of cosmological information. They are sensitive to cosmological parameters that determine the formation of cosmic structure, such as a combination of $\sigma_8$ and $\Omega_\m$, the sum of neutrino masses~\cite{2006PhRvDLes} and the presence of dark energy~\cite{2011Sherwin}. They are also a probe of the flatness of space, since curvature changes the relative efficiency of lensing events at different distances. Different to other probes of large-scale structure, CMB lensing is mostly sensitive to structures at relatively high redshifts ($z\approx2$), and and has the advantage of directly probing the total matter distribution.

Since the first detection of CMB lensing in cross correlations~\cite{2007Smith,2008Hirata}, CMB lensing measurements have matured from detections in CMB data alone~\cite{2011Das}, through increasingly significant detections in CMB temperature, polarization and cross correlations~\cite{2012SPT,2013Hanson,2014PolBear1,2014PolBear2,Bicep2Keck2016,2015StorySPT} to a compatible and complementary cosmological probe~\cite{2016PlanckLensing,2017ACT, 2017SPTParams}. Forecasts for current and future surveys~\cite{AdvACT,SimArr,SPT3G,CMB-S4} promise sample variance limited measurements of the CMB lensing power spectrum  up to  multipoles of $L\approx1000$ and a sensitivity of these measurements to the total mass of neutrinos of $\sigma_{\sum\m}\approx 30 \, \m\eV$ if combined with suitable other probes to break degeneracies with $\tau$ and $\Omega_\mathrm{m} h^2$. 

Common CMB lensing reconstruction uses a quadratic, weighted combination of CMB fields to recover the deflection field~\cite{HuQuadEsti,HuOkamPol}. Power spectrum measurements from this quadratic estimator extract lensing information from the lensed CMB 4-point function. The 4-point estimator for the CMB lensing power spectrum is a biased estimator. It is non-zero even in the absence of lensing and carries bias terms at all orders in the lensing power spectrum~\cite{Kesden2003,HansonN2}. 
Other sources of systematics in CMB lensing measurements are masking, anisotropic beam or noise properties~\cite{2009Hanson,HansonBeam} and foregrounds~\cite{2012JCAPFantaye,2014vanEngelen,2014Osborne,2018FerrHill}. Biases to power spectrum measurements can either be estimated and subtracted, or alleviated by suitable modifications to the lensing estimator~\cite{Lewis2011,Namikawa2013,Namikawa2014,2018MadHill}.

Recently, Ref.~\cite{N32} (hereafter BSS16) have identified a new bias to CMB lensing measurements, which arises as a consequence of the non-Gaussian structure of the lensing deflection field. BSS16 specifically considered the effect of a non-vanishing bispectrum of the lensing potential. In a purely analytic study, they found that the bispectrum that arises as a consequence of non-linear structure formation can change the amplitude of the CMB lensing power spectrum measured from CMB temperature data in current and future experiments at the percent level. Since most of these experiments rely primarily on temperature, a corrections to CMB lensing measurements of this magnitude would constitute a significant systematic and, if uncorrected, hinder the accurate estimation of cosmological parameters. However, the theory calculation in BSS16 made a number of non-trivial assumptions (see Appendix \ref{sec:N32} for details) and the actual size of the bias depends on their validity.

Motivated by this, we study the effect of non-Gaussianity on CMB lensing measurement in this work in a completely independent way with ray-traced lensing simulations. Specifically, we look at the difference between lensing power spectra measured with the standard 4-point estimator in two different sets of simulated noisy, lensed CMB maps: one set lensed with purely Gaussian deflection fields, the other with fully non-Gaussian deflections obtained from ray-tracing through snapshots of an N-body simulation.
By using the same unlensed CMB and detector noise realizations for both sets, any significant difference in the measured spectra is a consequence of the non-Gaussianity of the deflection field and can be interpreted as a non-Gaussian bias.
While the study with simulations provides less intuition about the specific source of a non-Gaussian bias, it is in some sense more complete than the theoretical analysis in BSS16, since it captures the full non-Gaussianity of the field, which can manifest itself in more ways than a non-zero bispectrum and relies on fewer simplifying assumptions.
We compare the measured non-Gaussian bias to the theoretical prediction of BSS16. For this, we update the theoretical prediction to also take into account the lensing bispectrum sourced by multiple correlated deflections (so-called post Born corrections). The importance of the post-Born bispectrum was recently pointed out by Ref.~\cite{postBornPratten} and we use their analytically derived expression to model it.

Post-Born corrections in CMB lensing were recently studied in simulations in Refs.~\cite{2017PhRvDPetri} and~\cite{2018JCAPFabbian}.
The effect of non-linear structure formation on lensing reconstructions, in particular its impact on the second order lensing bias $N^{(2)}$, was measured in ray-traced simulations for the interpretation of data from the South Pole Telescope~\cite{2012SPT}, but found to be irrelevant for this specific data set. Parallel to the work presented here, Beck et al. 2018 (in prep), have carried out a measurement of a non-Gaussian bias on an independent set of ray-traced lensing simulations.

This paper is organized as follows: we start with briefly reviewing CMB lensing and CMB lensing reconstruction in Section~\ref{sec:lensrecon}. In Sec.~\ref{sec:sims} we give a full overview of the production of mock CMB data maps: In subsections we describe the production of ray-traced lensing maps and their Gaussian counterparts (Sec.~\ref{sec:conv}), the generation of noisy, lensed CMB simulations (Sec.~\ref{sec:CMBmaps}) and the reconstruction from these mock data sets (Sec.~\ref{sec:recon}). Results and their comparison to theory are presented in Sec.~\ref{sec:res}. We conclude with a discussion of the results and a comment on the importance of the non-Gaussian bias for cross correlations with low-redshift tracers in Sec.~\ref{sec:dis}. For details on the theoretical bias model derived in BSS16, we refer the reader to Appendix~\ref{sec:N32} and Ref.~\cite{N32}.

\section{CMB lensing and CMB lensing reconstruction}
\label{sec:lensrecon}
Lensing distortions are a measure of the integrated mass distribution along the photons' trajectories. In a flat standard cosmology and under the Born approximation, the lensing convergence $\kappa(\VL)$ is related to the density contrast $\delta(\VL,\chi)$ through the line-of-sight integration
\beq
\label{kappa}
\kappa(\VL)= \frac{3}{2}\frac{\Omega_m H_0}{c^2} \int_0^{{\chiCMB}} \drm \chi \, W(\chi,\chiCMB)\ \delta\(\VL,\chi\)
\eeq
with lensing kernel
\beq
\label{eq:LensEff}
W \( \chi,\chiCMB\) =  \[1+z(\chi)\]  \frac{\chi\(\chiCMB-\chi\)}{\chiCMB}.
\eeq
Throughout this paper, we use the flat sky approximation, where $\VL$ denotes the wave vector of a 2D Fourier mode on the sky.

The mapping between unlensed CMB fields $(T,Q,U)$ and their lensed counterparts $(\tilde{T},\tilde{Q},\tilde{U})$ is determined by the lensing deflection angle $\va$,
\beq
\tilde{T}(\vx)=T\[\vx+\va(\vx)\],
\eeq
which is to good approximation curl-free and can be expressed in terms of a scalar lensing potential $\phi(\vx)$
\beq
\va(\vx)=\nabla\phi(\vx).
\eeq
Similar to overdensity and gravitational potential in three dimensions, the lensing convergence (Eq.~\ref{kappa}) and the lensing potential are related by the Poisson equation
\beq
\kappa(\vx)=-\frac{1}{2}\nabla^2\phi(\vx).
\eeq

CMB lensing reconstruction is the recovery of the lensing deflection field from lensed, noisy CMB data. It is commonly performed with an estimator that is quadratic in the lensed CMB~\cite{HuQuadEsti,HuOkamPol,1999PhRvDZal},
\beq
\label{QuadEsti}
\hat{\kappa}(\VL)=\frac{1}{2} L^2 A^{XY}_L \int_\vl g^{XY}_{\vl,\VL} \tilde{X}_\expt(\vl) \tilde{Y}_\expt(\VL-\vl).
\eeq
In Eq.~\ref{QuadEsti} $X$ and $Y$ represent either temperature ($T$) or polarization fields ($E/B$) and the subscript "$\expt$" labels noisy, beam-deconvolved data. The weight $g$ and the normalization $A_L$ depend on the fiducial lensed CMB power spectra as well as the beam and noise properties of the experiment (see Ref.~\cite{HuOkamPol} for the exact expressions\footnote{Ref.~\cite{HuOkamPol} uses unlensed power spectra in the lensing weights. Replacing them by their lensed counterparts partly removes higher order biases from the power spectrum estimate~\cite{Lewis2011,HansonN2,Anderes2013}.}).
Weight and normalization are chosen such that the estimator in Eq.~\ref{QuadEsti} has minimum variance and is unbiased in the absence of any source of mode-coupling other than lensing.

A few alternatives to the quadratic estimator have been proposed. Some are are based on maximizing the CMB lensing posterior\footnote{The quadratic estimator can be interpreted as a first order approximation to a maximum likelihood estimator for the lensing potential.} or sampling the joint distribution of lensing deflections and CMB~\cite{2003Hirata,2017Carron,2017Millea}. Other estimators are derived from a configuration-space perspective and use the magnification and shear of the lensed CMB fluctuations to estimate the lensing field~\cite{Bucher2012,Prince2018,SchaanFerr2018_2}. To date the quadratic estimator remains the most widely used and best understood estimator for the CMB lensing deflection field. 

Measurements of the CMB lensing power spectrum from the quadratic estimator are sensitive to the lensed CMB four-point function, 
\begin{align}
\label{FourEsti}
\nonumber
\hat{C}^{\kappa\kappa}_{WX,YZ}(L) = \frac{1}{4} L^4 A_L^{WX} A_L^{YZ} \int_{\vl_1,\vl_2}g^{WX}_{\vl_1,\VL}g^{YZ}_{\vl_2,\VL}\times \\ 
\la \tilde{W}_\expt(\vl_1) \tilde{X}_\expt(\VL-\vl_1) \tilde{Y}_\expt(-\vl_2) \tilde{Z}_\expt(\vl_2-\VL)\ra.
\end{align}

Since the response of the CMB to lensing is non-linear in the deflection, this four-point estimator gets contributions from terms at all orders in the lensing convergence. Only one of the contributing second order terms gives rise to the convergence power spectrum. The remaining terms are bias terms that need to be subtracted in order to  obtain an unbiased estimate for $C_L^{\kappa\kappa}$. They are commonly summarized and labeled by their power in the lensing power spectrum: $N^{(0)}_L$ for the bias that is sourced by Gaussian CMB fluctuations (this term is present even in the absence of lensing),  $N_L^{(1)}$ for all biases proportional to $C^{\kappa\kappa}_L$ and $N_L^{(2)}$ for biases proportional to $\(C^{\kappa\kappa}_L\)^2$~\cite{Kesden2003,HansonN2}. The $N_L^{(2)}$ bias can be greatly reduced by a slight modification to the lensing weights, see e.g. Refs.~\cite{HansonN2,Lewis2011,Anderes2013}. Adapting this notation, the expectation value of Eq.~\ref{FourEsti}, averaged over realizations of CMB and lensing deflections (and assuming that both are Gaussian fields), becomes
\beq
\label{bias_exp}
\la\hat{C}^{\kappa\kappa}_L\ra=N^{(0)}_L+C^{\kappa\kappa}_L+N^{(1)}_L+\mathcal{O}\[(C_L^{\kappa\kappa})^2\].
\eeq
Non-linear processes, such as non-linear structure formation and multiple correlated deflections, introduce a small, but detectable amount of non-Gaussianity to the lensing convergence~\cite{BkkkNamikawa,postBornPratten,postBornMarozzi,JiaPeaks2016}. In the limit of small density perturbations, the non-Gaussianity can be characterized by a lensing bispectrum.
A non-zero lensing bispectrum changes the lensed temperature four-point function and introduces an additional bias term to Eq.~\ref{bias_exp},
\beq
\label{bias_exp_N32}
\la\hat{C}^{\kappa\kappa}_L\ra=N^{(0)}_L+C^{\kappa\kappa}_L+N^{(1)}_L+{\color{red}N_L^{(3/2)}}+\mathcal{O}\[(C_L^{\kappa\kappa})^2\].
\eeq
This new bias was first identified in Ref.~\cite{N32}.
We will now compare this theoretically derived term with measurements of a non-Gaussian bias in simulations.

\section{Simulations}
\label{sec:sims}
\begin{figure*}
\includegraphics[width=0.95\textwidth]{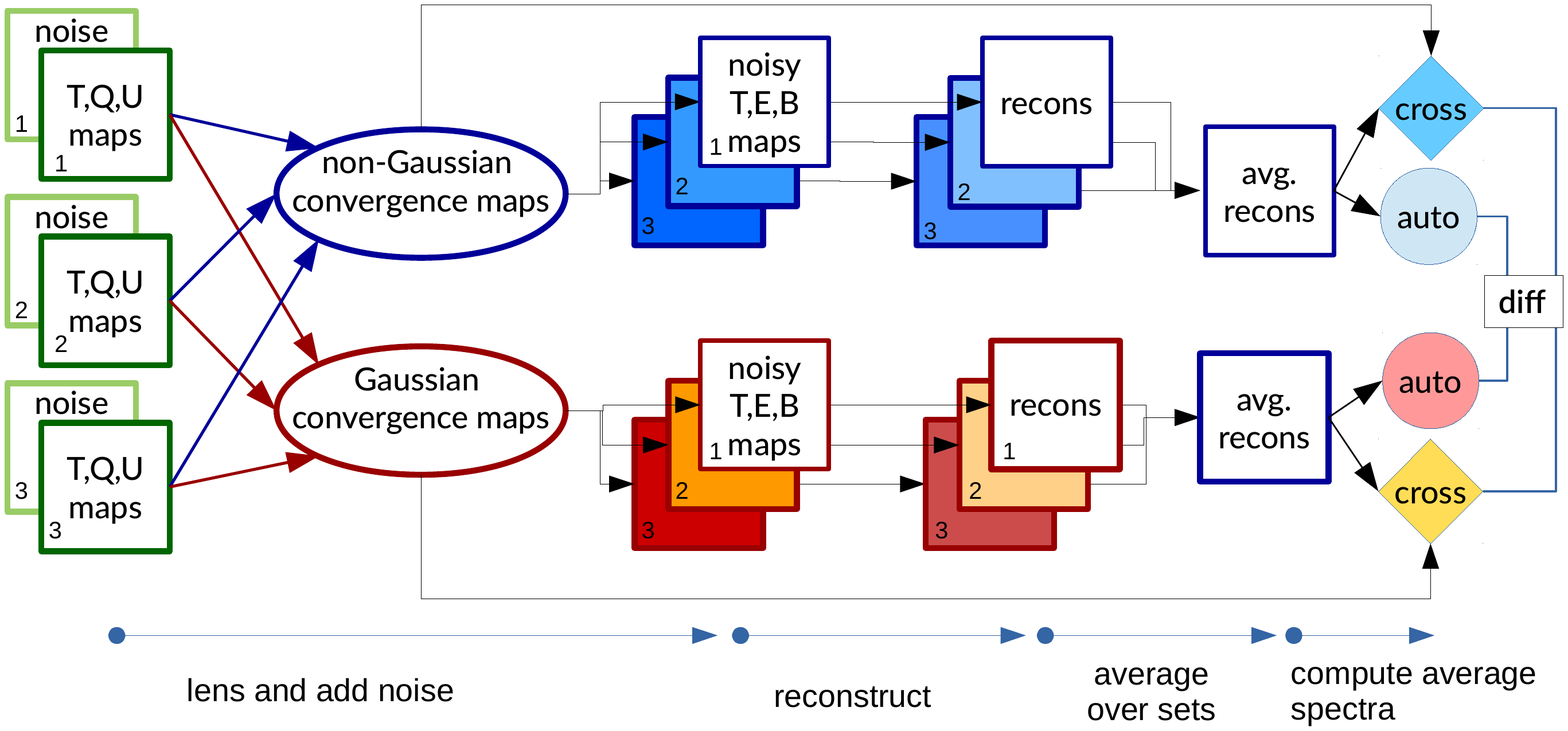}
\vspace{-3.5cm}
\caption{\label{fig:coolgraph} Schematic outline of the simulation pipeline: Squares and ellipses represent sets of 10240 simulations. We start by generating 3 sets of unlensed CMB and noise realizations. We then lens each CMB realization with both Gaussian and non-Gaussian convergence maps. This results in two times three sets of lensed CMB simulations, which we convolve with a Gaussian beam. We then add the same noise realization to each corresponding set in the Gaussian and non-Gaussian branch. After beam-deconvolution of the noisy maps, we run the standard quadratic estimator on each of them. We average the reconstruction results over the three sets in each branch to beat down the reconstruction noise originating from the CMB sample variance. This leaves us with one averaged non-Gaussian and another averaged Gaussian set of reconstructed convergence maps. We compute the average power spectrum in each of these sets as well as the average cross correlation with the true underlying realizations. Any significant difference between the average power spectra in the two branches is a non-Gaussian bias.}
\end{figure*}

The general work flow for isolating a non-Gaussian bias is simple:  we use a set of non-Gaussian convergence maps generated by ray-tracing through an N-body simulation and a corresponding set of Gaussian convergence maps with the same average power spectrum. The same CMB realizations are lensed with both Gaussian and non-Gaussian convergence maps, which results in two sets of lensed CMB maps. We convolve these maps with a Gaussian beam before we add the same realizations of white measurement noise to both sets. We beam-deconvolve the noisy maps before we apply the standard quadratic and four-point estimators. We then compare the results of the reconstructions between both sets and look for significant differences. In the following sections, we provide detailed descriptions and validations for each of these steps. The entire procedure is also illustrated in Fig.~\ref{fig:coolgraph}.

For all simulations (N-body and CMB), we use a standard $\Lambda$CDM cosmology with parameters, $H_0= 72\ \mathrm{km/s/Mpc}$, $\Omega_m=0.296$, $\sigma_8=0.786$, $w=-1$, $n_s= 0.96$ and $\Omega_b= 0.046$.

\subsection{Convergence maps}
\label{sec:conv}
We use a set of 10240 non-Gaussian convergence maps that was obtained from ray-tracing through snapshots of an N-body simulation. For a detailed description of their production we refer the reader to Ref.~\cite{JiaPeaks2016}. The underlying N-body simulation is based on the public Gadget-2 code~\cite{Gadget-2}, has a box size of $600\ \mathrm{Mpc}/h$ and is resolved by $\mathrm{N}=1024^3$ particles (corresponding to a mass resolution of $1.4\times10^{10}M_\odot/h$). The linear matter power spectrum for its initialization was computed with CAMB\footnote{\url{http://camb.info/}}~\cite{CAMB} and initial conditions at $z=100$ generated with N-GenIC. Snapshots were recorded between $z\approx45$ and $z=0$, a range which covers 99\% of the growth corrected lensing kernel $W(\chi,\chi^*)D(z)$. Convergence maps were computed with LensTools~\cite{LensTools} tracing $4096^2$ light rays and calculating their deflections on 3 planes per box. This procedure does not assume that the deflection angle is small or that the light rays follow unperturbed geodesics. Different realizations of the convergence maps were produced by randomly rotating and shifting the potential planes~\cite{Petri1026sampleVar}. The resulting maps are $12.25\ \mathrm{deg}^2$ in size and resolved by $2048^2$ pixels measuring $0.1025^2\ \arcmin^2$. We refer to this simulation set as non-Gaussian or N-body lensing simulations. Their non-Gaussianity is not only a consequence of non-linear structure formation in the N-body simulation, but also of the multiple deflections along the lens planes.
We do not measure or take into account the curl of the deflection field that is introduced by multiple deflection, because we do not expect a significant bias from bispectra involving the curl component (see Appendix~\ref{sec:curl} for details).

We further produce a second set of 10240 purely Gaussian convergence maps. These Gaussian simulations are generated by first measuring the average power spectrum of the non-Gaussian simulations and then drawing convergence realizations from a multivariate Gaussian with exactly this power spectrum.

\begin{figure}
\includegraphics[width=0.95\columnwidth]{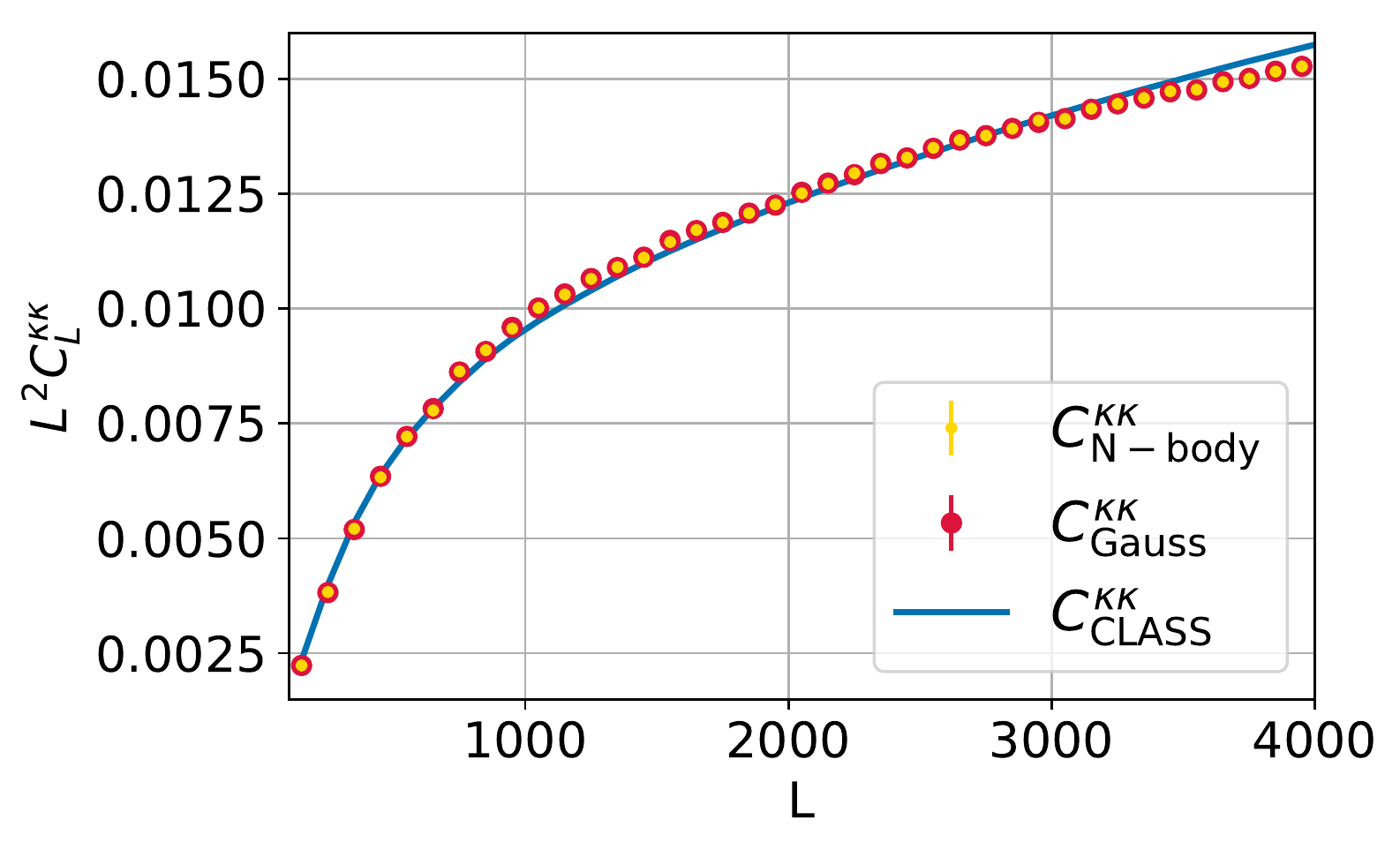}
\caption{\label{clkk}Power spectra measured from 10240 Gaussian (red) and ray-traced non-Gaussian (yellow) convergence maps closely follow the theory curve computed with CLASS (blue). For modeling non-linear effects in the matter power spectrum CLASS uses a version of HALOFIT~\cite{HALOFITT}. We use precision parameters tol\textunderscore perturb\textunderscore integration=1e-6, perturb\textunderscore sampling\textunderscore stepsize=0.01, k\textunderscore min\textunderscore tau0=0.002, k\textunderscore max\textunderscore tau0\textunderscore over\textunderscore l\textunderscore max=10., halofit\textunderscore k\textunderscore per\textunderscore decade=3000 and l\textunderscore max\textunderscore scalars=8000 to produce the theory curve. Missing power on small scales is owed to the finite resolution of the simulation.}
\end{figure}

\begin{figure}
\includegraphics[width=0.95\columnwidth]{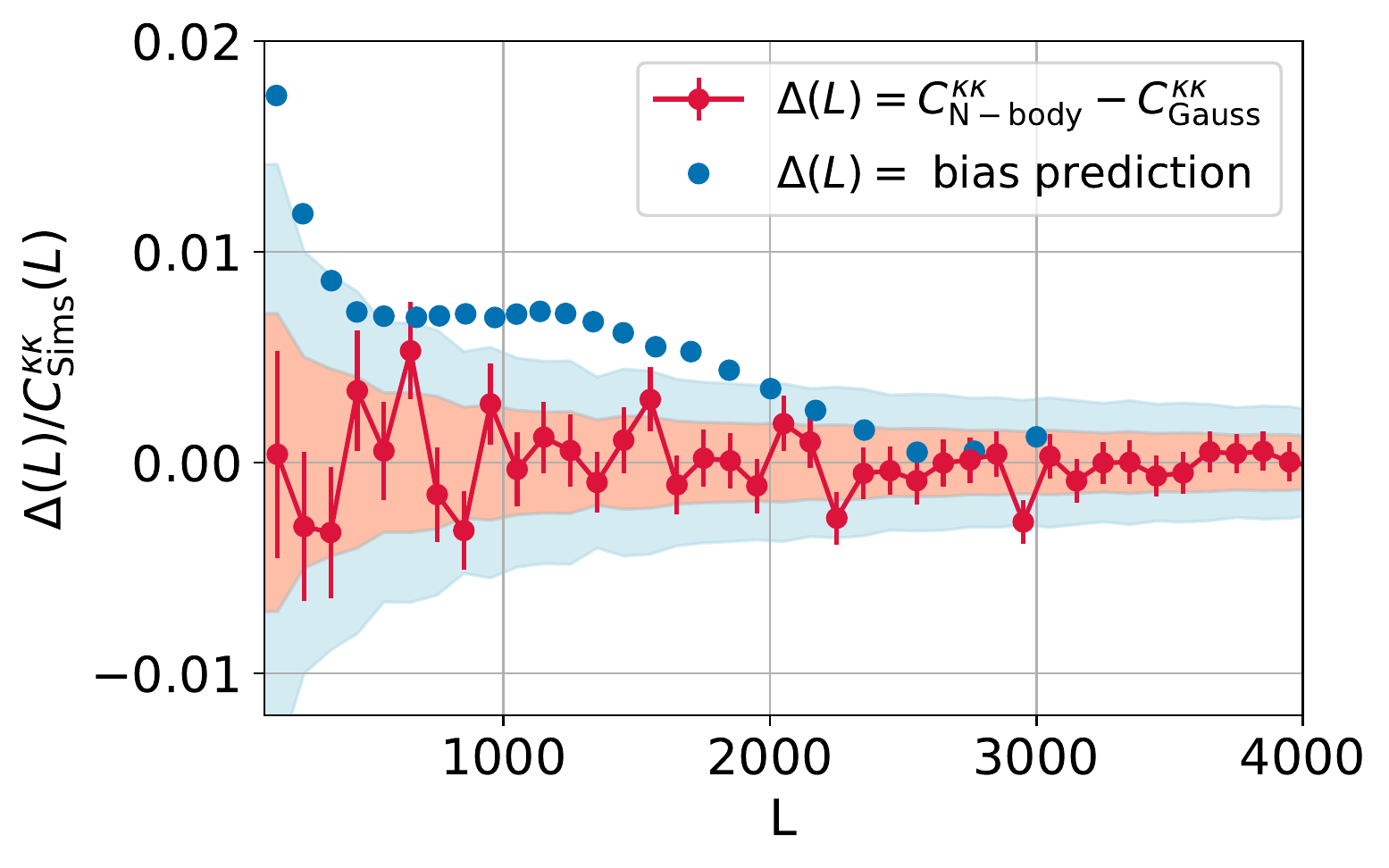}
\caption{\label{clkk_comp} For an accurate measurement of the non-linear bias in the reconstructions it is crucial that the power spectra of the original, non-reconstructed, Gaussian and non-Gaussian simulations are consistent within their sample variance. We show that this is indeed the case by checking that their difference is consistent with zero (red dots, $\chi^2/\nu=1.02$). We further require the sample variance of both sets to be small enough to allow a detection of a bias at the percent-level. To see this, we plot the combined sample variance of both sets as shaded regions ($\pm\sigma, \pm 2\sigma$) and compare it to the expected size of the non-linear bias (blue dots).}
\end{figure}

In Fig.~\ref{clkk} we compare the average power spectra of the Gaussian and non-Gaussian simulation set to a theory power spectrum computed with the anisotropy solver CLASS\footnote{\url{http://class-code.net/}}~\cite{CLASSII}. The missing power on the small-scale end, $L>3000$, in the simulations is caused by the finite resolution of the N-body simulation~\cite{JiaPeaks2016}. On the large-scale end, the power is slightly suppressed because of the finite size of the simulation box.
To allow for an accurate and unbiased detection of the non-Gaussian bias, we require an excellent agreement of the average power spectra in both simulation sets. Any significant difference between the power spectra could result in a false detection of a non-Gaussian bias. We find that the power spectrum of the Gaussian set agrees with the spectrum of the non-Gaussian simulations, as expected, within the sample variance (Fig.~\ref{clkk_comp}, red curve). We further compare the combined standard deviation of the average power in both simulation sets to the size of the bias predicted by BSS16 (Fig.~\ref{clkk_comp} shaded region and blue dots). The comparison shows that the sample variance in the simulation sets is low enough to allow for a detection of a bias at the percent level (which corresponds to the the magnitude predicted by BSS16).

To get a sense of the non-Gaussianity of the ray traced convergence maps, we measure their skewness, $\left \langle \kappa(\vx)^3 \right \rangle$ after smoothing them with a Gaussian kernel on different scales. The skewness is an integrated measure of the bispectrum~\footnote{Note that the quadratic estimator results in an additional skewness in the measured maps, i.e. the measured maps have non-zero skewness even if the underlying field is Gaussian~\cite{JiaPeaks2016}. We measure the skewness in the true noiseless convergence maps and not in the reconstructions, since we are interested in quantifying the bispectrum introduced by non-linear physics. }. By comparing the measurement with the theoretical prediction, 
\begin{align} 
\label{skew}
\nonumber
\left \langle \kappa(\vx)^3 \right \rangle=\int_\vl \int_\VL W_R(L)W_R(l)W_R(|-\VL-\vl|)\\
\times B^{\kappa\kappa\kappa}(L,l,|-\VL-\vl|)\\
W_R(l)=\mathrm{exp}\(-l^2 R^2/2\)
\end{align}
we can determine the most suitable theoretical bispectrum model for computing the bias following BSS16.

\begin{figure}
\includegraphics[width=0.95\columnwidth]{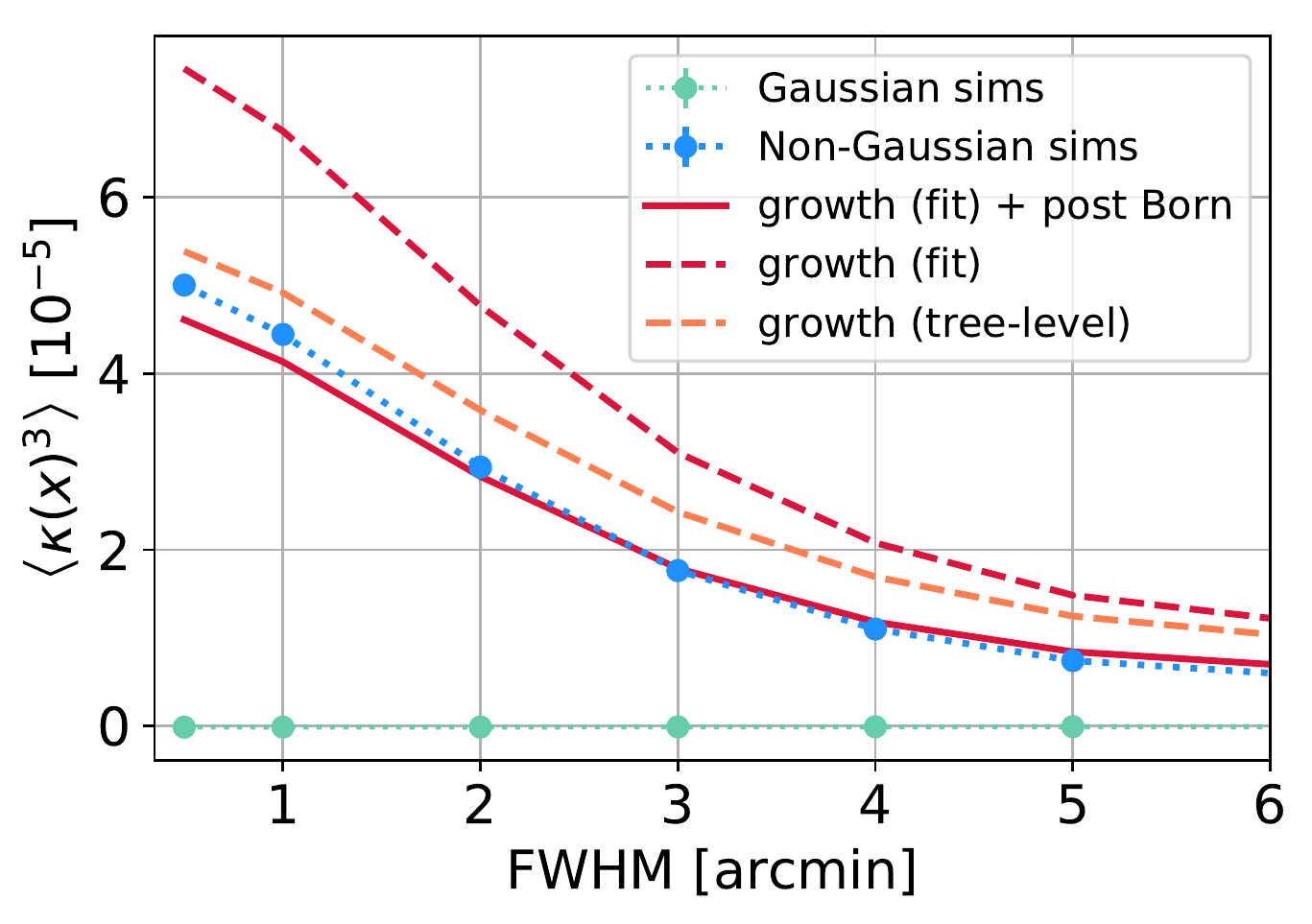}
\caption{\label{skewness} The skewness measured on different scales provides some information on the bispectrum of the non-Gaussian convergence maps. We find that we can accurately model the skewness by assuming that the bispectrum consists of a non-linear growth induced and post-Born induced contribution. The growth-induced part is best described by using a simulation-calibrated fit to the matter bispectrum. The convergence maps were smoothed with a Gaussian kernel with FWHMs indicated on the x-axis and filtered to exclude modes with $L>4000$. For the theory curves we impose cut offs at $k_{\min}=0.0105 \left[h/\mathrm{Mpc}\right]$ corresponding to the box size of the simulation and $k_{\max}=50 \left[h/\Mpc\right]$. Outside of these bounds we set the matter bispectrum (and matter power spectrum in the computation of the post Born terms) to zero. From the comparison with a theory curve computed with $k_{\max}=100 \left[h/\Mpc\right]$, we find that the results are not sensitive to the $k_{\max}$ cut off. The error bars correspond to the standard deviation of the mean and are smaller than the marker size.}
\end{figure}

We expect the bispectrum to have two contributions: one from non-linear structure formation, where the convergence bispectrum is an integrated measure of the bispectrum of large-scales structure, and a second contribution from  post-Born effects~\cite{postBornPratten}. In the squeezed limit these two contributions have opposite sign and partly cancel each other.
We compare two different models for the convergence bispectrum induced by non-linear structure formation; one in which we model the matter bispectrum in tree-level perturbation theory\footnote{In this model we replace the linear matter power spectrum by its HALOFIT counterpart} and one in which we use a simulation-calibrated fit to the matter bispectrum~\cite{GilMarin2011}\footnote{This bispectrum model has 9 free parameters which are assumed to be independent from cosmology and have been measured and fixed in Ref.~\cite{GilMarin2011}. It also depends on cosmological parameters through a direct appearance of $\sigma_8$ and indirectly by its dependence on the non-linear scale and the non-linear matter power spectrum. We adapt these quantities to agree with the cosmology of the simulation.} The results of the skewness measurement together with the different theoretical models are shown in Fig.~\ref{skewness}. We find that the theory curve computed from a combination of structure formation induced and post-born bispectra agrees well with the measurement on smoothing scales $\FWHM>2$ arcmin. On smaller scales we find a slight discrepancy, with the measurement lying above the theory prediction. We also find that simulation-calibrated fit to the matter bispectrum leads to better agreement with the simulation than the tree-level perturbation theory model. We use this best fitting model (red line in Fig.~\ref{skewness}) in the following sections to compute the theoretical prediction for the non-linear bias.

\subsection{CMB simulations}
\label{sec:CMBmaps}
\begin{figure}
\includegraphics[width=0.95\columnwidth]{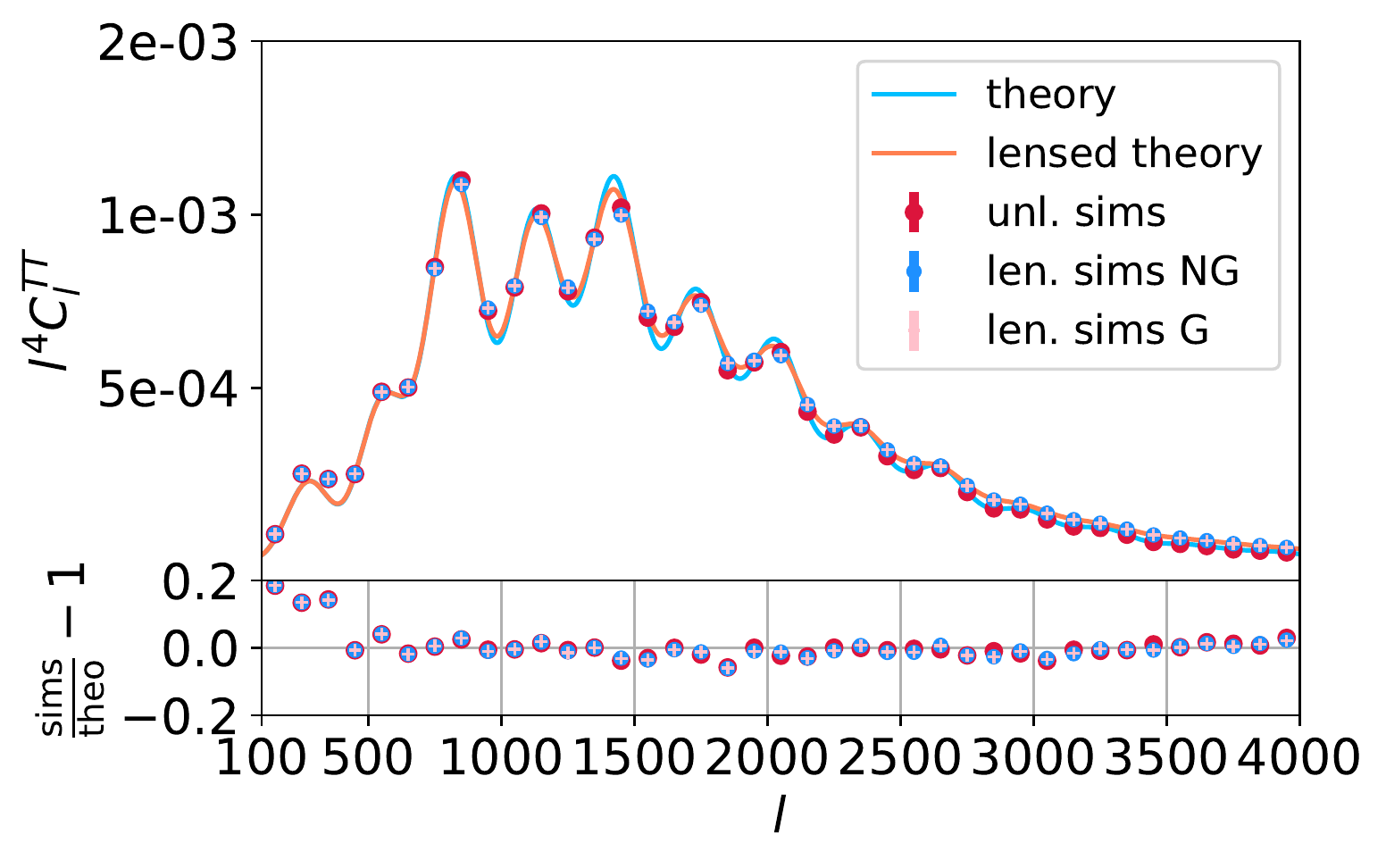}
\caption{\label{cltt} The average power spectra of the lensed temperature maps in Gaussian and non-Gaussian simulation branches agree well with each other (lower panel). The agreement of lensed and unlensed realizations with the theory prediction is good except for large scales, where we find a significant deviation at $l<500$. We exclude these scales from the reconstruction.}
\end{figure}

\begin{figure}
\includegraphics[width=0.95\columnwidth]{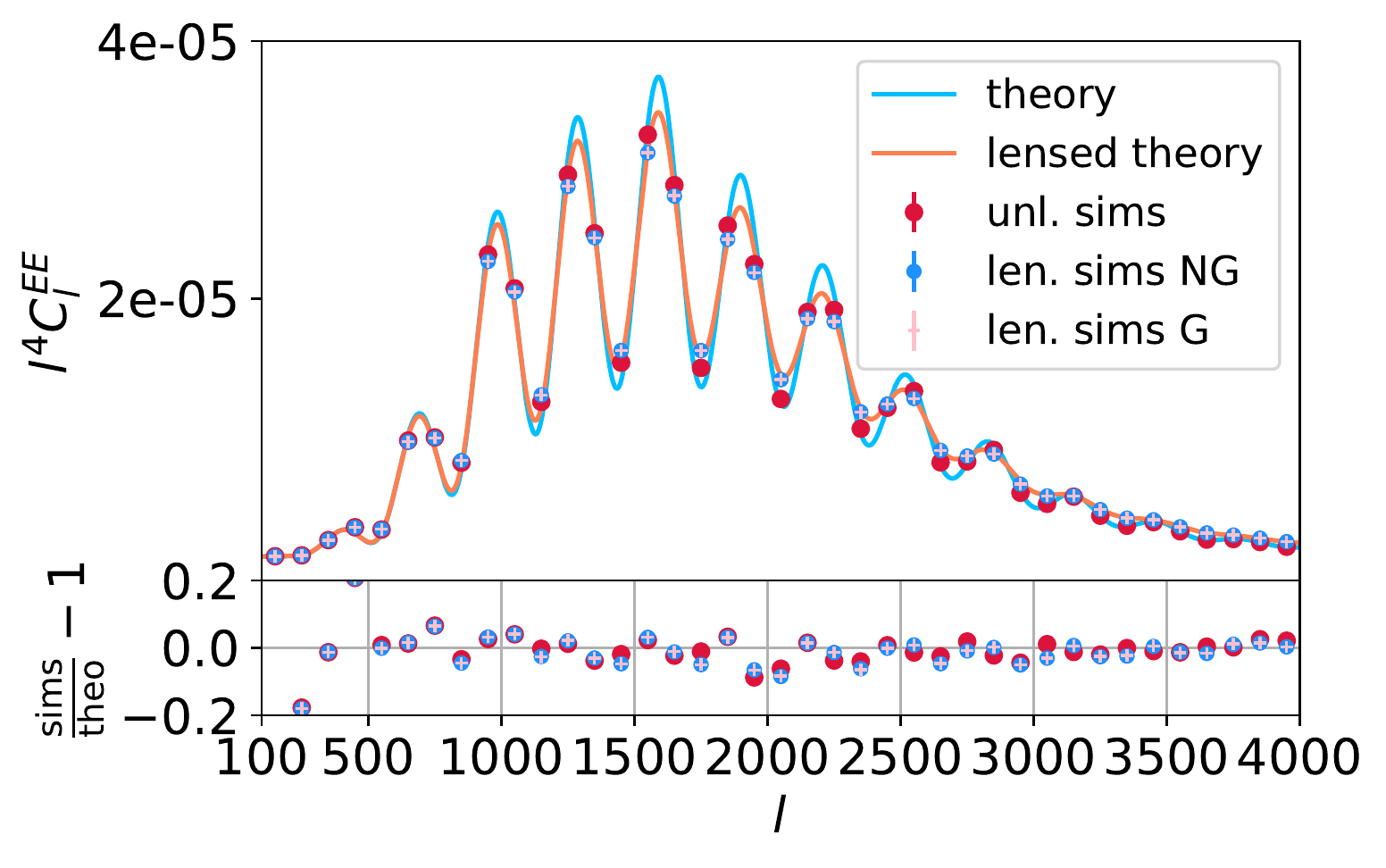}
\caption{\label{clee} The average power spectra of the unlensed  and lensed polarization E-modes agree with the theory prediction except for large scales, where we find a significant deviation at $l<500$. We exclude these scales from the reconstruction.}
\end{figure}

We produce three sets of $10240$ unlensed CMB realization in temperature $(T)$ and polarization ($Q,U$) based on power spectra computed with CAMB. We use each Gaussian and non-Gaussian convergence map to lens the same three CMB maps (one map from each set). Using the same lenses for three background CMBs and averaging over their lensing measurements reduces the Gaussian reconstruction noise and thus the noise of the bias measurements.
The lensing algorithm is described in detail in Ref.~\cite{LouisLens2013}\footnote{We include terms up to fifth order in this algorithm.}. We apply a filter that removes modes with $L>6000$ from the convergence maps prior to the lensing. This step is necessary for numerical stability and to remove unphysical effects that are caused by the finite resolution of the simulations. The power spectra of the lensed maps agree well with theory (as shown in Figs.~\ref{cltt} and ~\ref{clee}).

After lensing, we convolve the lensed CMB maps with a Gaussian beam of width $\FWHM = 1\, \arcmin$ and add the same white noise realization with a noise level of $\sigma_T= 1\, \mu \K{-}\arcmin$ in temperature and $\sigma_{\mathrm{pol}}=\sqrt{2}\sigma_T$ in polarization to each corresponding Gaussian and non-Gaussian set. The noise configurations are chosen to roughly match prospective CMB surveys.

The entire procedure leaves us with 3 times 2 sets of 10240 mock CMB measurements, where corresponding maps in each of the three pairs have same CMB and noise realizations and differ only in the underlying convergence field.

\subsection{Lensing Reconstruction}
\label{sec:recon}
We apply a quadratic estimator to all six sets of noisy, beam-deconvolved, lensed CMB maps in $(TT)$ and $(EB)$ to obtain noisy estimates of the underlying convergence fields. The reconstruction pipeline is described in detail in ~\cite{2017ACT}. 

We filter scales with $l<500$ from the CMB maps prior to reconstruction since Fig.~\ref{cltt} indicates some inconsistency between the power spectra of the lensed simulations and the theory power spectra on these scales. We also filter out any multipoles with $l>4000$. Realistic CMB temperature data could already be contaminated by extragalactic foregrounds at lower multipoles. Using $l<3000$ reduces the theoretically predicted size of the bias by approx. a factor of $0.5$ (see Fig.~\ref{fig:n32}). 

Since three realizations in each branch have the same underlying convergence field, we average over their reconstructed power maps to recover a set of 10240 power measurements corresponding to the 10240 input maps in each simulation branch. These averaged measurements have reduced noise compared to measurements with only one CMB realization. 

We proceed by computing the average power spectrum of the reconstructed lensing maps in the Gaussian and non-Gaussian branches for $(TT)$ and $(EB)$ reconstructions. 
By construction, we expect all lensing biases that are sensitive to the convergence power spectrum and the lensed or unlensed CMB power spectra (c.p. Eq.~\ref{bias_exp}) to be identical in the Gaussian and non-Gaussian simulations~\footnote{The bispectrum of the convergence also changes the lensed CMB power spectra but this is a sub-percent effect and not detectable in our simulations~\cite{BiSpecCMBPow}}. Since we are only interested in the difference of the reconstructed convergence power spectra, in which these biases cancel out, we do not compute and remove them.
Apart from the auto power spectra, we also compute the average power in the cross correlation between input maps and reconstructed maps. This cross correlation is not an actual observable, but can serve as a proxy for the non-Gaussian bias in cross correlations with other tracers of large-scale structure. Also, measurements of the cross power are not affected by the $N^{(0)}$ and $N^{(1)}$ bias and have lower noise. The theory prediction for the bias in cross-correlations is $N_{\NG}^\cross \approx 1/2 N_{\NG}^\auto$~\cite{N32}.

\section{Results}
\label{sec:res}
As discussed above we measure the non-Gaussian bias from the difference between
\begin{align}
\label{eq:measuredbias}
\nonumber
N^{\mathrm{auto}}_\mathrm{NG}(L)&=\hat{C}^{\hkappa\hkappa}_{\mathrm{NG}}(L)-\hat{C}^{\hkappa\hkappa}_{\mathrm{G}}(L)\\
N^{\mathrm{cross}}_\mathrm{NG}(L)&=\hat{C}^{\hkappa\kappa}_{\mathrm{NG}}(L)-\hat{C}^{\hkappa\kappa}_{\mathrm{G}}(L).
\end{align}
We start by examining this difference in the temperature based reconstructions. 
\begin{figure}
\includegraphics[width=0.95\columnwidth]
{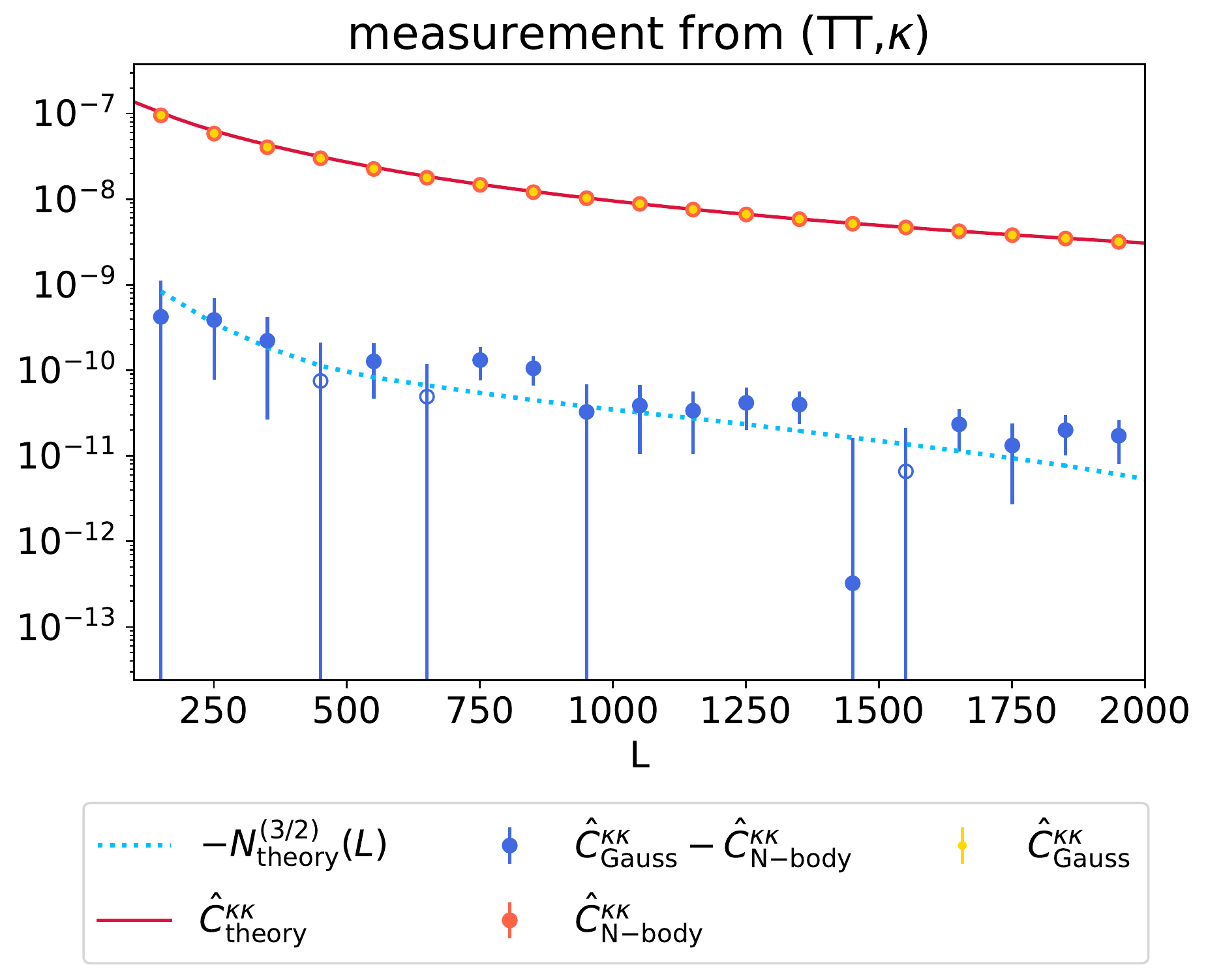}
\caption{\label{fig:N32biasXfull} Average power in the cross correlation between convergence maps reconstructed from the lensed temperature maps and the input convergence field. The measured power follows the theory curve (red line) for both Gaussian and non-Gaussian simulations (yellow and orange points). The difference between the Gaussian and non-Gaussian reconstructions is shown as blue points (circles if they have negative sign). It is consistent with the theory prediction of BSS16 (light blue line).}
\end{figure} 
\begin{figure}
\includegraphics[width=0.95\columnwidth]
{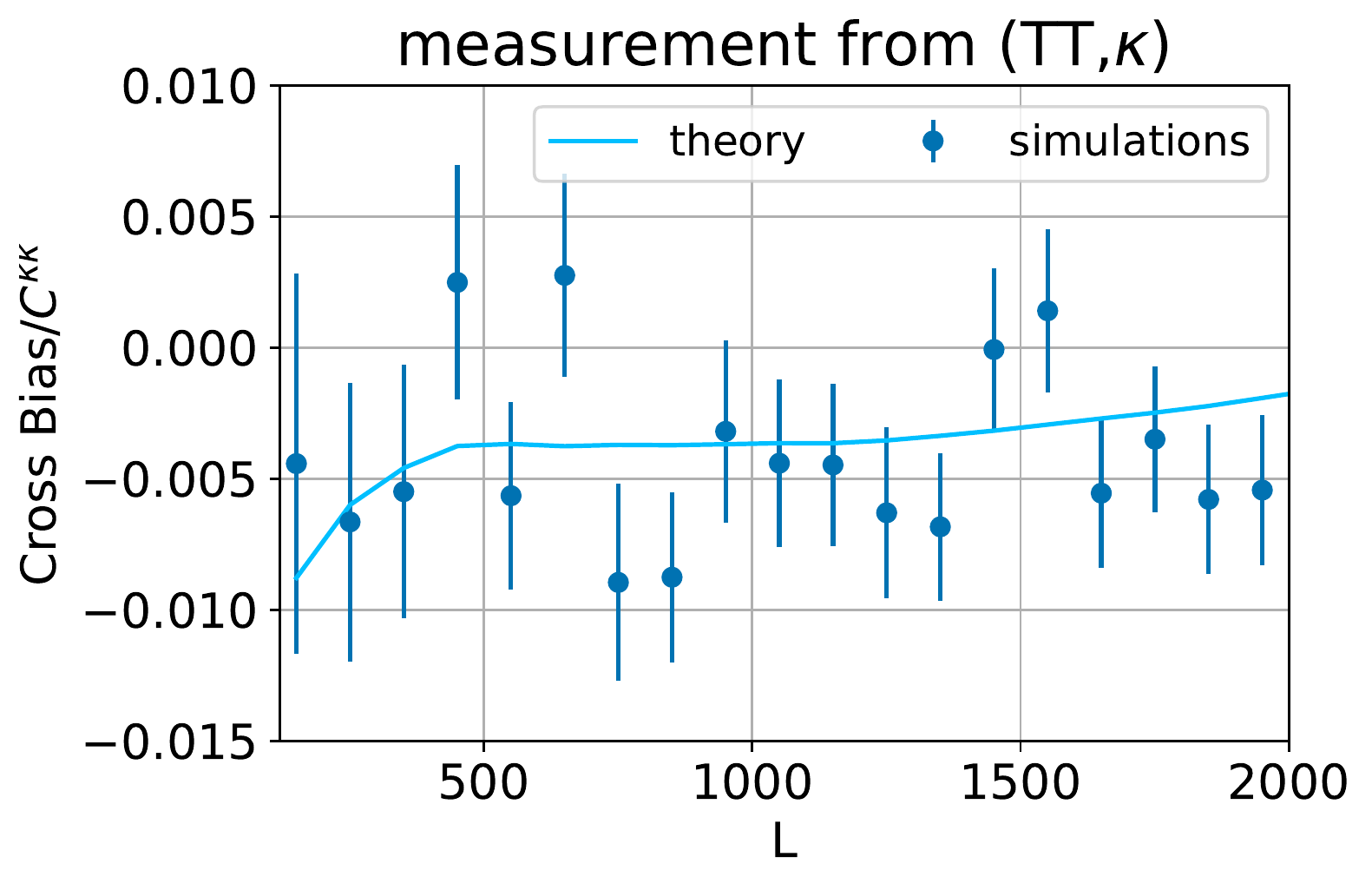}
\caption{\label{fig:N32biasX} We detect a non-Gaussian bias of the size predicted by BSS16 ($0.5\%$ of the signal) in the cross correlation of temperature-based reconstructions and input maps with a significance of $5.2 \sigma$. The p-value of the measurement for a no-bias null-hypothesis is 0.0003. The reduced $\chi^2$ between prediction and measurement is 1.02.}
\end{figure} 
\begin{figure}
\includegraphics[width=0.95\columnwidth]{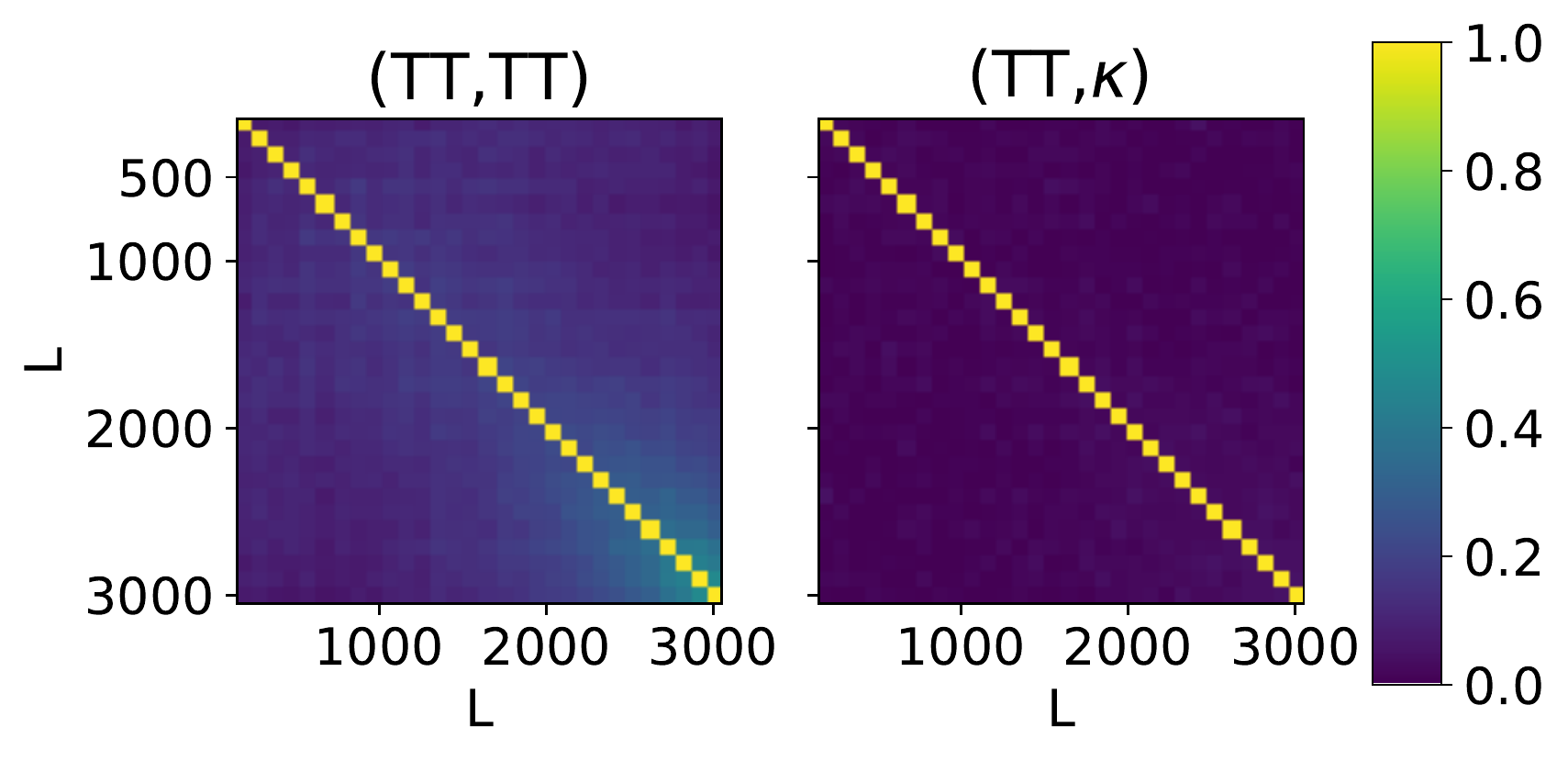}
\caption{\label{fig:cov} Covariance matrices of the power spectrum measurements in units of the variance. The error bars in the measurement of the power in the cross correlation are to good approximation uncorrelated. The measurements of the auto power show some expected degree of correlation between the bins.}
\end{figure} 

In Fig.~\ref{fig:N32biasXfull} we show the measured bias from cross correlating reconstructions from $(TT)$ with the input maps and plot the reconstructed power spectra for comparison.  Fig.~\ref{fig:N32biasX} shows the measured bias in units of the signal. The theory prediction from BSS16 is plotted in light blue for comparison.
The p-value of the data points assuming no-bias is $0.0003$, while a non-Gaussian bias of the size predicted by BSS16 is detected with a significance of $5.2\sigma$. Measuring the covariance between the data points (Fig.~\ref{fig:cov}, right panel), shows that the measurements in different bins can to good approximation be treated as uncorrelated. 

\begin{figure}
\includegraphics[width=0.95\columnwidth]
{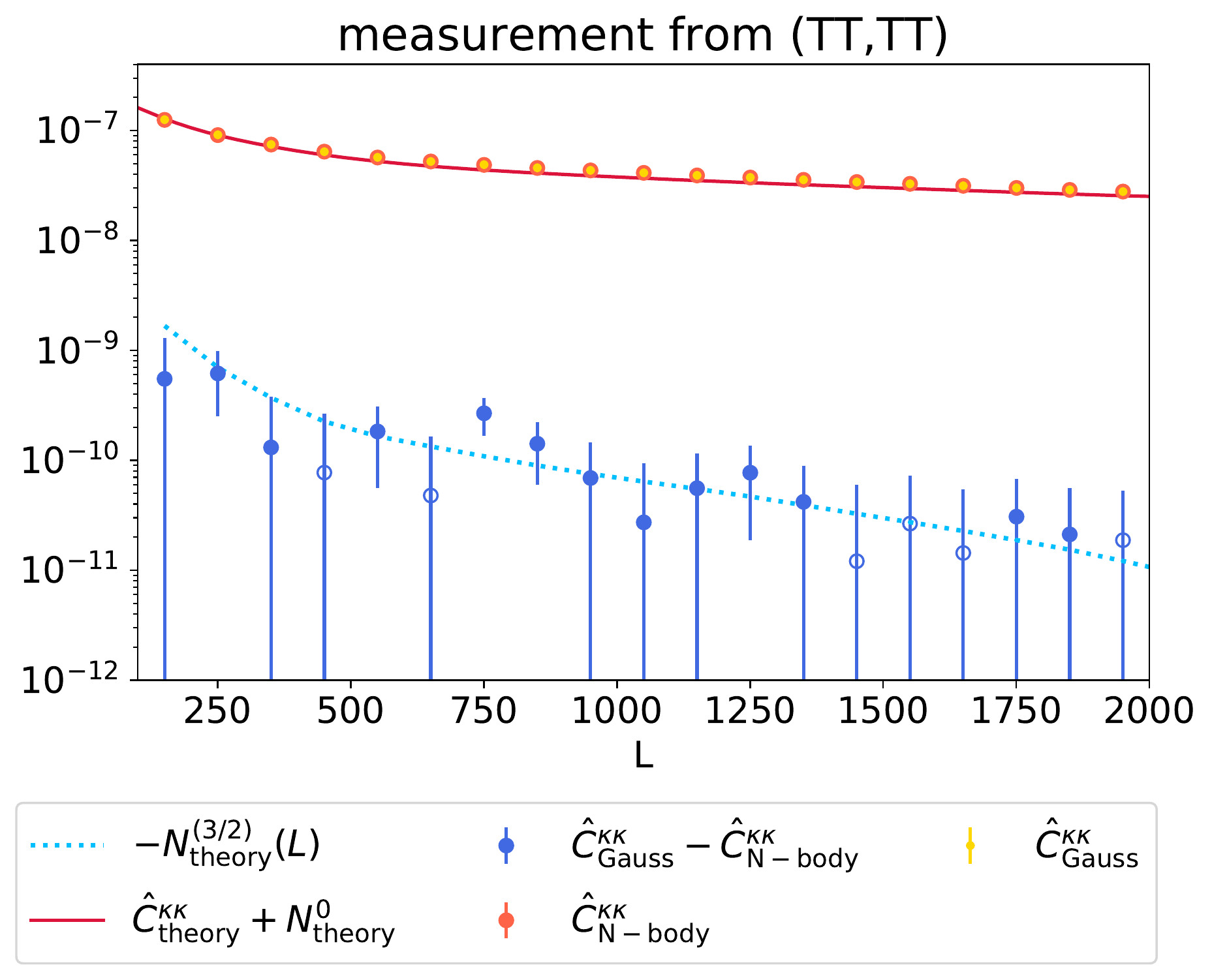}
\caption{\label{fig:N32biasfull} Measured non-Gaussian bias in the CMB lensing power spectrum measurement from the temperature four-point function. The reconstruction agrees well with the theory prediction for the sum of convergence power spectrum and $N^{(0)}$ reconstruction noise. The measured bias is consistent with the theory prediction of BSS16, but the null-hypothesis of no bias cannot be excluded with high statistical significance.}
\end{figure}

\begin{figure}
\includegraphics[width=0.95\columnwidth]
{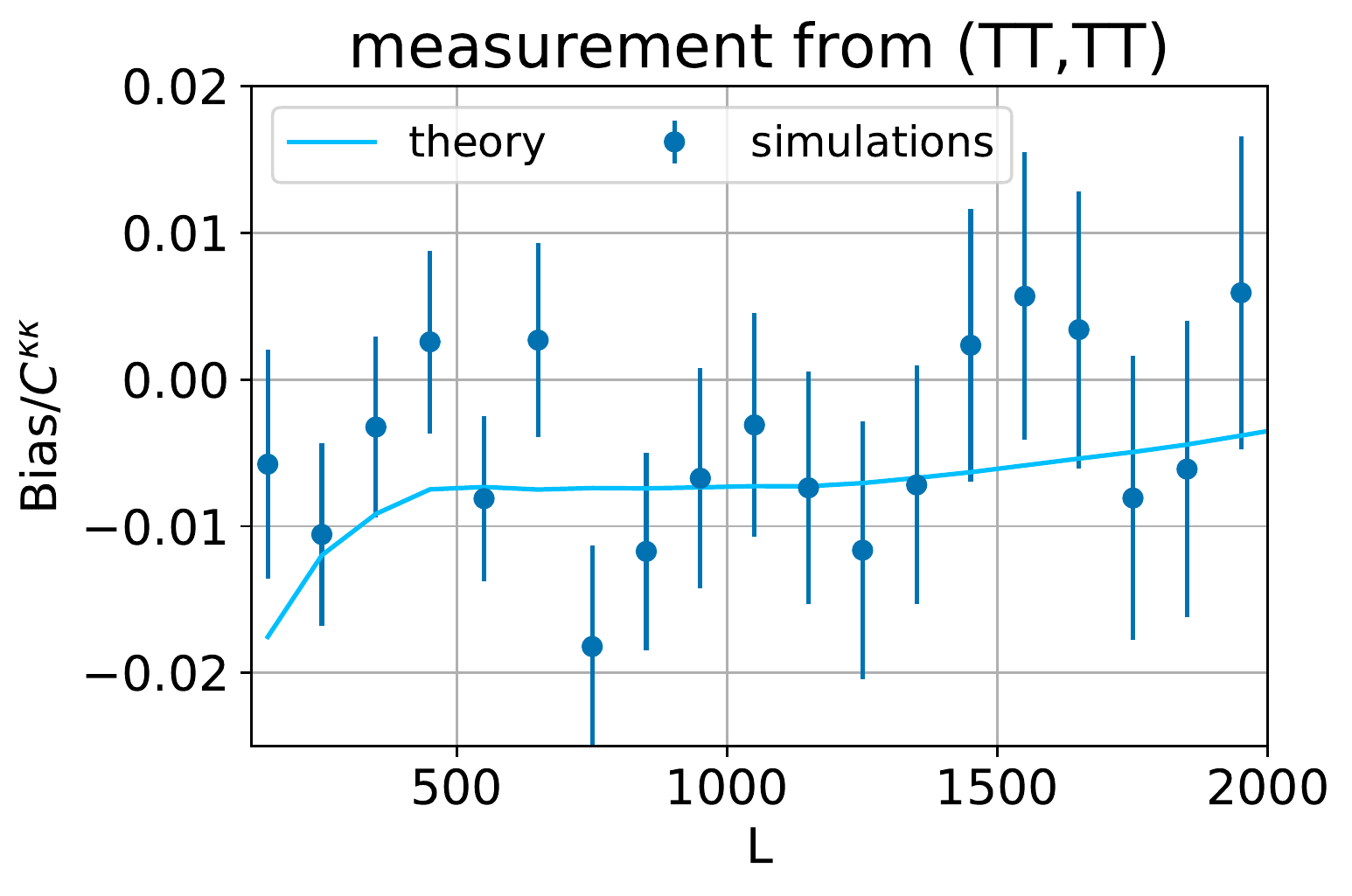}
\caption{\label{fig:N32bias} The non-Gaussian bias in temperature-based CMB lensing power spectrum measurements in units of the signal. The points are consistent with the theory predictions, the bias is detected with a significance of 2.84$\sigma$.}
\end{figure}

The bias in the auto power measured from $(TT,TT)$ is detected with a lower significance of $2.84 \sigma$ (Figs.~\ref{fig:N32biasfull},\ref{fig:N32bias}). The p-value of the measurement for a null-hypothesis of no bias is $p=0.266$. The errors in the auto spectrum are slightly correlated as the left panel of Fig.~\ref{fig:cov} shows. The correlation is below 20\% for most of the bins and below $47\%$ for all bins. This correlation is expected since the covariance of the four-point estimator is non-diagonal~\cite{HansonN2,2013Schmittfull,2017Peloton}.

\begin{figure}
\includegraphics[width=0.95\columnwidth]
{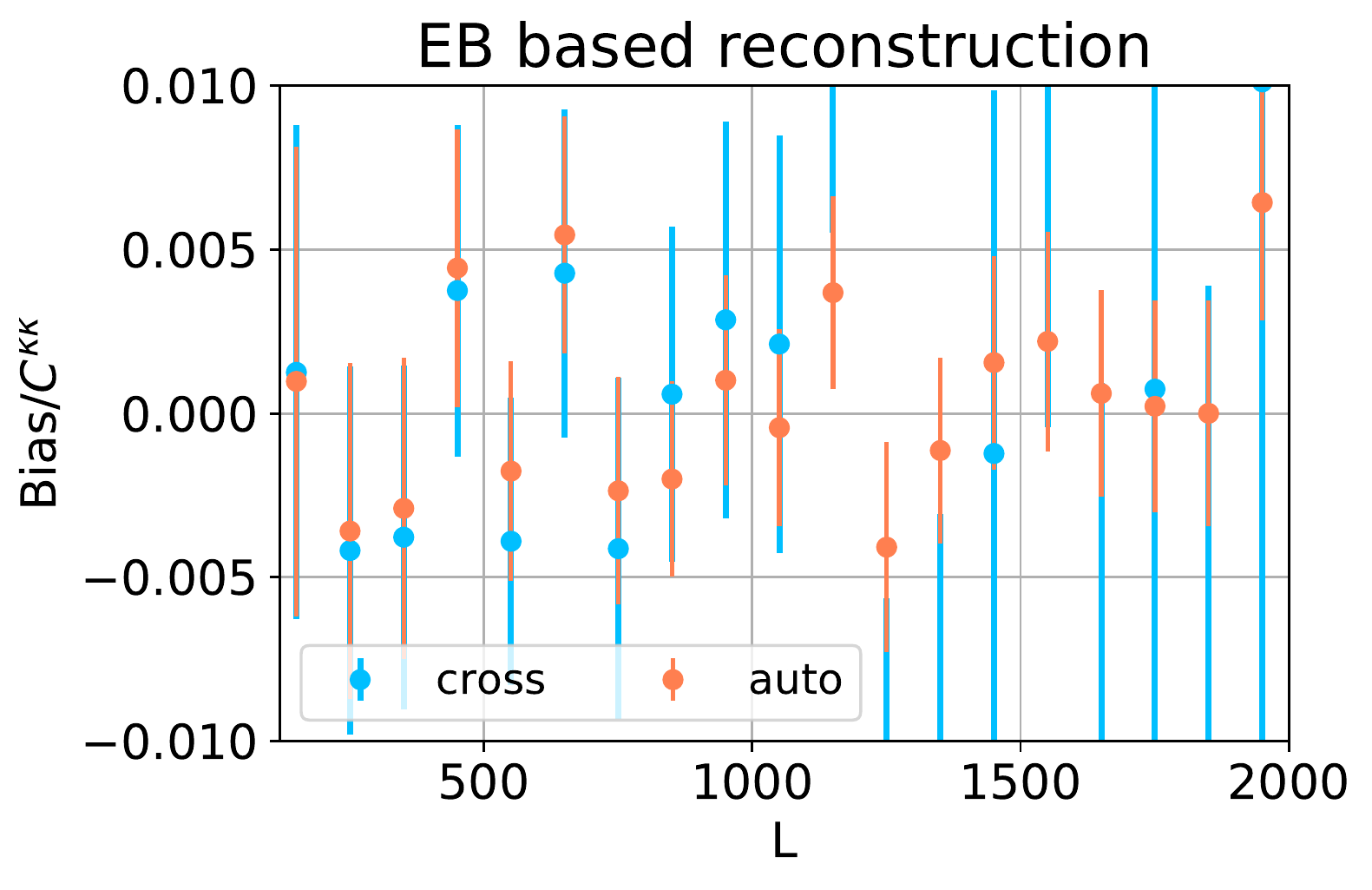}
\caption{\label{fig:N32biasEB} The non-Gaussian bias to CMB lensing measurements from $(EB,EB)$ is consistent with zero in both auto and cross correlations.}
\end{figure}

We do not find any indication for a non-Gaussian bias in the polarization-based reconstruction (Fig.~\ref{fig:N32biasEB}). This agrees with the intuition gained from its functional form: additional angular dependencies (as compared to the bias in temperature-based reconstruction) reduce the support of the contributing integrals (see App.~\ref{sec:N32} and Ref.~\cite{N32}).

\begin{figure}
\includegraphics[width=0.95\columnwidth]
{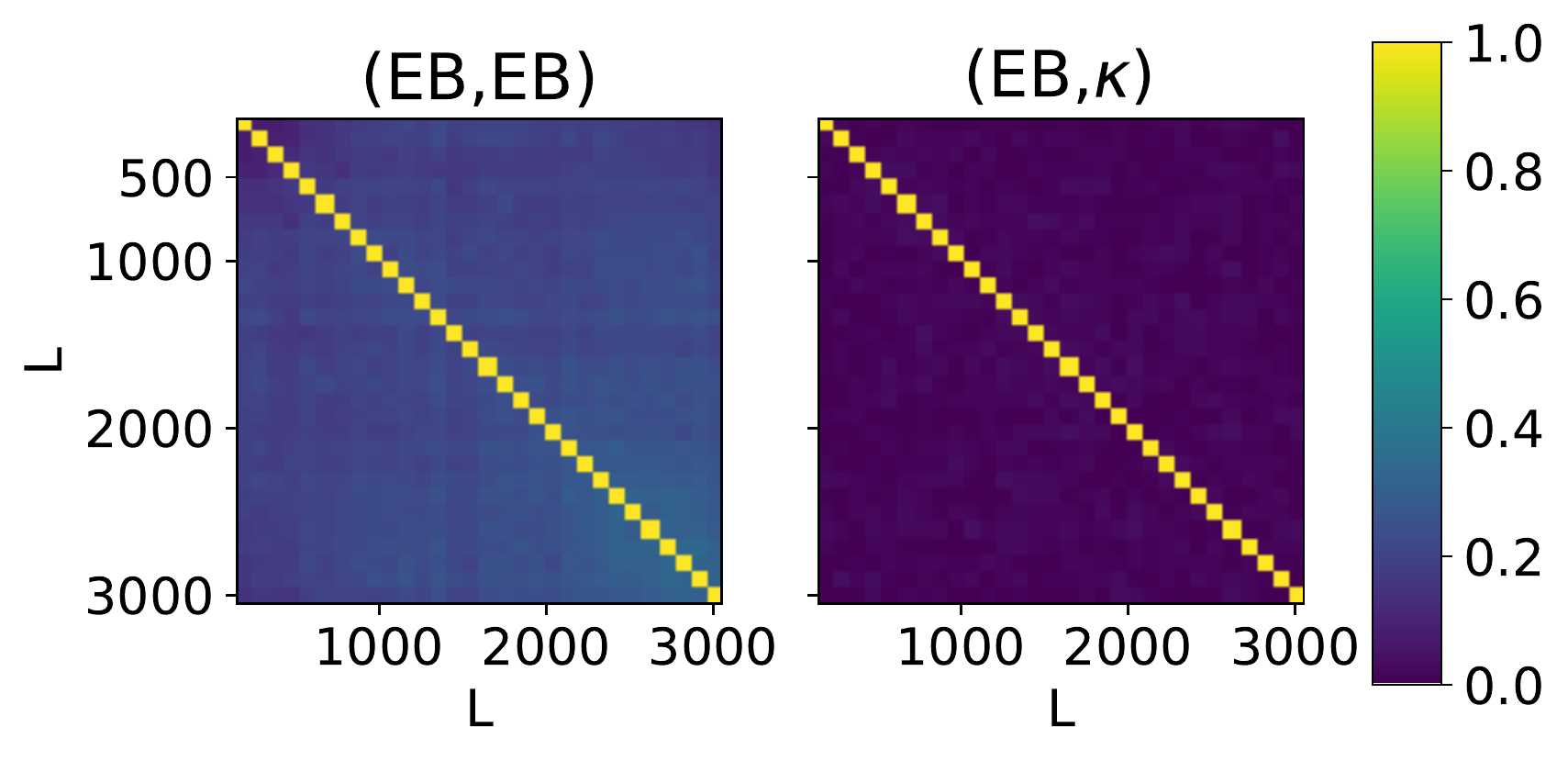}
\caption{\label{fig:covEB} Covariance matrices of the power spectrum measurements from $(EB,EB)$ in units of the variance. The errorbars in the measurement of the power in the cross correlation are to good approximation uncorrelated. The measurements of the auto power show some expected degree of correlation between the bins.}
\end{figure}

Again, we find correlations between the data points in the auto power measurement (Fig.~\ref{fig:covEB}).

\begin{figure}
\includegraphics[width=0.95\columnwidth]{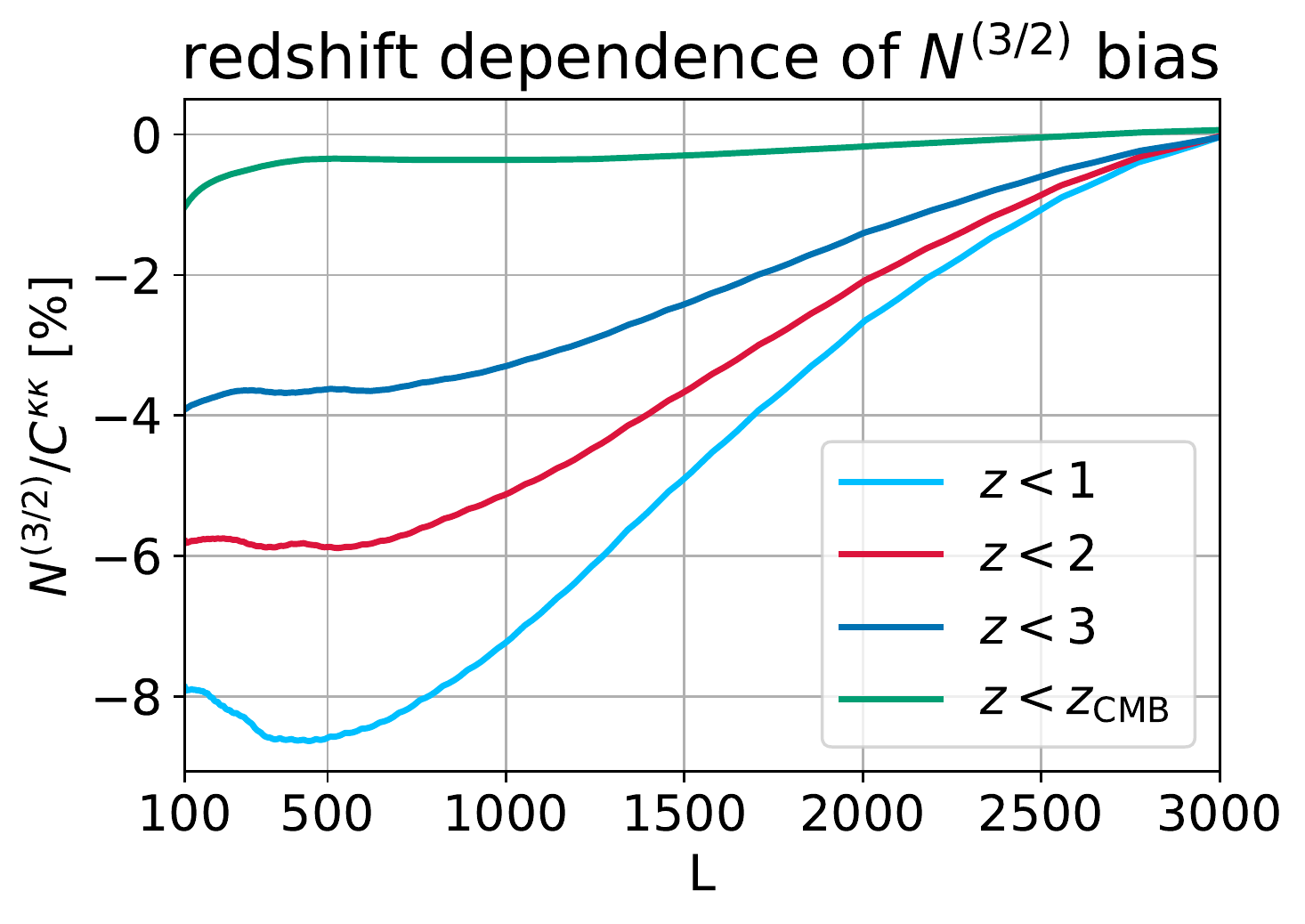}
\caption{\label{fig:n32z} The relative size of the non-Gaussian bias increases if we only consider lenses at low redshifts $z_\max<z_\CMB$. This is a consequence of the different redshift-scalings of the competing terms in the lensing bispectrum from non-linear structure formation and post Born effects. To illustrate this we ignore any contributions to the lensing bispectrum and power spectrum with $z>z_{\max}$ and plot the ratio of the resulting cross bias to the power spectrum. These results suggest that the non-Gaussian bias could be more important for measurements of cross correlations of CMB lensing with low-redshift tracers. We note that the curves shown here are still preliminary and should be seen as a motivation to investigate the bias on cross correlations in future work (B\"ohm et al. in prep).}
\end{figure} 

\section{Discussion}
\label{sec:dis}
By comparing lensing measurements from CMB simulations lensed with Gaussian and non-Gaussian convergence fields, we find strong indication for the existence of a non-Gaussian bias to CMB lensing measurements from temperature data. The bias is at the $ 1\%$ level, which agrees with the theoretical prediction for a bispectrum-induced bias of Ref.~\cite{N32} if we take into account two sources for the lensing bispectrum, non-linear structure formation and multiple correlated deflections.

By measuring the bias in the cross correlations of reconstructed lensing maps with the true underlying lensing fields, we detect the theoretically predicted bias in the simulations at the 5 $\sigma$ significance level. We detect the non-Gaussian bias in the auto correlation with a significance of $\sim 3\sigma$. The measured bias in power spectrum measurements from a combination of E-and B-mode polarization, $(EB,EB)$, is consistent with zero. We note that lensing B-modes at intermediate scales are more sensitive to smaller scales in the deflection field than lensed E-modes or temperature. A non-zero bias in $EB,EB$ could therefore be present in real data, if it was generated by scales that are not accurately modeled in the simulation due to its finite resolution.

We point out that our results have been independently confirmed by Beck et al. 2018 (in prep), who use a completely different simulation set on the full sky.

The good agreement between the simulations and theory suggests that the assumptions that entered into the theory calculation are valid and that we can rely on it to make predictions for different experiments. Theoretical bias predictions for different experimental configuration are shown in Fig.~\ref{fig:n32}. 

The non-Gaussian bias is likely to affect lensing measurements from CMB-experiments that are dominated by temperature reconstruction. This includes current and upcoming experiments such as AdvACT~\cite{AdvACT} and Simons Observatory\footnote{https://simonsobservatory.org/}. An uncorrected non-Gaussian bias at the percent level degrades the accuracy with which these experiments can measure cosmological parameters. The non-Gaussian bias is unlikely to affect experiments that are polarization-dominated, such as the ground based SPT-3G\cite{SPT3G} and CMB-S4 experiments~\cite{CMB-S4} and space-based missions like LiteBird~\cite{2018LiteBird} or Pico~\cite{2018Pico}. 

It is further important to note that the smallness of the bias is a consequence of a somewhat coincidental cancellation: The bias is mostly sensitive to elongated bispectrum configurations. For these shapes, the bispectra from non-linear structure formation and multiple correlated deflections have opposite sign. The fact that they are in addition of similar magnitude is only true for sources at high redshifts. If we consider sources at low redshifts or restrict contributions to the bispectra to low redshifts, we expect this cancellation to be much less efficient. The bias could therefore be more important in cross-correlations of CMB lensing with low-redshift tracers (B\"ohm et al. in prep., Ref.~\cite{2017Merkel}). We illustrate this by plotting preliminary results for the non-Gaussian bias (corrected by a factor of $1/2$ as it is expected for cross correlations) in units of the CMB lensing signal when only allowing lenses at low redshift to contribute to both ($z_\max \leq z_\CMB, z_\source=z_\CMB$) in Fig.~\ref{fig:n32z}. These results suggest that the bias could be of the order of several percent for cross correlation measurements from temperature data.
With these measurements getting most of their signal from high multipoles where the $TT$ estimator performs best, this could make  this bias relevant for most future wide-field surveys. We caution at this point that Fig.~\ref{fig:n32z} should only be seen as a motivation to investigate the non-linear bias for cross correlations. By setting all contribution to the post-Born bispectrum above a certain redshift to zero, it becomes negligible. For realistic cross correlations the expression for the post-Born bispectrum is more complicated and its contribution to the bias could be more important.

The results shown in this work only apply to power spectrum estimates with a quadratic estimator but similar biases could arise for alternative estimators if they are derived under the assumption of a Gaussian deflection field. 

Recently, Ref.~\cite{SchaanFerr2018_2} pointed out that a shear estimator~\cite{Bucher2012,Prince2018} is to good approximation robust against contamination from isotropic foregrounds (at the cost of lower signal to noise in the reconstruction). The fact that we find no bias in the reconstruction from $EB,EB$, for which the quadratic estimator corresponds to a shear estimator, suggests that a shear-only estimator could also be less sensitive to the non-linear bias (this can also be seen analytically, since the shear estimator has an additional angular dependence, which should lead to additional cancellations in the bias integrals in Eq.~\ref{biasterms}) This possibility could be easily tested on simulations and could be explored in future work. We note, however, that since the non-linear bias can be modeled theoretically, a mitigation at the cost of lower signal-to-noise is not crucial. 

\section*{Acknowledgements}
We thank Dominic Beck and Giulio Fabbian for cross checking some of the results with their simulations and Antony Lewis and Geraint Pratten for sharing their numerical implementations of post Born terms. VB thanks Emmanuel Schaan, Dominic Beck and Giulio Fabbian for useful discussions. BDS was supported by an STFC Ernest Rutherford Fellowship and an Isaac Newton Trust Early Career Grant. JL is supported by an NSF Astronomy and Astrophysics Postdoctoral Fellowship under award AST-1602663. JCH is supported by the Friends of the Institute for Advanced Study. MS was supported by the Bezos fund. This work used the Extreme Science and Engineering Discovery Environment (XSEDE), which is supported by NSF grant ACI-1053575. This research used resources of the National Energy Research Scientific Computing Center, a DOE Office of Science User Facility supported by the Office of Science of the U.S. Department of Energy under Contract No. DE-AC02-05CH11231. Part of this work used computational resources at the Max Planck Computing and Data Facility (MPCDF).

\appendix
\section{Analytic prediction for a bispectrum-induced CMB lensing bias}
\label{sec:N32}
\begin{figure}
\includegraphics[width=0.95\columnwidth]{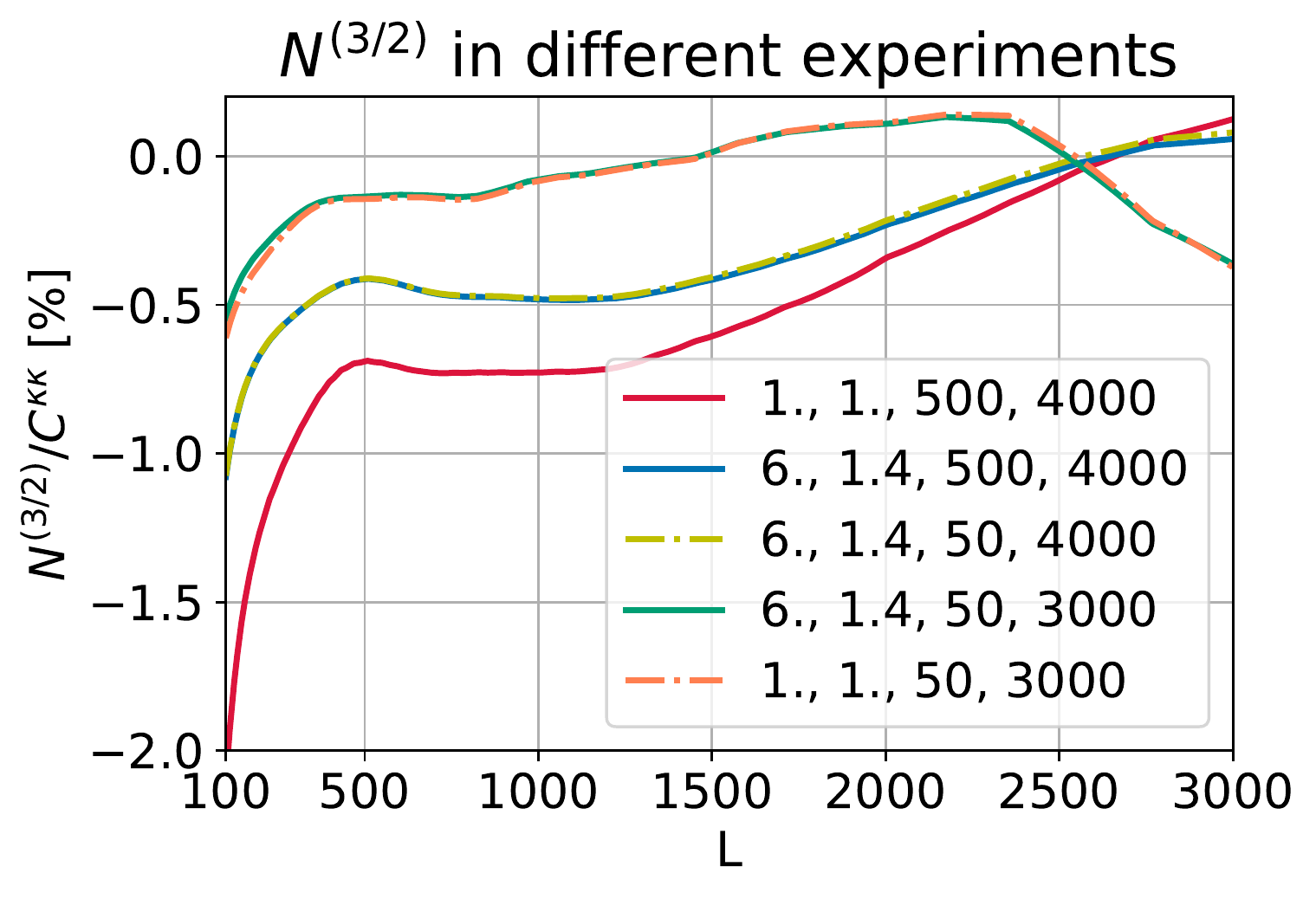}
\caption{\label{fig:n32} Bispectrum-induced $N^{(3/2)}$ bias in temperature-based measurements for different experimental configurations. The lines are labeled by noise in $\mu K{-}\arcmin$, beam FWHM in arcmin, $l_{\min}$ and $l_{\max}$. The size of the bias is sensitive to the maximal, signal-dominated CMB scale that is used in the reconstruction.}
\end{figure} 

All CMB lensing analyses to date assume that the lensing convergence is a Gaussian field. However, non-linear structure formation and multiple correlated lenses introduce a small, but detectable amount of non-Gaussianity~\cite{BkkkNamikawa,postBornPratten,postBornMarozzi,JiaPeaks2016}. In the limit of small density perturbations, the non-Gaussian structure can be characterized by a hierarchy of connected correlation functions. To lowest order, the lensing convergence acquires a bispectrum. 

The lensing bispectrum introduces an additional term to the standard four-point estimator (compare Eq.~\ref{bias_exp_N32})
This new bias was first identified in Ref.~\cite{N32} (BSS16). Its name follows from the naming convention for CMB lensing biases, where biases are labeled by their power in the lensing power spectrum. The $N^{3/2}$ bias arises because the lensing bispectrum changes the lensed temperature four-point function. 

BSS16 found that the $N^{3/2}$ bias can change the measured lensing power spectrum in temperature-based CMB lensing analyses at the percent level. This corresponds to a 1-2 $\sigma$ effect (per L-bin) for current and upcoming CMB experiments.

The estimation of the size of the bias in BSS16 is based on the numerical evaluation of analytically derived expressions. This evaluation relies on a number of assumptions:
\begin{enumerate}
\item The lensing bispectrum contributes to the lensed temperature four-point function with 8 terms. Due to the complicated structure of these terms (they involve 6-dimensional coupled integrals over reconstruction weights $g$, the lensing bispectrum and CMB power spectra.), only two of these terms were evaluated. These two terms were chosen because they factor maximally under the reconstruction weights (one of them can even be split into a product of 3 two-dimensional integrals). Their structure suggests that these terms are the dominant contributions to the bispectrum-induced bias. 
For temperature-only reconstruction, the two terms read
\begin{align}
\nonumber
N^{(3/2)}_{1}(L) = -4 A_L^2  S_L \int_{\vl_1,\vl} g_{\vl_1,\VL}
 \[\vl\cdot(\vl_1{-}\vl)\]\\
 \nonumber
 \times  \[\vl\cdot(\VL{-}(\vl_1{-}\vl))\] C^{TT}_l\, B_\phi\[\vl_1{-}\vl,\VL-(\vl_1{-}\vl),{-}\VL\]  \\
 \nonumber
N^{(3/2)}_{2}(L) = 4 A_L^2 S_L\int_{\vl_1,\vl} g_{\vl_1,\VL}\,(\vl_1 \cdot \vl ) \left[\vl_1 \cdot(\VL{-}\vl) \right]\\
 \times C_{l_1}^{TT} B_\phi(\vl,\VL{-}\vl,-\VL),
 \label{biasterms}
\end{align}
with
\beq
S_L = \int_{\vl_2} g_{\vl_2,\VL} (\vl_2\cdot\VL) C_{l_2}^{TT}\approx \frac{1}{2}A_L^{-1}.
\eeq
For polarization-based reconstruction, the structure of the terms is similar, but with additional angular dependencies (see Ref.~\cite{N32} for details).
In cross-correlations, all other terms vanish and the bias depends only on the two terms above.

\item The evaluation in BSS16 only considered non-linear structure formation as a source of the lensing bispectrum. Recently Ref.~\cite{postBornPratten} pointed out that an additional lensing bispectrum arises from multiple correlated lensing deflections. The effect of multiple deflections is commonly ignored in the Born approximation. Both effects, post Born corrections and non-linear structure formation, lead to lensing bispectra of the same order of magnitude, but for certain triangle configurations of opposite sign.
\item The modeling of the bispectrum from non-linear structure formation in BSS16 relied on tree-level perturbation theory, which breaks down on small scales (and the bias was shown to be sensitive to replacing the linear matter power spectrum by its non-linear (HALOFIT~\cite{HALOFITT}) counterpart in the matter bispectrum model).
\item The theoretical modeling of the bias relied on a Taylor series expansion of the lensed CMB in the deflection angle, and thus on the assumption of small deflection angles.
\end{enumerate}

In this work, we use an updated analytical prediction for the bias, which still assumes that all but two terms are negligible, but that takes into account the bispectrum from post-Born effects and uses an extended, simulation-calibrated, semi-analytic model for the matter bispectrum~\cite{GilMarin2011}. Updated theoretical results are shown in Fig.~\ref{fig:n32} for different experimental set-ups and together with the measurement of the bias in Sec.~\ref{sec:res}.

\section{Non-linear bias from bispectra involving the curl of the lensing deflection}
\label{sec:curl}
Allowing for multiple deflections introduces an additional degree of freedom, $\omega$, to the linear mapping between the lensed and unlensed image of a source, which describes a rotation of the image~\cite{1996Stebbins}. With this additional dof, the lensing deflection angle is no longer a pure gradient field, but acquires an additional curl component,
\beq
\alpha(\vx)=\nabla \phi(\vx) + * \nabla \Omega(\vx),
\eeq
sourced by the curl potential $\Omega$~\cite{2003PhRvDHirata}. We use a $*$ to denote a rotation by 90 degrees, and, for notational simplicity, also abbreviate the combination of rotation and scalar product, $\cdot *$, in the following by $*$.

Being second order in the gravitational potential, the rotation is suppressed compared to the first order convergence and shear distortions to the image. We thus expect the largest bias that involves the curl potential to be sourced by a "cross" bispectrum of the form $B^{\Omega,\phi,\phi}(\VL,\vl, -\VL-\vl)$~\cite{postBornPratten}. The curl potential can be treated in complete analogy to the scalar lensing potential $\phi$. E.g., when expressing the effect of lensing on the CMB in terms of a small perturbation to the unlensed CMB, we can write~\cite{2005PhRvDCooray}
\beq
\label{OmegaLens}
\tilde{T}=T+\delta_\Omega T+\delta_\phi T+\delta^2_\Omega T + \delta^2_\phi T + \mathcal{O}(\phi^3,\Omega^3).
\eeq
Adapting the flat sky approximation, the first two terms are given in harmonic space by
\begin{align}
\delta_\Omega T(\vl) &=\int_{\vl'} \[\vl'*(\vl'-\vl)\] T(\vl') \Omega(\vl-\vl')\\
\delta_\phi T(\vl)  &=\int_{\vl'} \[\vl'\cdot(\vl'-\vl)\] T(\vl') \phi(\vl-\vl').
\end{align}
Using this perturbative framework to model the lensed temperature 4-point function, Ref.~\cite{N32} show that the two dominant terms in the $N^{(3/2)}$ bias are sourced by contractions of the following expectation values over $\phi$ and $T$~\cite{N32}\footnote{By contractions we mean the terms that arise from taking the expectation value over unlensed CMB realizations. Assuming that the unlensed CMB is Gaussian, each expectation value can be split into a sum of three terms. We only consider the bias arising from one of these three terms, which BSS16 identified as the dominant one. Also, for readability, we do not write the symmetry factors that arise from permutations of $T,\delta T$ and $\delta^2 T$ that leave the result invariant.}
\begin{align}
\nonumber
N^{(3/2)}_1 \[\phi^3\] &\leftarrow \langle \delta_\phi T \delta_\phi T \delta_\phi T' T' \rangle \\
N^{(3/2)}_2  \[\phi^3\] &\leftarrow \langle \delta_\phi T T \delta_\phi^2 T' T' \rangle.
\end{align}
A bias sourced by the cross bispectrum $B^{\Omega,\kappa,\kappa}(\VL,\vl, -\VL-\vl)$ (we refer to it as $\tilde{N}^{(3/2)}$) should therefore be dominated by contractions of the following expectation values
\begin{align}
\nonumber
\tilde{N}^{(3/2)}_1  \[\phi^2 \Omega\] & \leftarrow \langle \delta_\phi T \delta_\phi T \delta_\Omega T' T' \rangle_{1a} + \langle \delta_\Omega T \delta_\phi T \delta_\phi T' T' \rangle_{1b} \\
\tilde{N}^{(3/2)}_2  \[\phi^2 \Omega\] & \leftarrow \langle \delta_\Omega T T \delta_\phi^2 T' T' \rangle_{2}.
\end{align}
The expressions for the dominant contractions arising from $1a$ and $2$ are identical to the auto bias (Eq.~\ref{biasterms}), but with $B^{\phi^3}$ replaced by $B^{\phi^2\Omega}$ and $S_L$ replaced by 
\beq
\label{Scross}
S^{\times}_L = \int_{\vl_2} g_{\vl_2,\VL} (\vl_2 * \VL) C_{l_2}^{TT}=0.
\eeq
This integral vanishes because the integrand is uneven under the angular integration.
The remaining dominant contraction from $1b$ is of the form
\begin{align}
\nonumber
\tilde{N}^{(3/2)}_{1b}(L) = -4 A_L^2  S_L \int_{\vl_1,\vl} g_{\vl_1,\VL}
 \[\vl* \vl_1\]\\
\times  \[\vl\cdot(\VL{-}(\vl_1{-}\vl))\] C^{TT}_l\, B_{\phi,\phi,\Omega}\[\vl_1{-}\vl,\VL-(\vl_1{-}\vl),{-}\VL\].
\label{1b}
\end{align}
Because of the mixing of sines and cosines in the angular integrations in Eq.~\ref{1b}, we expect this contribution to be strongly suppressed compared to the corresponding term in the bias from the auto bispectrum.

This short calculation suggests that biases from bispectra involving the curl component are likely to be negligible for current and upcoming CMB experiments. 

\bibliography{CMBLens,CosmoSoft,LSS}

\begin{thebibliography}{66}%
\makeatletter
\providecommand \@ifxundefined [1]{%
 \@ifx{#1\undefined}
}%
\providecommand \@ifnum [1]{%
 \ifnum #1\expandafter \@firstoftwo
 \else \expandafter \@secondoftwo
 \fi
}%
\providecommand \@ifx [1]{%
 \ifx #1\expandafter \@firstoftwo
 \else \expandafter \@secondoftwo
 \fi
}%
\providecommand \natexlab [1]{#1}%
\providecommand \enquote  [1]{``#1''}%
\providecommand \bibnamefont  [1]{#1}%
\providecommand \bibfnamefont [1]{#1}%
\providecommand \citenamefont [1]{#1}%
\providecommand \href@noop [0]{\@secondoftwo}%
\providecommand \href [0]{\begingroup \@sanitize@url \@href}%
\providecommand \@href[1]{\@@startlink{#1}\@@href}%
\providecommand \@@href[1]{\endgroup#1\@@endlink}%
\providecommand \@sanitize@url [0]{\catcode `\\12\catcode `\$12\catcode
  `\&12\catcode `\#12\catcode `\^12\catcode `\_12\catcode `\%12\relax}%
\providecommand \@@startlink[1]{}%
\providecommand \@@endlink[0]{}%
\providecommand \url  [0]{\begingroup\@sanitize@url \@url }%
\providecommand \@url [1]{\endgroup\@href {#1}{\urlprefix }}%
\providecommand \urlprefix  [0]{URL }%
\providecommand \Eprint [0]{\href }%
\providecommand \doibase [0]{http://dx.doi.org/}%
\providecommand \selectlanguage [0]{\@gobble}%
\providecommand \bibinfo  [0]{\@secondoftwo}%
\providecommand \bibfield  [0]{\@secondoftwo}%
\providecommand \translation [1]{[#1]}%
\providecommand \BibitemOpen [0]{}%
\providecommand \bibitemStop [0]{}%
\providecommand \bibitemNoStop [0]{.\EOS\space}%
\providecommand \EOS [0]{\spacefactor3000\relax}%
\providecommand \BibitemShut  [1]{\csname bibitem#1\endcsname}%
\let\auto@bib@innerbib\@empty
\bibitem [{\citenamefont {{Lewis}}\ and\ \citenamefont
  {{Challinor}}(2006)}]{CMBLensRev1}%
  \BibitemOpen
  \bibfield  {author} {\bibinfo {author} {\bibfnamefont {A.}~\bibnamefont
  {{Lewis}}}\ and\ \bibinfo {author} {\bibfnamefont {A.}~\bibnamefont
  {{Challinor}}},\ }\href {\doibase 10.1016/j.physrep.2006.03.002} {\bibfield
  {journal} {\bibinfo  {journal} {\physrep}\ }\textbf {\bibinfo {volume}
  {429}},\ \bibinfo {pages} {1} (\bibinfo {year} {2006})},\ \Eprint
  {http://arxiv.org/abs/astro-ph/0601594} {astro-ph/0601594} \BibitemShut
  {NoStop}%
\bibitem [{\citenamefont {{Hanson}}\ \emph
  {et~al.}(2010{\natexlab{a}})\citenamefont {{Hanson}}, \citenamefont
  {{Challinor}},\ and\ \citenamefont {{Lewis}}}]{CMBLensRev2}%
  \BibitemOpen
  \bibfield  {author} {\bibinfo {author} {\bibfnamefont {D.}~\bibnamefont
  {{Hanson}}}, \bibinfo {author} {\bibfnamefont {A.}~\bibnamefont
  {{Challinor}}}, \ and\ \bibinfo {author} {\bibfnamefont {A.}~\bibnamefont
  {{Lewis}}},\ }\href {\doibase 10.1007/s10714-010-1036-y} {\bibfield
  {journal} {\bibinfo  {journal} {General Relativity and Gravitation}\ }\textbf
  {\bibinfo {volume} {42}},\ \bibinfo {pages} {2197} (\bibinfo {year}
  {2010}{\natexlab{a}})},\ \Eprint {http://arxiv.org/abs/0911.0612}
  {arXiv:0911.0612} \BibitemShut {NoStop}%
\bibitem [{\citenamefont {{Lesgourgues}}\ \emph {et~al.}(2006)\citenamefont
  {{Lesgourgues}}, \citenamefont {{Perotto}}, \citenamefont {{Pastor}},\ and\
  \citenamefont {{Piat}}}]{2006PhRvDLes}%
  \BibitemOpen
  \bibfield  {author} {\bibinfo {author} {\bibfnamefont {J.}~\bibnamefont
  {{Lesgourgues}}}, \bibinfo {author} {\bibfnamefont {L.}~\bibnamefont
  {{Perotto}}}, \bibinfo {author} {\bibfnamefont {S.}~\bibnamefont {{Pastor}}},
  \ and\ \bibinfo {author} {\bibfnamefont {M.}~\bibnamefont {{Piat}}},\ }\href
  {\doibase 10.1103/PhysRevD.73.045021} {\bibfield  {journal} {\bibinfo
  {journal} {\prd}\ }\textbf {\bibinfo {volume} {73}},\ \bibinfo {eid} {045021}
  (\bibinfo {year} {2006})},\ \Eprint {http://arxiv.org/abs/astro-ph/0511735}
  {astro-ph/0511735} \BibitemShut {NoStop}%
\bibitem [{\citenamefont {{Sherwin}}\ \emph {et~al.}(2011)\citenamefont
  {{Sherwin}}, \citenamefont {{Dunkley}}, \citenamefont {{Das}}, \citenamefont
  {{Appel}}, \citenamefont {{Bond}}, \citenamefont {{Carvalho}}, \citenamefont
  {{Devlin}}, \citenamefont {{D{\"u}nner}}, \citenamefont {{Essinger-Hileman}},
  \citenamefont {{Fowler}}, \citenamefont {{Hajian}}, \citenamefont
  {{Halpern}}, \citenamefont {{Hasselfield}}, \citenamefont {{Hincks}},
  \citenamefont {{Hlozek}}, \citenamefont {{Hughes}}, \citenamefont {{Irwin}},
  \citenamefont {{Klein}}, \citenamefont {{Kosowsky}}, \citenamefont
  {{Marriage}}, \citenamefont {{Marsden}}, \citenamefont {{Moodley}},
  \citenamefont {{Menanteau}}, \citenamefont {{Niemack}}, \citenamefont
  {{Nolta}}, \citenamefont {{Page}}, \citenamefont {{Parker}}, \citenamefont
  {{Reese}}, \citenamefont {{Schmitt}}, \citenamefont {{Sehgal}}, \citenamefont
  {{Sievers}}, \citenamefont {{Spergel}}, \citenamefont {{Staggs}},
  \citenamefont {{Swetz}}, \citenamefont {{Switzer}}, \citenamefont
  {{Thornton}}, \citenamefont {{Visnjic}},\ and\ \citenamefont
  {{Wollack}}}]{2011Sherwin}%
  \BibitemOpen
  \bibfield  {author} {\bibinfo {author} {\bibfnamefont {B.~D.}\ \bibnamefont
  {{Sherwin}}}, \bibinfo {author} {\bibfnamefont {J.}~\bibnamefont
  {{Dunkley}}}, \bibinfo {author} {\bibfnamefont {S.}~\bibnamefont {{Das}}},
  \bibinfo {author} {\bibfnamefont {J.~W.}\ \bibnamefont {{Appel}}}, \bibinfo
  {author} {\bibfnamefont {J.~R.}\ \bibnamefont {{Bond}}}, \bibinfo {author}
  {\bibfnamefont {C.~S.}\ \bibnamefont {{Carvalho}}}, \bibinfo {author}
  {\bibfnamefont {M.~J.}\ \bibnamefont {{Devlin}}}, \bibinfo {author}
  {\bibfnamefont {R.}~\bibnamefont {{D{\"u}nner}}}, \bibinfo {author}
  {\bibfnamefont {T.}~\bibnamefont {{Essinger-Hileman}}}, \bibinfo {author}
  {\bibfnamefont {J.~W.}\ \bibnamefont {{Fowler}}}, \bibinfo {author}
  {\bibfnamefont {A.}~\bibnamefont {{Hajian}}}, \bibinfo {author}
  {\bibfnamefont {M.}~\bibnamefont {{Halpern}}}, \bibinfo {author}
  {\bibfnamefont {M.}~\bibnamefont {{Hasselfield}}}, \bibinfo {author}
  {\bibfnamefont {A.~D.}\ \bibnamefont {{Hincks}}}, \bibinfo {author}
  {\bibfnamefont {R.}~\bibnamefont {{Hlozek}}}, \bibinfo {author}
  {\bibfnamefont {J.~P.}\ \bibnamefont {{Hughes}}}, \bibinfo {author}
  {\bibfnamefont {K.~D.}\ \bibnamefont {{Irwin}}}, \bibinfo {author}
  {\bibfnamefont {J.}~\bibnamefont {{Klein}}}, \bibinfo {author} {\bibfnamefont
  {A.}~\bibnamefont {{Kosowsky}}}, \bibinfo {author} {\bibfnamefont {T.~A.}\
  \bibnamefont {{Marriage}}}, \bibinfo {author} {\bibfnamefont
  {D.}~\bibnamefont {{Marsden}}}, \bibinfo {author} {\bibfnamefont
  {K.}~\bibnamefont {{Moodley}}}, \bibinfo {author} {\bibfnamefont
  {F.}~\bibnamefont {{Menanteau}}}, \bibinfo {author} {\bibfnamefont {M.~D.}\
  \bibnamefont {{Niemack}}}, \bibinfo {author} {\bibfnamefont {M.~R.}\
  \bibnamefont {{Nolta}}}, \bibinfo {author} {\bibfnamefont {L.~A.}\
  \bibnamefont {{Page}}}, \bibinfo {author} {\bibfnamefont {L.}~\bibnamefont
  {{Parker}}}, \bibinfo {author} {\bibfnamefont {E.~D.}\ \bibnamefont
  {{Reese}}}, \bibinfo {author} {\bibfnamefont {B.~L.}\ \bibnamefont
  {{Schmitt}}}, \bibinfo {author} {\bibfnamefont {N.}~\bibnamefont {{Sehgal}}},
  \bibinfo {author} {\bibfnamefont {J.}~\bibnamefont {{Sievers}}}, \bibinfo
  {author} {\bibfnamefont {D.~N.}\ \bibnamefont {{Spergel}}}, \bibinfo {author}
  {\bibfnamefont {S.~T.}\ \bibnamefont {{Staggs}}}, \bibinfo {author}
  {\bibfnamefont {D.~S.}\ \bibnamefont {{Swetz}}}, \bibinfo {author}
  {\bibfnamefont {E.~R.}\ \bibnamefont {{Switzer}}}, \bibinfo {author}
  {\bibfnamefont {R.}~\bibnamefont {{Thornton}}}, \bibinfo {author}
  {\bibfnamefont {K.}~\bibnamefont {{Visnjic}}}, \ and\ \bibinfo {author}
  {\bibfnamefont {E.}~\bibnamefont {{Wollack}}},\ }\href {\doibase
  10.1103/PhysRevLett.107.021302} {\bibfield  {journal} {\bibinfo  {journal}
  {Physical Review Letters}\ }\textbf {\bibinfo {volume} {107}},\ \bibinfo
  {eid} {021302} (\bibinfo {year} {2011})},\ \Eprint
  {http://arxiv.org/abs/1105.0419} {arXiv:1105.0419 [astro-ph.CO]} \BibitemShut
  {NoStop}%
\bibitem [{\citenamefont {{Smith}}\ \emph {et~al.}(2007)\citenamefont
  {{Smith}}, \citenamefont {{Zahn}},\ and\ \citenamefont
  {{Dor{\'e}}}}]{2007Smith}%
  \BibitemOpen
  \bibfield  {author} {\bibinfo {author} {\bibfnamefont {K.~M.}\ \bibnamefont
  {{Smith}}}, \bibinfo {author} {\bibfnamefont {O.}~\bibnamefont {{Zahn}}}, \
  and\ \bibinfo {author} {\bibfnamefont {O.}~\bibnamefont {{Dor{\'e}}}},\
  }\href {\doibase 10.1103/PhysRevD.76.043510} {\bibfield  {journal} {\bibinfo
  {journal} {\prd}\ }\textbf {\bibinfo {volume} {76}},\ \bibinfo {eid} {043510}
  (\bibinfo {year} {2007})},\ \Eprint {http://arxiv.org/abs/0705.3980}
  {arXiv:0705.3980} \BibitemShut {NoStop}%
\bibitem [{\citenamefont {{Hirata}}\ \emph {et~al.}(2008)\citenamefont
  {{Hirata}}, \citenamefont {{Ho}}, \citenamefont {{Padmanabhan}},
  \citenamefont {{Seljak}},\ and\ \citenamefont {{Bahcall}}}]{2008Hirata}%
  \BibitemOpen
  \bibfield  {author} {\bibinfo {author} {\bibfnamefont {C.~M.}\ \bibnamefont
  {{Hirata}}}, \bibinfo {author} {\bibfnamefont {S.}~\bibnamefont {{Ho}}},
  \bibinfo {author} {\bibfnamefont {N.}~\bibnamefont {{Padmanabhan}}}, \bibinfo
  {author} {\bibfnamefont {U.}~\bibnamefont {{Seljak}}}, \ and\ \bibinfo
  {author} {\bibfnamefont {N.~A.}\ \bibnamefont {{Bahcall}}},\ }\href {\doibase
  10.1103/PhysRevD.78.043520} {\bibfield  {journal} {\bibinfo  {journal}
  {\prd}\ }\textbf {\bibinfo {volume} {78}},\ \bibinfo {eid} {043520} (\bibinfo
  {year} {2008})},\ \Eprint {http://arxiv.org/abs/0801.0644} {arXiv:0801.0644}
  \BibitemShut {NoStop}%
\bibitem [{\citenamefont {{Das}}\ \emph {et~al.}(2011)\citenamefont {{Das}},
  \citenamefont {{Sherwin}}, \citenamefont {{Aguirre}}, \citenamefont
  {{Appel}}, \citenamefont {{Bond}}, \citenamefont {{Carvalho}}, \citenamefont
  {{Devlin}}, \citenamefont {{Dunkley}}, \citenamefont {{D{\"u}nner}},
  \citenamefont {{Essinger-Hileman}}, \citenamefont {{Fowler}}, \citenamefont
  {{Hajian}}, \citenamefont {{Halpern}}, \citenamefont {{Hasselfield}},
  \citenamefont {{Hincks}}, \citenamefont {{Hlozek}}, \citenamefont
  {{Huffenberger}}, \citenamefont {{Hughes}}, \citenamefont {{Irwin}},
  \citenamefont {{Klein}}, \citenamefont {{Kosowsky}}, \citenamefont
  {{Lupton}}, \citenamefont {{Marriage}}, \citenamefont {{Marsden}},
  \citenamefont {{Menanteau}}, \citenamefont {{Moodley}}, \citenamefont
  {{Niemack}}, \citenamefont {{Nolta}}, \citenamefont {{Page}}, \citenamefont
  {{Parker}}, \citenamefont {{Reese}}, \citenamefont {{Schmitt}}, \citenamefont
  {{Sehgal}}, \citenamefont {{Sievers}}, \citenamefont {{Spergel}},
  \citenamefont {{Staggs}}, \citenamefont {{Swetz}}, \citenamefont {{Switzer}},
  \citenamefont {{Thornton}}, \citenamefont {{Visnjic}},\ and\ \citenamefont
  {{Wollack}}}]{2011Das}%
  \BibitemOpen
  \bibfield  {author} {\bibinfo {author} {\bibfnamefont {S.}~\bibnamefont
  {{Das}}}, \bibinfo {author} {\bibfnamefont {B.~D.}\ \bibnamefont
  {{Sherwin}}}, \bibinfo {author} {\bibfnamefont {P.}~\bibnamefont
  {{Aguirre}}}, \bibinfo {author} {\bibfnamefont {J.~W.}\ \bibnamefont
  {{Appel}}}, \bibinfo {author} {\bibfnamefont {J.~R.}\ \bibnamefont {{Bond}}},
  \bibinfo {author} {\bibfnamefont {C.~S.}\ \bibnamefont {{Carvalho}}},
  \bibinfo {author} {\bibfnamefont {M.~J.}\ \bibnamefont {{Devlin}}}, \bibinfo
  {author} {\bibfnamefont {J.}~\bibnamefont {{Dunkley}}}, \bibinfo {author}
  {\bibfnamefont {R.}~\bibnamefont {{D{\"u}nner}}}, \bibinfo {author}
  {\bibfnamefont {T.}~\bibnamefont {{Essinger-Hileman}}}, \bibinfo {author}
  {\bibfnamefont {J.~W.}\ \bibnamefont {{Fowler}}}, \bibinfo {author}
  {\bibfnamefont {A.}~\bibnamefont {{Hajian}}}, \bibinfo {author}
  {\bibfnamefont {M.}~\bibnamefont {{Halpern}}}, \bibinfo {author}
  {\bibfnamefont {M.}~\bibnamefont {{Hasselfield}}}, \bibinfo {author}
  {\bibfnamefont {A.~D.}\ \bibnamefont {{Hincks}}}, \bibinfo {author}
  {\bibfnamefont {R.}~\bibnamefont {{Hlozek}}}, \bibinfo {author}
  {\bibfnamefont {K.~M.}\ \bibnamefont {{Huffenberger}}}, \bibinfo {author}
  {\bibfnamefont {J.~P.}\ \bibnamefont {{Hughes}}}, \bibinfo {author}
  {\bibfnamefont {K.~D.}\ \bibnamefont {{Irwin}}}, \bibinfo {author}
  {\bibfnamefont {J.}~\bibnamefont {{Klein}}}, \bibinfo {author} {\bibfnamefont
  {A.}~\bibnamefont {{Kosowsky}}}, \bibinfo {author} {\bibfnamefont {R.~H.}\
  \bibnamefont {{Lupton}}}, \bibinfo {author} {\bibfnamefont {T.~A.}\
  \bibnamefont {{Marriage}}}, \bibinfo {author} {\bibfnamefont
  {D.}~\bibnamefont {{Marsden}}}, \bibinfo {author} {\bibfnamefont
  {F.}~\bibnamefont {{Menanteau}}}, \bibinfo {author} {\bibfnamefont
  {K.}~\bibnamefont {{Moodley}}}, \bibinfo {author} {\bibfnamefont {M.~D.}\
  \bibnamefont {{Niemack}}}, \bibinfo {author} {\bibfnamefont {M.~R.}\
  \bibnamefont {{Nolta}}}, \bibinfo {author} {\bibfnamefont {L.~A.}\
  \bibnamefont {{Page}}}, \bibinfo {author} {\bibfnamefont {L.}~\bibnamefont
  {{Parker}}}, \bibinfo {author} {\bibfnamefont {E.~D.}\ \bibnamefont
  {{Reese}}}, \bibinfo {author} {\bibfnamefont {B.~L.}\ \bibnamefont
  {{Schmitt}}}, \bibinfo {author} {\bibfnamefont {N.}~\bibnamefont {{Sehgal}}},
  \bibinfo {author} {\bibfnamefont {J.}~\bibnamefont {{Sievers}}}, \bibinfo
  {author} {\bibfnamefont {D.~N.}\ \bibnamefont {{Spergel}}}, \bibinfo {author}
  {\bibfnamefont {S.~T.}\ \bibnamefont {{Staggs}}}, \bibinfo {author}
  {\bibfnamefont {D.~S.}\ \bibnamefont {{Swetz}}}, \bibinfo {author}
  {\bibfnamefont {E.~R.}\ \bibnamefont {{Switzer}}}, \bibinfo {author}
  {\bibfnamefont {R.}~\bibnamefont {{Thornton}}}, \bibinfo {author}
  {\bibfnamefont {K.}~\bibnamefont {{Visnjic}}}, \ and\ \bibinfo {author}
  {\bibfnamefont {E.}~\bibnamefont {{Wollack}}},\ }\href {\doibase
  10.1103/PhysRevLett.107.021301} {\bibfield  {journal} {\bibinfo  {journal}
  {Physical Review Letters}\ }\textbf {\bibinfo {volume} {107}},\ \bibinfo
  {eid} {021301} (\bibinfo {year} {2011})},\ \Eprint
  {http://arxiv.org/abs/1103.2124} {arXiv:1103.2124} \BibitemShut {NoStop}%
\bibitem [{\citenamefont {{van Engelen}}\ \emph {et~al.}(2012)\citenamefont
  {{van Engelen}}, \citenamefont {{Keisler}}, \citenamefont {{Zahn}},
  \citenamefont {{Aird}}, \citenamefont {{Benson}}, \citenamefont {{Bleem}},
  \citenamefont {{Carlstrom}}, \citenamefont {{Chang}}, \citenamefont {{Cho}},
  \citenamefont {{Crawford}}, \citenamefont {{Crites}}, \citenamefont {{de
  Haan}}, \citenamefont {{Dobbs}}, \citenamefont {{Dudley}}, \citenamefont
  {{George}}, \citenamefont {{Halverson}}, \citenamefont {{Holder}},
  \citenamefont {{Holzapfel}}, \citenamefont {{Hoover}}, \citenamefont {{Hou}},
  \citenamefont {{Hrubes}}, \citenamefont {{Joy}}, \citenamefont {{Knox}},
  \citenamefont {{Lee}}, \citenamefont {{Leitch}}, \citenamefont {{Lueker}},
  \citenamefont {{Luong-Van}}, \citenamefont {{McMahon}}, \citenamefont
  {{Mehl}}, \citenamefont {{Meyer}}, \citenamefont {{Millea}}, \citenamefont
  {{Mohr}}, \citenamefont {{Montroy}}, \citenamefont {{Natoli}}, \citenamefont
  {{Padin}}, \citenamefont {{Plagge}}, \citenamefont {{Pryke}}, \citenamefont
  {{Reichardt}}, \citenamefont {{Ruhl}}, \citenamefont {{Sayre}}, \citenamefont
  {{Schaffer}}, \citenamefont {{Shaw}}, \citenamefont {{Shirokoff}},
  \citenamefont {{Spieler}}, \citenamefont {{Staniszewski}}, \citenamefont
  {{Stark}}, \citenamefont {{Story}}, \citenamefont {{Vanderlinde}},
  \citenamefont {{Vieira}},\ and\ \citenamefont {{Williamson}}}]{2012SPT}%
  \BibitemOpen
  \bibfield  {author} {\bibinfo {author} {\bibfnamefont {A.}~\bibnamefont {{van
  Engelen}}}, \bibinfo {author} {\bibfnamefont {R.}~\bibnamefont {{Keisler}}},
  \bibinfo {author} {\bibfnamefont {O.}~\bibnamefont {{Zahn}}}, \bibinfo
  {author} {\bibfnamefont {K.~A.}\ \bibnamefont {{Aird}}}, \bibinfo {author}
  {\bibfnamefont {B.~A.}\ \bibnamefont {{Benson}}}, \bibinfo {author}
  {\bibfnamefont {L.~E.}\ \bibnamefont {{Bleem}}}, \bibinfo {author}
  {\bibfnamefont {J.~E.}\ \bibnamefont {{Carlstrom}}}, \bibinfo {author}
  {\bibfnamefont {C.~L.}\ \bibnamefont {{Chang}}}, \bibinfo {author}
  {\bibfnamefont {H.~M.}\ \bibnamefont {{Cho}}}, \bibinfo {author}
  {\bibfnamefont {T.~M.}\ \bibnamefont {{Crawford}}}, \bibinfo {author}
  {\bibfnamefont {A.~T.}\ \bibnamefont {{Crites}}}, \bibinfo {author}
  {\bibfnamefont {T.}~\bibnamefont {{de Haan}}}, \bibinfo {author}
  {\bibfnamefont {M.~A.}\ \bibnamefont {{Dobbs}}}, \bibinfo {author}
  {\bibfnamefont {J.}~\bibnamefont {{Dudley}}}, \bibinfo {author}
  {\bibfnamefont {E.~M.}\ \bibnamefont {{George}}}, \bibinfo {author}
  {\bibfnamefont {N.~W.}\ \bibnamefont {{Halverson}}}, \bibinfo {author}
  {\bibfnamefont {G.~P.}\ \bibnamefont {{Holder}}}, \bibinfo {author}
  {\bibfnamefont {W.~L.}\ \bibnamefont {{Holzapfel}}}, \bibinfo {author}
  {\bibfnamefont {S.}~\bibnamefont {{Hoover}}}, \bibinfo {author}
  {\bibfnamefont {Z.}~\bibnamefont {{Hou}}}, \bibinfo {author} {\bibfnamefont
  {J.~D.}\ \bibnamefont {{Hrubes}}}, \bibinfo {author} {\bibfnamefont
  {M.}~\bibnamefont {{Joy}}}, \bibinfo {author} {\bibfnamefont
  {L.}~\bibnamefont {{Knox}}}, \bibinfo {author} {\bibfnamefont {A.~T.}\
  \bibnamefont {{Lee}}}, \bibinfo {author} {\bibfnamefont {E.~M.}\ \bibnamefont
  {{Leitch}}}, \bibinfo {author} {\bibfnamefont {M.}~\bibnamefont {{Lueker}}},
  \bibinfo {author} {\bibfnamefont {D.}~\bibnamefont {{Luong-Van}}}, \bibinfo
  {author} {\bibfnamefont {J.~J.}\ \bibnamefont {{McMahon}}}, \bibinfo {author}
  {\bibfnamefont {J.}~\bibnamefont {{Mehl}}}, \bibinfo {author} {\bibfnamefont
  {S.~S.}\ \bibnamefont {{Meyer}}}, \bibinfo {author} {\bibfnamefont
  {M.}~\bibnamefont {{Millea}}}, \bibinfo {author} {\bibfnamefont {J.~J.}\
  \bibnamefont {{Mohr}}}, \bibinfo {author} {\bibfnamefont {T.~E.}\
  \bibnamefont {{Montroy}}}, \bibinfo {author} {\bibfnamefont {T.}~\bibnamefont
  {{Natoli}}}, \bibinfo {author} {\bibfnamefont {S.}~\bibnamefont {{Padin}}},
  \bibinfo {author} {\bibfnamefont {T.}~\bibnamefont {{Plagge}}}, \bibinfo
  {author} {\bibfnamefont {C.}~\bibnamefont {{Pryke}}}, \bibinfo {author}
  {\bibfnamefont {C.~L.}\ \bibnamefont {{Reichardt}}}, \bibinfo {author}
  {\bibfnamefont {J.~E.}\ \bibnamefont {{Ruhl}}}, \bibinfo {author}
  {\bibfnamefont {J.~T.}\ \bibnamefont {{Sayre}}}, \bibinfo {author}
  {\bibfnamefont {K.~K.}\ \bibnamefont {{Schaffer}}}, \bibinfo {author}
  {\bibfnamefont {L.}~\bibnamefont {{Shaw}}}, \bibinfo {author} {\bibfnamefont
  {E.}~\bibnamefont {{Shirokoff}}}, \bibinfo {author} {\bibfnamefont {H.~G.}\
  \bibnamefont {{Spieler}}}, \bibinfo {author} {\bibfnamefont {Z.}~\bibnamefont
  {{Staniszewski}}}, \bibinfo {author} {\bibfnamefont {A.~A.}\ \bibnamefont
  {{Stark}}}, \bibinfo {author} {\bibfnamefont {K.}~\bibnamefont {{Story}}},
  \bibinfo {author} {\bibfnamefont {K.}~\bibnamefont {{Vanderlinde}}}, \bibinfo
  {author} {\bibfnamefont {J.~D.}\ \bibnamefont {{Vieira}}}, \ and\ \bibinfo
  {author} {\bibfnamefont {R.}~\bibnamefont {{Williamson}}},\ }\href {\doibase
  10.1088/0004-637X/756/2/142} {\bibfield  {journal} {\bibinfo  {journal}
  {\apj}\ }\textbf {\bibinfo {volume} {756}},\ \bibinfo {eid} {142} (\bibinfo
  {year} {2012})},\ \Eprint {http://arxiv.org/abs/1202.0546} {arXiv:1202.0546}
  \BibitemShut {NoStop}%
\bibitem [{\citenamefont {{Hanson}}\ \emph {et~al.}(2013)\citenamefont
  {{Hanson}}, \citenamefont {{Hoover}}, \citenamefont {{Crites}}, \citenamefont
  {{Ade}}, \citenamefont {{Aird}}, \citenamefont {{Austermann}}, \citenamefont
  {{Beall}}, \citenamefont {{Bender}}, \citenamefont {{Benson}}, \citenamefont
  {{Bleem}}, \citenamefont {{Bock}}, \citenamefont {{Carlstrom}}, \citenamefont
  {{Chang}}, \citenamefont {{Chiang}}, \citenamefont {{Cho}}, \citenamefont
  {{Conley}}, \citenamefont {{Crawford}}, \citenamefont {{de Haan}},
  \citenamefont {{Dobbs}}, \citenamefont {{Everett}}, \citenamefont
  {{Gallicchio}}, \citenamefont {{Gao}}, \citenamefont {{George}},
  \citenamefont {{Halverson}}, \citenamefont {{Harrington}}, \citenamefont
  {{Henning}}, \citenamefont {{Hilton}}, \citenamefont {{Holder}},
  \citenamefont {{Holzapfel}}, \citenamefont {{Hrubes}}, \citenamefont
  {{Huang}}, \citenamefont {{Hubmayr}}, \citenamefont {{Irwin}}, \citenamefont
  {{Keisler}}, \citenamefont {{Knox}}, \citenamefont {{Lee}}, \citenamefont
  {{Leitch}}, \citenamefont {{Li}}, \citenamefont {{Liang}}, \citenamefont
  {{Luong-Van}}, \citenamefont {{Marsden}}, \citenamefont {{McMahon}},
  \citenamefont {{Mehl}}, \citenamefont {{Meyer}}, \citenamefont {{Mocanu}},
  \citenamefont {{Montroy}}, \citenamefont {{Natoli}}, \citenamefont
  {{Nibarger}}, \citenamefont {{Novosad}}, \citenamefont {{Padin}},
  \citenamefont {{Pryke}}, \citenamefont {{Reichardt}}, \citenamefont {{Ruhl}},
  \citenamefont {{Saliwanchik}}, \citenamefont {{Sayre}}, \citenamefont
  {{Schaffer}}, \citenamefont {{Schulz}}, \citenamefont {{Smecher}},
  \citenamefont {{Stark}}, \citenamefont {{Story}}, \citenamefont {{Tucker}},
  \citenamefont {{Vanderlinde}}, \citenamefont {{Vieira}}, \citenamefont
  {{Viero}}, \citenamefont {{Wang}}, \citenamefont {{Yefremenko}},
  \citenamefont {{Zahn}},\ and\ \citenamefont {{Zemcov}}}]{2013Hanson}%
  \BibitemOpen
  \bibfield  {author} {\bibinfo {author} {\bibfnamefont {D.}~\bibnamefont
  {{Hanson}}}, \bibinfo {author} {\bibfnamefont {S.}~\bibnamefont {{Hoover}}},
  \bibinfo {author} {\bibfnamefont {A.}~\bibnamefont {{Crites}}}, \bibinfo
  {author} {\bibfnamefont {P.~A.~R.}\ \bibnamefont {{Ade}}}, \bibinfo {author}
  {\bibfnamefont {K.~A.}\ \bibnamefont {{Aird}}}, \bibinfo {author}
  {\bibfnamefont {J.~E.}\ \bibnamefont {{Austermann}}}, \bibinfo {author}
  {\bibfnamefont {J.~A.}\ \bibnamefont {{Beall}}}, \bibinfo {author}
  {\bibfnamefont {A.~N.}\ \bibnamefont {{Bender}}}, \bibinfo {author}
  {\bibfnamefont {B.~A.}\ \bibnamefont {{Benson}}}, \bibinfo {author}
  {\bibfnamefont {L.~E.}\ \bibnamefont {{Bleem}}}, \bibinfo {author}
  {\bibfnamefont {J.~J.}\ \bibnamefont {{Bock}}}, \bibinfo {author}
  {\bibfnamefont {J.~E.}\ \bibnamefont {{Carlstrom}}}, \bibinfo {author}
  {\bibfnamefont {C.~L.}\ \bibnamefont {{Chang}}}, \bibinfo {author}
  {\bibfnamefont {H.~C.}\ \bibnamefont {{Chiang}}}, \bibinfo {author}
  {\bibfnamefont {H.-M.}\ \bibnamefont {{Cho}}}, \bibinfo {author}
  {\bibfnamefont {A.}~\bibnamefont {{Conley}}}, \bibinfo {author}
  {\bibfnamefont {T.~M.}\ \bibnamefont {{Crawford}}}, \bibinfo {author}
  {\bibfnamefont {T.}~\bibnamefont {{de Haan}}}, \bibinfo {author}
  {\bibfnamefont {M.~A.}\ \bibnamefont {{Dobbs}}}, \bibinfo {author}
  {\bibfnamefont {W.}~\bibnamefont {{Everett}}}, \bibinfo {author}
  {\bibfnamefont {J.}~\bibnamefont {{Gallicchio}}}, \bibinfo {author}
  {\bibfnamefont {J.}~\bibnamefont {{Gao}}}, \bibinfo {author} {\bibfnamefont
  {E.~M.}\ \bibnamefont {{George}}}, \bibinfo {author} {\bibfnamefont {N.~W.}\
  \bibnamefont {{Halverson}}}, \bibinfo {author} {\bibfnamefont
  {N.}~\bibnamefont {{Harrington}}}, \bibinfo {author} {\bibfnamefont {J.~W.}\
  \bibnamefont {{Henning}}}, \bibinfo {author} {\bibfnamefont {G.~C.}\
  \bibnamefont {{Hilton}}}, \bibinfo {author} {\bibfnamefont {G.~P.}\
  \bibnamefont {{Holder}}}, \bibinfo {author} {\bibfnamefont {W.~L.}\
  \bibnamefont {{Holzapfel}}}, \bibinfo {author} {\bibfnamefont {J.~D.}\
  \bibnamefont {{Hrubes}}}, \bibinfo {author} {\bibfnamefont {N.}~\bibnamefont
  {{Huang}}}, \bibinfo {author} {\bibfnamefont {J.}~\bibnamefont {{Hubmayr}}},
  \bibinfo {author} {\bibfnamefont {K.~D.}\ \bibnamefont {{Irwin}}}, \bibinfo
  {author} {\bibfnamefont {R.}~\bibnamefont {{Keisler}}}, \bibinfo {author}
  {\bibfnamefont {L.}~\bibnamefont {{Knox}}}, \bibinfo {author} {\bibfnamefont
  {A.~T.}\ \bibnamefont {{Lee}}}, \bibinfo {author} {\bibfnamefont
  {E.}~\bibnamefont {{Leitch}}}, \bibinfo {author} {\bibfnamefont
  {D.}~\bibnamefont {{Li}}}, \bibinfo {author} {\bibfnamefont {C.}~\bibnamefont
  {{Liang}}}, \bibinfo {author} {\bibfnamefont {D.}~\bibnamefont
  {{Luong-Van}}}, \bibinfo {author} {\bibfnamefont {G.}~\bibnamefont
  {{Marsden}}}, \bibinfo {author} {\bibfnamefont {J.~J.}\ \bibnamefont
  {{McMahon}}}, \bibinfo {author} {\bibfnamefont {J.}~\bibnamefont {{Mehl}}},
  \bibinfo {author} {\bibfnamefont {S.~S.}\ \bibnamefont {{Meyer}}}, \bibinfo
  {author} {\bibfnamefont {L.}~\bibnamefont {{Mocanu}}}, \bibinfo {author}
  {\bibfnamefont {T.~E.}\ \bibnamefont {{Montroy}}}, \bibinfo {author}
  {\bibfnamefont {T.}~\bibnamefont {{Natoli}}}, \bibinfo {author}
  {\bibfnamefont {J.~P.}\ \bibnamefont {{Nibarger}}}, \bibinfo {author}
  {\bibfnamefont {V.}~\bibnamefont {{Novosad}}}, \bibinfo {author}
  {\bibfnamefont {S.}~\bibnamefont {{Padin}}}, \bibinfo {author} {\bibfnamefont
  {C.}~\bibnamefont {{Pryke}}}, \bibinfo {author} {\bibfnamefont {C.~L.}\
  \bibnamefont {{Reichardt}}}, \bibinfo {author} {\bibfnamefont {J.~E.}\
  \bibnamefont {{Ruhl}}}, \bibinfo {author} {\bibfnamefont {B.~R.}\
  \bibnamefont {{Saliwanchik}}}, \bibinfo {author} {\bibfnamefont {J.~T.}\
  \bibnamefont {{Sayre}}}, \bibinfo {author} {\bibfnamefont {K.~K.}\
  \bibnamefont {{Schaffer}}}, \bibinfo {author} {\bibfnamefont
  {B.}~\bibnamefont {{Schulz}}}, \bibinfo {author} {\bibfnamefont
  {G.}~\bibnamefont {{Smecher}}}, \bibinfo {author} {\bibfnamefont {A.~A.}\
  \bibnamefont {{Stark}}}, \bibinfo {author} {\bibfnamefont {K.~T.}\
  \bibnamefont {{Story}}}, \bibinfo {author} {\bibfnamefont {C.}~\bibnamefont
  {{Tucker}}}, \bibinfo {author} {\bibfnamefont {K.}~\bibnamefont
  {{Vanderlinde}}}, \bibinfo {author} {\bibfnamefont {J.~D.}\ \bibnamefont
  {{Vieira}}}, \bibinfo {author} {\bibfnamefont {M.~P.}\ \bibnamefont
  {{Viero}}}, \bibinfo {author} {\bibfnamefont {G.}~\bibnamefont {{Wang}}},
  \bibinfo {author} {\bibfnamefont {V.}~\bibnamefont {{Yefremenko}}}, \bibinfo
  {author} {\bibfnamefont {O.}~\bibnamefont {{Zahn}}}, \ and\ \bibinfo {author}
  {\bibfnamefont {M.}~\bibnamefont {{Zemcov}}},\ }\href {\doibase
  10.1103/PhysRevLett.111.141301} {\bibfield  {journal} {\bibinfo  {journal}
  {Physical Review Letters}\ }\textbf {\bibinfo {volume} {111}},\ \bibinfo
  {eid} {141301} (\bibinfo {year} {2013})},\ \Eprint
  {http://arxiv.org/abs/1307.5830} {arXiv:1307.5830 [astro-ph.CO]} \BibitemShut
  {NoStop}%
\bibitem [{\citenamefont {{Ade}}\ \emph
  {et~al.}(2014{\natexlab{a}})\citenamefont {{Ade}}, \citenamefont {{Akiba}},
  \citenamefont {{Anthony}}, \citenamefont {{Arnold}}, \citenamefont {{Atlas}},
  \citenamefont {{Barron}}, \citenamefont {{Boettger}}, \citenamefont
  {{Borrill}}, \citenamefont {{Chapman}}, \citenamefont {{Chinone}},
  \citenamefont {{Dobbs}}, \citenamefont {{Elleflot}}, \citenamefont
  {{Errard}}, \citenamefont {{Fabbian}}, \citenamefont {{Feng}}, \citenamefont
  {{Flanigan}}, \citenamefont {{Gilbert}}, \citenamefont {{Grainger}},
  \citenamefont {{Halverson}}, \citenamefont {{Hasegawa}}, \citenamefont
  {{Hattori}}, \citenamefont {{Hazumi}}, \citenamefont {{Holzapfel}},
  \citenamefont {{Hori}}, \citenamefont {{Howard}}, \citenamefont {{Hyland}},
  \citenamefont {{Inoue}}, \citenamefont {{Jaehnig}}, \citenamefont {{Jaffe}},
  \citenamefont {{Keating}}, \citenamefont {{Kermish}}, \citenamefont
  {{Keskitalo}}, \citenamefont {{Kisner}}, \citenamefont {{Le Jeune}},
  \citenamefont {{Lee}}, \citenamefont {{Linder}}, \citenamefont {{Leitch}},
  \citenamefont {{Lungu}}, \citenamefont {{Matsuda}}, \citenamefont
  {{Matsumura}}, \citenamefont {{Meng}}, \citenamefont {{Miller}},
  \citenamefont {{Morii}}, \citenamefont {{Moyerman}}, \citenamefont {{Myers}},
  \citenamefont {{Navaroli}}, \citenamefont {{Nishino}}, \citenamefont
  {{Paar}}, \citenamefont {{Peloton}}, \citenamefont {{Quealy}}, \citenamefont
  {{Rebeiz}}, \citenamefont {{Reichardt}}, \citenamefont {{Richards}},
  \citenamefont {{Ross}}, \citenamefont {{Schanning}}, \citenamefont
  {{Schenck}}, \citenamefont {{Sherwin}}, \citenamefont {{Shimizu}},
  \citenamefont {{Shimmin}}, \citenamefont {{Shimon}}, \citenamefont
  {{Siritanasak}}, \citenamefont {{Smecher}}, \citenamefont {{Spieler}},
  \citenamefont {{Stebor}}, \citenamefont {{Steinbach}}, \citenamefont
  {{Stompor}}, \citenamefont {{Suzuki}}, \citenamefont {{Takakura}},
  \citenamefont {{Tomaru}}, \citenamefont {{Wilson}}, \citenamefont {{Yadav}},
  \citenamefont {{Zahn}},\ and\ \citenamefont {{Polarbear
  Collaboration}}}]{2014PolBear1}%
  \BibitemOpen
  \bibfield  {author} {\bibinfo {author} {\bibfnamefont {P.~A.~R.}\
  \bibnamefont {{Ade}}}, \bibinfo {author} {\bibfnamefont {Y.}~\bibnamefont
  {{Akiba}}}, \bibinfo {author} {\bibfnamefont {A.~E.}\ \bibnamefont
  {{Anthony}}}, \bibinfo {author} {\bibfnamefont {K.}~\bibnamefont {{Arnold}}},
  \bibinfo {author} {\bibfnamefont {M.}~\bibnamefont {{Atlas}}}, \bibinfo
  {author} {\bibfnamefont {D.}~\bibnamefont {{Barron}}}, \bibinfo {author}
  {\bibfnamefont {D.}~\bibnamefont {{Boettger}}}, \bibinfo {author}
  {\bibfnamefont {J.}~\bibnamefont {{Borrill}}}, \bibinfo {author}
  {\bibfnamefont {S.}~\bibnamefont {{Chapman}}}, \bibinfo {author}
  {\bibfnamefont {Y.}~\bibnamefont {{Chinone}}}, \bibinfo {author}
  {\bibfnamefont {M.}~\bibnamefont {{Dobbs}}}, \bibinfo {author} {\bibfnamefont
  {T.}~\bibnamefont {{Elleflot}}}, \bibinfo {author} {\bibfnamefont
  {J.}~\bibnamefont {{Errard}}}, \bibinfo {author} {\bibfnamefont
  {G.}~\bibnamefont {{Fabbian}}}, \bibinfo {author} {\bibfnamefont
  {C.}~\bibnamefont {{Feng}}}, \bibinfo {author} {\bibfnamefont
  {D.}~\bibnamefont {{Flanigan}}}, \bibinfo {author} {\bibfnamefont
  {A.}~\bibnamefont {{Gilbert}}}, \bibinfo {author} {\bibfnamefont
  {W.}~\bibnamefont {{Grainger}}}, \bibinfo {author} {\bibfnamefont {N.~W.}\
  \bibnamefont {{Halverson}}}, \bibinfo {author} {\bibfnamefont
  {M.}~\bibnamefont {{Hasegawa}}}, \bibinfo {author} {\bibfnamefont
  {K.}~\bibnamefont {{Hattori}}}, \bibinfo {author} {\bibfnamefont
  {M.}~\bibnamefont {{Hazumi}}}, \bibinfo {author} {\bibfnamefont {W.~L.}\
  \bibnamefont {{Holzapfel}}}, \bibinfo {author} {\bibfnamefont
  {Y.}~\bibnamefont {{Hori}}}, \bibinfo {author} {\bibfnamefont
  {J.}~\bibnamefont {{Howard}}}, \bibinfo {author} {\bibfnamefont
  {P.}~\bibnamefont {{Hyland}}}, \bibinfo {author} {\bibfnamefont
  {Y.}~\bibnamefont {{Inoue}}}, \bibinfo {author} {\bibfnamefont {G.~C.}\
  \bibnamefont {{Jaehnig}}}, \bibinfo {author} {\bibfnamefont {A.}~\bibnamefont
  {{Jaffe}}}, \bibinfo {author} {\bibfnamefont {B.}~\bibnamefont {{Keating}}},
  \bibinfo {author} {\bibfnamefont {Z.}~\bibnamefont {{Kermish}}}, \bibinfo
  {author} {\bibfnamefont {R.}~\bibnamefont {{Keskitalo}}}, \bibinfo {author}
  {\bibfnamefont {T.}~\bibnamefont {{Kisner}}}, \bibinfo {author}
  {\bibfnamefont {M.}~\bibnamefont {{Le Jeune}}}, \bibinfo {author}
  {\bibfnamefont {A.~T.}\ \bibnamefont {{Lee}}}, \bibinfo {author}
  {\bibfnamefont {E.}~\bibnamefont {{Linder}}}, \bibinfo {author}
  {\bibfnamefont {E.~M.}\ \bibnamefont {{Leitch}}}, \bibinfo {author}
  {\bibfnamefont {M.}~\bibnamefont {{Lungu}}}, \bibinfo {author} {\bibfnamefont
  {F.}~\bibnamefont {{Matsuda}}}, \bibinfo {author} {\bibfnamefont
  {T.}~\bibnamefont {{Matsumura}}}, \bibinfo {author} {\bibfnamefont
  {X.}~\bibnamefont {{Meng}}}, \bibinfo {author} {\bibfnamefont {N.~J.}\
  \bibnamefont {{Miller}}}, \bibinfo {author} {\bibfnamefont {H.}~\bibnamefont
  {{Morii}}}, \bibinfo {author} {\bibfnamefont {S.}~\bibnamefont {{Moyerman}}},
  \bibinfo {author} {\bibfnamefont {M.~J.}\ \bibnamefont {{Myers}}}, \bibinfo
  {author} {\bibfnamefont {M.}~\bibnamefont {{Navaroli}}}, \bibinfo {author}
  {\bibfnamefont {H.}~\bibnamefont {{Nishino}}}, \bibinfo {author}
  {\bibfnamefont {H.}~\bibnamefont {{Paar}}}, \bibinfo {author} {\bibfnamefont
  {J.}~\bibnamefont {{Peloton}}}, \bibinfo {author} {\bibfnamefont
  {E.}~\bibnamefont {{Quealy}}}, \bibinfo {author} {\bibfnamefont
  {G.}~\bibnamefont {{Rebeiz}}}, \bibinfo {author} {\bibfnamefont {C.~L.}\
  \bibnamefont {{Reichardt}}}, \bibinfo {author} {\bibfnamefont {P.~L.}\
  \bibnamefont {{Richards}}}, \bibinfo {author} {\bibfnamefont
  {C.}~\bibnamefont {{Ross}}}, \bibinfo {author} {\bibfnamefont
  {I.}~\bibnamefont {{Schanning}}}, \bibinfo {author} {\bibfnamefont {D.~E.}\
  \bibnamefont {{Schenck}}}, \bibinfo {author} {\bibfnamefont {B.}~\bibnamefont
  {{Sherwin}}}, \bibinfo {author} {\bibfnamefont {A.}~\bibnamefont
  {{Shimizu}}}, \bibinfo {author} {\bibfnamefont {C.}~\bibnamefont
  {{Shimmin}}}, \bibinfo {author} {\bibfnamefont {M.}~\bibnamefont {{Shimon}}},
  \bibinfo {author} {\bibfnamefont {P.}~\bibnamefont {{Siritanasak}}}, \bibinfo
  {author} {\bibfnamefont {G.}~\bibnamefont {{Smecher}}}, \bibinfo {author}
  {\bibfnamefont {H.}~\bibnamefont {{Spieler}}}, \bibinfo {author}
  {\bibfnamefont {N.}~\bibnamefont {{Stebor}}}, \bibinfo {author}
  {\bibfnamefont {B.}~\bibnamefont {{Steinbach}}}, \bibinfo {author}
  {\bibfnamefont {R.}~\bibnamefont {{Stompor}}}, \bibinfo {author}
  {\bibfnamefont {A.}~\bibnamefont {{Suzuki}}}, \bibinfo {author}
  {\bibfnamefont {S.}~\bibnamefont {{Takakura}}}, \bibinfo {author}
  {\bibfnamefont {T.}~\bibnamefont {{Tomaru}}}, \bibinfo {author}
  {\bibfnamefont {B.}~\bibnamefont {{Wilson}}}, \bibinfo {author}
  {\bibfnamefont {A.}~\bibnamefont {{Yadav}}}, \bibinfo {author} {\bibfnamefont
  {O.}~\bibnamefont {{Zahn}}}, \ and\ \bibinfo {author} {\bibnamefont
  {{Polarbear Collaboration}}},\ }\href {\doibase
  10.1103/PhysRevLett.113.021301} {\bibfield  {journal} {\bibinfo  {journal}
  {Physical Review Letters}\ }\textbf {\bibinfo {volume} {113}},\ \bibinfo
  {eid} {021301} (\bibinfo {year} {2014}{\natexlab{a}})},\ \Eprint
  {http://arxiv.org/abs/1312.6646} {arXiv:1312.6646} \BibitemShut {NoStop}%
\bibitem [{\citenamefont {{Ade}}\ \emph
  {et~al.}(2014{\natexlab{b}})\citenamefont {{Ade}}, \citenamefont {{Akiba}},
  \citenamefont {{Anthony}}, \citenamefont {{Arnold}}, \citenamefont {{Atlas}},
  \citenamefont {{Barron}}, \citenamefont {{Boettger}}, \citenamefont
  {{Borrill}}, \citenamefont {{Chapman}}, \citenamefont {{Chinone}},
  \citenamefont {{Dobbs}}, \citenamefont {{Elleflot}}, \citenamefont
  {{Errard}}, \citenamefont {{Fabbian}}, \citenamefont {{Feng}}, \citenamefont
  {{Flanigan}}, \citenamefont {{Gilbert}}, \citenamefont {{Grainger}},
  \citenamefont {{Halverson}}, \citenamefont {{Hasegawa}}, \citenamefont
  {{Hattori}}, \citenamefont {{Hazumi}}, \citenamefont {{Holzapfel}},
  \citenamefont {{Hori}}, \citenamefont {{Howard}}, \citenamefont {{Hyland}},
  \citenamefont {{Inoue}}, \citenamefont {{Jaehnig}}, \citenamefont {{Jaffe}},
  \citenamefont {{Keating}}, \citenamefont {{Kermish}}, \citenamefont
  {{Keskitalo}}, \citenamefont {{Kisner}}, \citenamefont {{Le Jeune}},
  \citenamefont {{Lee}}, \citenamefont {{Linder}}, \citenamefont {{Leitch}},
  \citenamefont {{Lungu}}, \citenamefont {{Matsuda}}, \citenamefont
  {{Matsumura}}, \citenamefont {{Meng}}, \citenamefont {{Miller}},
  \citenamefont {{Morii}}, \citenamefont {{Moyerman}}, \citenamefont {{Myers}},
  \citenamefont {{Navaroli}}, \citenamefont {{Nishino}}, \citenamefont
  {{Paar}}, \citenamefont {{Peloton}}, \citenamefont {{Quealy}}, \citenamefont
  {{Rebeiz}}, \citenamefont {{Reichardt}}, \citenamefont {{Richards}},
  \citenamefont {{Ross}}, \citenamefont {{Schanning}}, \citenamefont
  {{Schenck}}, \citenamefont {{Sherwin}}, \citenamefont {{Shimizu}},
  \citenamefont {{Shimmin}}, \citenamefont {{Shimon}}, \citenamefont
  {{Siritanasak}}, \citenamefont {{Smecher}}, \citenamefont {{Spieler}},
  \citenamefont {{Stebor}}, \citenamefont {{Steinbach}}, \citenamefont
  {{Stompor}}, \citenamefont {{Suzuki}}, \citenamefont {{Takakura}},
  \citenamefont {{Tomaru}}, \citenamefont {{Wilson}}, \citenamefont {{Yadav}},
  \citenamefont {{Zahn}},\ and\ \citenamefont {{Polarbear
  Collaboration}}}]{2014PolBear2}%
  \BibitemOpen
  \bibfield  {author} {\bibinfo {author} {\bibfnamefont {P.~A.~R.}\
  \bibnamefont {{Ade}}}, \bibinfo {author} {\bibfnamefont {Y.}~\bibnamefont
  {{Akiba}}}, \bibinfo {author} {\bibfnamefont {A.~E.}\ \bibnamefont
  {{Anthony}}}, \bibinfo {author} {\bibfnamefont {K.}~\bibnamefont {{Arnold}}},
  \bibinfo {author} {\bibfnamefont {M.}~\bibnamefont {{Atlas}}}, \bibinfo
  {author} {\bibfnamefont {D.}~\bibnamefont {{Barron}}}, \bibinfo {author}
  {\bibfnamefont {D.}~\bibnamefont {{Boettger}}}, \bibinfo {author}
  {\bibfnamefont {J.}~\bibnamefont {{Borrill}}}, \bibinfo {author}
  {\bibfnamefont {S.}~\bibnamefont {{Chapman}}}, \bibinfo {author}
  {\bibfnamefont {Y.}~\bibnamefont {{Chinone}}}, \bibinfo {author}
  {\bibfnamefont {M.}~\bibnamefont {{Dobbs}}}, \bibinfo {author} {\bibfnamefont
  {T.}~\bibnamefont {{Elleflot}}}, \bibinfo {author} {\bibfnamefont
  {J.}~\bibnamefont {{Errard}}}, \bibinfo {author} {\bibfnamefont
  {G.}~\bibnamefont {{Fabbian}}}, \bibinfo {author} {\bibfnamefont
  {C.}~\bibnamefont {{Feng}}}, \bibinfo {author} {\bibfnamefont
  {D.}~\bibnamefont {{Flanigan}}}, \bibinfo {author} {\bibfnamefont
  {A.}~\bibnamefont {{Gilbert}}}, \bibinfo {author} {\bibfnamefont
  {W.}~\bibnamefont {{Grainger}}}, \bibinfo {author} {\bibfnamefont {N.~W.}\
  \bibnamefont {{Halverson}}}, \bibinfo {author} {\bibfnamefont
  {M.}~\bibnamefont {{Hasegawa}}}, \bibinfo {author} {\bibfnamefont
  {K.}~\bibnamefont {{Hattori}}}, \bibinfo {author} {\bibfnamefont
  {M.}~\bibnamefont {{Hazumi}}}, \bibinfo {author} {\bibfnamefont {W.~L.}\
  \bibnamefont {{Holzapfel}}}, \bibinfo {author} {\bibfnamefont
  {Y.}~\bibnamefont {{Hori}}}, \bibinfo {author} {\bibfnamefont
  {J.}~\bibnamefont {{Howard}}}, \bibinfo {author} {\bibfnamefont
  {P.}~\bibnamefont {{Hyland}}}, \bibinfo {author} {\bibfnamefont
  {Y.}~\bibnamefont {{Inoue}}}, \bibinfo {author} {\bibfnamefont {G.~C.}\
  \bibnamefont {{Jaehnig}}}, \bibinfo {author} {\bibfnamefont {A.}~\bibnamefont
  {{Jaffe}}}, \bibinfo {author} {\bibfnamefont {B.}~\bibnamefont {{Keating}}},
  \bibinfo {author} {\bibfnamefont {Z.}~\bibnamefont {{Kermish}}}, \bibinfo
  {author} {\bibfnamefont {R.}~\bibnamefont {{Keskitalo}}}, \bibinfo {author}
  {\bibfnamefont {T.}~\bibnamefont {{Kisner}}}, \bibinfo {author}
  {\bibfnamefont {M.}~\bibnamefont {{Le Jeune}}}, \bibinfo {author}
  {\bibfnamefont {A.~T.}\ \bibnamefont {{Lee}}}, \bibinfo {author}
  {\bibfnamefont {E.}~\bibnamefont {{Linder}}}, \bibinfo {author}
  {\bibfnamefont {E.~M.}\ \bibnamefont {{Leitch}}}, \bibinfo {author}
  {\bibfnamefont {M.}~\bibnamefont {{Lungu}}}, \bibinfo {author} {\bibfnamefont
  {F.}~\bibnamefont {{Matsuda}}}, \bibinfo {author} {\bibfnamefont
  {T.}~\bibnamefont {{Matsumura}}}, \bibinfo {author} {\bibfnamefont
  {X.}~\bibnamefont {{Meng}}}, \bibinfo {author} {\bibfnamefont {N.~J.}\
  \bibnamefont {{Miller}}}, \bibinfo {author} {\bibfnamefont {H.}~\bibnamefont
  {{Morii}}}, \bibinfo {author} {\bibfnamefont {S.}~\bibnamefont {{Moyerman}}},
  \bibinfo {author} {\bibfnamefont {M.~J.}\ \bibnamefont {{Myers}}}, \bibinfo
  {author} {\bibfnamefont {M.}~\bibnamefont {{Navaroli}}}, \bibinfo {author}
  {\bibfnamefont {H.}~\bibnamefont {{Nishino}}}, \bibinfo {author}
  {\bibfnamefont {H.}~\bibnamefont {{Paar}}}, \bibinfo {author} {\bibfnamefont
  {J.}~\bibnamefont {{Peloton}}}, \bibinfo {author} {\bibfnamefont
  {E.}~\bibnamefont {{Quealy}}}, \bibinfo {author} {\bibfnamefont
  {G.}~\bibnamefont {{Rebeiz}}}, \bibinfo {author} {\bibfnamefont {C.~L.}\
  \bibnamefont {{Reichardt}}}, \bibinfo {author} {\bibfnamefont {P.~L.}\
  \bibnamefont {{Richards}}}, \bibinfo {author} {\bibfnamefont
  {C.}~\bibnamefont {{Ross}}}, \bibinfo {author} {\bibfnamefont
  {I.}~\bibnamefont {{Schanning}}}, \bibinfo {author} {\bibfnamefont {D.~E.}\
  \bibnamefont {{Schenck}}}, \bibinfo {author} {\bibfnamefont {B.}~\bibnamefont
  {{Sherwin}}}, \bibinfo {author} {\bibfnamefont {A.}~\bibnamefont
  {{Shimizu}}}, \bibinfo {author} {\bibfnamefont {C.}~\bibnamefont
  {{Shimmin}}}, \bibinfo {author} {\bibfnamefont {M.}~\bibnamefont {{Shimon}}},
  \bibinfo {author} {\bibfnamefont {P.}~\bibnamefont {{Siritanasak}}}, \bibinfo
  {author} {\bibfnamefont {G.}~\bibnamefont {{Smecher}}}, \bibinfo {author}
  {\bibfnamefont {H.}~\bibnamefont {{Spieler}}}, \bibinfo {author}
  {\bibfnamefont {N.}~\bibnamefont {{Stebor}}}, \bibinfo {author}
  {\bibfnamefont {B.}~\bibnamefont {{Steinbach}}}, \bibinfo {author}
  {\bibfnamefont {R.}~\bibnamefont {{Stompor}}}, \bibinfo {author}
  {\bibfnamefont {A.}~\bibnamefont {{Suzuki}}}, \bibinfo {author}
  {\bibfnamefont {S.}~\bibnamefont {{Takakura}}}, \bibinfo {author}
  {\bibfnamefont {T.}~\bibnamefont {{Tomaru}}}, \bibinfo {author}
  {\bibfnamefont {B.}~\bibnamefont {{Wilson}}}, \bibinfo {author}
  {\bibfnamefont {A.}~\bibnamefont {{Yadav}}}, \bibinfo {author} {\bibfnamefont
  {O.}~\bibnamefont {{Zahn}}}, \ and\ \bibinfo {author} {\bibnamefont
  {{Polarbear Collaboration}}},\ }\href {\doibase
  10.1103/PhysRevLett.113.021301} {\bibfield  {journal} {\bibinfo  {journal}
  {Physical Review Letters}\ }\textbf {\bibinfo {volume} {113}},\ \bibinfo
  {eid} {021301} (\bibinfo {year} {2014}{\natexlab{b}})},\ \Eprint
  {http://arxiv.org/abs/1312.6646} {arXiv:1312.6646} \BibitemShut {NoStop}%
\bibitem [{\citenamefont {{BICEP2 Collaboration}}\ \emph
  {et~al.}(2016)\citenamefont {{BICEP2 Collaboration}}, \citenamefont {{Keck
  Array Collaboration}}, \citenamefont {{Ade}}, \citenamefont {{Ahmed}},
  \citenamefont {{Aikin}}, \citenamefont {{Alexander}}, \citenamefont
  {{Barkats}}, \citenamefont {{Benton}}, \citenamefont {{Bischoff}},
  \citenamefont {{Bock}}, \citenamefont {{Bowens-Rubin}}, \citenamefont
  {{Brevik}}, \citenamefont {{Buder}}, \citenamefont {{Bullock}}, \citenamefont
  {{Buza}}, \citenamefont {{Connors}}, \citenamefont {{Crill}}, \citenamefont
  {{Duband}}, \citenamefont {{Dvorkin}}, \citenamefont {{Filippini}},
  \citenamefont {{Fliescher}}, \citenamefont {{Grayson}}, \citenamefont
  {{Halpern}}, \citenamefont {{Harrison}}, \citenamefont {{Hildebrandt}},
  \citenamefont {{Hilton}}, \citenamefont {{Hui}}, \citenamefont {{Irwin}},
  \citenamefont {{Kang}}, \citenamefont {{Karkare}}, \citenamefont {{Karpel}},
  \citenamefont {{Kaufman}}, \citenamefont {{Keating}}, \citenamefont
  {{Kefeli}}, \citenamefont {{Kernasovskiy}}, \citenamefont {{Kovac}},
  \citenamefont {{Kuo}}, \citenamefont {{Leitch}}, \citenamefont {{Lueker}},
  \citenamefont {{Megerian}}, \citenamefont {{Namikawa}}, \citenamefont
  {{Netterfield}}, \citenamefont {{Nguyen}}, \citenamefont {{O'Brient}},
  \citenamefont {{Ogburn}}, \citenamefont {{Orlando}}, \citenamefont {{Pryke}},
  \citenamefont {{Richter}}, \citenamefont {{Schwarz}}, \citenamefont
  {{Sheehy}}, \citenamefont {{Staniszewski}}, \citenamefont {{Steinbach}},
  \citenamefont {{Sudiwala}}, \citenamefont {{Teply}}, \citenamefont
  {{Thompson}}, \citenamefont {{Tolan}}, \citenamefont {{Tucker}},
  \citenamefont {{Turner}}, \citenamefont {{Vieregg}}, \citenamefont {{Weber}},
  \citenamefont {{Wiebe}}, \citenamefont {{Willmert}}, \citenamefont {{Wong}},
  \citenamefont {{Wu}},\ and\ \citenamefont {{Yoon}}}]{Bicep2Keck2016}%
  \BibitemOpen
  \bibfield  {author} {\bibinfo {author} {\bibnamefont {{BICEP2
  Collaboration}}}, \bibinfo {author} {\bibnamefont {{Keck Array
  Collaboration}}}, \bibinfo {author} {\bibfnamefont {P.~A.~R.}\ \bibnamefont
  {{Ade}}}, \bibinfo {author} {\bibfnamefont {Z.}~\bibnamefont {{Ahmed}}},
  \bibinfo {author} {\bibfnamefont {R.~W.}\ \bibnamefont {{Aikin}}}, \bibinfo
  {author} {\bibfnamefont {K.~D.}\ \bibnamefont {{Alexander}}}, \bibinfo
  {author} {\bibfnamefont {D.}~\bibnamefont {{Barkats}}}, \bibinfo {author}
  {\bibfnamefont {S.~J.}\ \bibnamefont {{Benton}}}, \bibinfo {author}
  {\bibfnamefont {C.~A.}\ \bibnamefont {{Bischoff}}}, \bibinfo {author}
  {\bibfnamefont {J.~J.}\ \bibnamefont {{Bock}}}, \bibinfo {author}
  {\bibfnamefont {R.}~\bibnamefont {{Bowens-Rubin}}}, \bibinfo {author}
  {\bibfnamefont {J.~A.}\ \bibnamefont {{Brevik}}}, \bibinfo {author}
  {\bibfnamefont {I.}~\bibnamefont {{Buder}}}, \bibinfo {author} {\bibfnamefont
  {E.}~\bibnamefont {{Bullock}}}, \bibinfo {author} {\bibfnamefont
  {V.}~\bibnamefont {{Buza}}}, \bibinfo {author} {\bibfnamefont
  {J.}~\bibnamefont {{Connors}}}, \bibinfo {author} {\bibfnamefont {B.~P.}\
  \bibnamefont {{Crill}}}, \bibinfo {author} {\bibfnamefont {L.}~\bibnamefont
  {{Duband}}}, \bibinfo {author} {\bibfnamefont {C.}~\bibnamefont {{Dvorkin}}},
  \bibinfo {author} {\bibfnamefont {J.~P.}\ \bibnamefont {{Filippini}}},
  \bibinfo {author} {\bibfnamefont {S.}~\bibnamefont {{Fliescher}}}, \bibinfo
  {author} {\bibfnamefont {J.}~\bibnamefont {{Grayson}}}, \bibinfo {author}
  {\bibfnamefont {M.}~\bibnamefont {{Halpern}}}, \bibinfo {author}
  {\bibfnamefont {S.}~\bibnamefont {{Harrison}}}, \bibinfo {author}
  {\bibfnamefont {S.~R.}\ \bibnamefont {{Hildebrandt}}}, \bibinfo {author}
  {\bibfnamefont {G.~C.}\ \bibnamefont {{Hilton}}}, \bibinfo {author}
  {\bibfnamefont {H.}~\bibnamefont {{Hui}}}, \bibinfo {author} {\bibfnamefont
  {K.~D.}\ \bibnamefont {{Irwin}}}, \bibinfo {author} {\bibfnamefont
  {J.}~\bibnamefont {{Kang}}}, \bibinfo {author} {\bibfnamefont {K.~S.}\
  \bibnamefont {{Karkare}}}, \bibinfo {author} {\bibfnamefont {E.}~\bibnamefont
  {{Karpel}}}, \bibinfo {author} {\bibfnamefont {J.~P.}\ \bibnamefont
  {{Kaufman}}}, \bibinfo {author} {\bibfnamefont {B.~G.}\ \bibnamefont
  {{Keating}}}, \bibinfo {author} {\bibfnamefont {S.}~\bibnamefont {{Kefeli}}},
  \bibinfo {author} {\bibfnamefont {S.~A.}\ \bibnamefont {{Kernasovskiy}}},
  \bibinfo {author} {\bibfnamefont {J.~M.}\ \bibnamefont {{Kovac}}}, \bibinfo
  {author} {\bibfnamefont {C.~L.}\ \bibnamefont {{Kuo}}}, \bibinfo {author}
  {\bibfnamefont {E.~M.}\ \bibnamefont {{Leitch}}}, \bibinfo {author}
  {\bibfnamefont {M.}~\bibnamefont {{Lueker}}}, \bibinfo {author}
  {\bibfnamefont {K.~G.}\ \bibnamefont {{Megerian}}}, \bibinfo {author}
  {\bibfnamefont {T.}~\bibnamefont {{Namikawa}}}, \bibinfo {author}
  {\bibfnamefont {C.~B.}\ \bibnamefont {{Netterfield}}}, \bibinfo {author}
  {\bibfnamefont {H.~T.}\ \bibnamefont {{Nguyen}}}, \bibinfo {author}
  {\bibfnamefont {R.}~\bibnamefont {{O'Brient}}}, \bibinfo {author}
  {\bibfnamefont {R.~W.}\ \bibnamefont {{Ogburn}}, \bibfnamefont {IV}},
  \bibinfo {author} {\bibfnamefont {A.}~\bibnamefont {{Orlando}}}, \bibinfo
  {author} {\bibfnamefont {C.}~\bibnamefont {{Pryke}}}, \bibinfo {author}
  {\bibfnamefont {S.}~\bibnamefont {{Richter}}}, \bibinfo {author}
  {\bibfnamefont {R.}~\bibnamefont {{Schwarz}}}, \bibinfo {author}
  {\bibfnamefont {C.~D.}\ \bibnamefont {{Sheehy}}}, \bibinfo {author}
  {\bibfnamefont {Z.~K.}\ \bibnamefont {{Staniszewski}}}, \bibinfo {author}
  {\bibfnamefont {B.}~\bibnamefont {{Steinbach}}}, \bibinfo {author}
  {\bibfnamefont {R.~V.}\ \bibnamefont {{Sudiwala}}}, \bibinfo {author}
  {\bibfnamefont {G.~P.}\ \bibnamefont {{Teply}}}, \bibinfo {author}
  {\bibfnamefont {K.~L.}\ \bibnamefont {{Thompson}}}, \bibinfo {author}
  {\bibfnamefont {J.~E.}\ \bibnamefont {{Tolan}}}, \bibinfo {author}
  {\bibfnamefont {C.}~\bibnamefont {{Tucker}}}, \bibinfo {author}
  {\bibfnamefont {A.~D.}\ \bibnamefont {{Turner}}}, \bibinfo {author}
  {\bibfnamefont {A.~G.}\ \bibnamefont {{Vieregg}}}, \bibinfo {author}
  {\bibfnamefont {A.~C.}\ \bibnamefont {{Weber}}}, \bibinfo {author}
  {\bibfnamefont {D.~V.}\ \bibnamefont {{Wiebe}}}, \bibinfo {author}
  {\bibfnamefont {J.}~\bibnamefont {{Willmert}}}, \bibinfo {author}
  {\bibfnamefont {C.~L.}\ \bibnamefont {{Wong}}}, \bibinfo {author}
  {\bibfnamefont {W.~L.~K.}\ \bibnamefont {{Wu}}}, \ and\ \bibinfo {author}
  {\bibfnamefont {K.~W.}\ \bibnamefont {{Yoon}}},\ }\href {\doibase
  10.3847/1538-4357/833/2/228} {\bibfield  {journal} {\bibinfo  {journal}
  {\apj}\ }\textbf {\bibinfo {volume} {833}},\ \bibinfo {eid} {228} (\bibinfo
  {year} {2016})},\ \Eprint {http://arxiv.org/abs/1606.01968}
  {arXiv:1606.01968} \BibitemShut {NoStop}%
\bibitem [{\citenamefont {{Story}}\ \emph {et~al.}(2015)\citenamefont
  {{Story}}, \citenamefont {{Hanson}}, \citenamefont {{Ade}}, \citenamefont
  {{Aird}}, \citenamefont {{Austermann}}, \citenamefont {{Beall}},
  \citenamefont {{Bender}}, \citenamefont {{Benson}}, \citenamefont {{Bleem}},
  \citenamefont {{Carlstrom}}, \citenamefont {{Chang}}, \citenamefont
  {{Chiang}}, \citenamefont {{Cho}}, \citenamefont {{Citron}}, \citenamefont
  {{Crawford}}, \citenamefont {{Crites}}, \citenamefont {{de Haan}},
  \citenamefont {{Dobbs}}, \citenamefont {{Everett}}, \citenamefont
  {{Gallicchio}}, \citenamefont {{Gao}}, \citenamefont {{George}},
  \citenamefont {{Gilbert}}, \citenamefont {{Halverson}}, \citenamefont
  {{Harrington}}, \citenamefont {{Henning}}, \citenamefont {{Hilton}},
  \citenamefont {{Holder}}, \citenamefont {{Holzapfel}}, \citenamefont
  {{Hoover}}, \citenamefont {{Hou}}, \citenamefont {{Hrubes}}, \citenamefont
  {{Huang}}, \citenamefont {{Hubmayr}}, \citenamefont {{Irwin}}, \citenamefont
  {{Keisler}}, \citenamefont {{Knox}}, \citenamefont {{Lee}}, \citenamefont
  {{Leitch}}, \citenamefont {{Li}}, \citenamefont {{Liang}}, \citenamefont
  {{Luong-Van}}, \citenamefont {{McMahon}}, \citenamefont {{Mehl}},
  \citenamefont {{Meyer}}, \citenamefont {{Mocanu}}, \citenamefont {{Montroy}},
  \citenamefont {{Natoli}}, \citenamefont {{Nibarger}}, \citenamefont
  {{Novosad}}, \citenamefont {{Padin}}, \citenamefont {{Pryke}}, \citenamefont
  {{Reichardt}}, \citenamefont {{Ruhl}}, \citenamefont {{Saliwanchik}},
  \citenamefont {{Sayre}}, \citenamefont {{Schaffer}}, \citenamefont
  {{Smecher}}, \citenamefont {{Stark}}, \citenamefont {{Tucker}}, \citenamefont
  {{Vanderlinde}}, \citenamefont {{Vieira}}, \citenamefont {{Wang}},
  \citenamefont {{Whitehorn}}, \citenamefont {{Yefremenko}},\ and\
  \citenamefont {{Zahn}}}]{2015StorySPT}%
  \BibitemOpen
  \bibfield  {author} {\bibinfo {author} {\bibfnamefont {K.~T.}\ \bibnamefont
  {{Story}}}, \bibinfo {author} {\bibfnamefont {D.}~\bibnamefont {{Hanson}}},
  \bibinfo {author} {\bibfnamefont {P.~A.~R.}\ \bibnamefont {{Ade}}}, \bibinfo
  {author} {\bibfnamefont {K.~A.}\ \bibnamefont {{Aird}}}, \bibinfo {author}
  {\bibfnamefont {J.~E.}\ \bibnamefont {{Austermann}}}, \bibinfo {author}
  {\bibfnamefont {J.~A.}\ \bibnamefont {{Beall}}}, \bibinfo {author}
  {\bibfnamefont {A.~N.}\ \bibnamefont {{Bender}}}, \bibinfo {author}
  {\bibfnamefont {B.~A.}\ \bibnamefont {{Benson}}}, \bibinfo {author}
  {\bibfnamefont {L.~E.}\ \bibnamefont {{Bleem}}}, \bibinfo {author}
  {\bibfnamefont {J.~E.}\ \bibnamefont {{Carlstrom}}}, \bibinfo {author}
  {\bibfnamefont {C.~L.}\ \bibnamefont {{Chang}}}, \bibinfo {author}
  {\bibfnamefont {H.~C.}\ \bibnamefont {{Chiang}}}, \bibinfo {author}
  {\bibfnamefont {H.-M.}\ \bibnamefont {{Cho}}}, \bibinfo {author}
  {\bibfnamefont {R.}~\bibnamefont {{Citron}}}, \bibinfo {author}
  {\bibfnamefont {T.~M.}\ \bibnamefont {{Crawford}}}, \bibinfo {author}
  {\bibfnamefont {A.~T.}\ \bibnamefont {{Crites}}}, \bibinfo {author}
  {\bibfnamefont {T.}~\bibnamefont {{de Haan}}}, \bibinfo {author}
  {\bibfnamefont {M.~A.}\ \bibnamefont {{Dobbs}}}, \bibinfo {author}
  {\bibfnamefont {W.}~\bibnamefont {{Everett}}}, \bibinfo {author}
  {\bibfnamefont {J.}~\bibnamefont {{Gallicchio}}}, \bibinfo {author}
  {\bibfnamefont {J.}~\bibnamefont {{Gao}}}, \bibinfo {author} {\bibfnamefont
  {E.~M.}\ \bibnamefont {{George}}}, \bibinfo {author} {\bibfnamefont
  {A.}~\bibnamefont {{Gilbert}}}, \bibinfo {author} {\bibfnamefont {N.~W.}\
  \bibnamefont {{Halverson}}}, \bibinfo {author} {\bibfnamefont
  {N.}~\bibnamefont {{Harrington}}}, \bibinfo {author} {\bibfnamefont {J.~W.}\
  \bibnamefont {{Henning}}}, \bibinfo {author} {\bibfnamefont {G.~C.}\
  \bibnamefont {{Hilton}}}, \bibinfo {author} {\bibfnamefont {G.~P.}\
  \bibnamefont {{Holder}}}, \bibinfo {author} {\bibfnamefont {W.~L.}\
  \bibnamefont {{Holzapfel}}}, \bibinfo {author} {\bibfnamefont
  {S.}~\bibnamefont {{Hoover}}}, \bibinfo {author} {\bibfnamefont
  {Z.}~\bibnamefont {{Hou}}}, \bibinfo {author} {\bibfnamefont {J.~D.}\
  \bibnamefont {{Hrubes}}}, \bibinfo {author} {\bibfnamefont {N.}~\bibnamefont
  {{Huang}}}, \bibinfo {author} {\bibfnamefont {J.}~\bibnamefont {{Hubmayr}}},
  \bibinfo {author} {\bibfnamefont {K.~D.}\ \bibnamefont {{Irwin}}}, \bibinfo
  {author} {\bibfnamefont {R.}~\bibnamefont {{Keisler}}}, \bibinfo {author}
  {\bibfnamefont {L.}~\bibnamefont {{Knox}}}, \bibinfo {author} {\bibfnamefont
  {A.~T.}\ \bibnamefont {{Lee}}}, \bibinfo {author} {\bibfnamefont {E.~M.}\
  \bibnamefont {{Leitch}}}, \bibinfo {author} {\bibfnamefont {D.}~\bibnamefont
  {{Li}}}, \bibinfo {author} {\bibfnamefont {C.}~\bibnamefont {{Liang}}},
  \bibinfo {author} {\bibfnamefont {D.}~\bibnamefont {{Luong-Van}}}, \bibinfo
  {author} {\bibfnamefont {J.~J.}\ \bibnamefont {{McMahon}}}, \bibinfo {author}
  {\bibfnamefont {J.}~\bibnamefont {{Mehl}}}, \bibinfo {author} {\bibfnamefont
  {S.~S.}\ \bibnamefont {{Meyer}}}, \bibinfo {author} {\bibfnamefont
  {L.}~\bibnamefont {{Mocanu}}}, \bibinfo {author} {\bibfnamefont {T.~E.}\
  \bibnamefont {{Montroy}}}, \bibinfo {author} {\bibfnamefont {T.}~\bibnamefont
  {{Natoli}}}, \bibinfo {author} {\bibfnamefont {J.~P.}\ \bibnamefont
  {{Nibarger}}}, \bibinfo {author} {\bibfnamefont {V.}~\bibnamefont
  {{Novosad}}}, \bibinfo {author} {\bibfnamefont {S.}~\bibnamefont {{Padin}}},
  \bibinfo {author} {\bibfnamefont {C.}~\bibnamefont {{Pryke}}}, \bibinfo
  {author} {\bibfnamefont {C.~L.}\ \bibnamefont {{Reichardt}}}, \bibinfo
  {author} {\bibfnamefont {J.~E.}\ \bibnamefont {{Ruhl}}}, \bibinfo {author}
  {\bibfnamefont {B.~R.}\ \bibnamefont {{Saliwanchik}}}, \bibinfo {author}
  {\bibfnamefont {J.~T.}\ \bibnamefont {{Sayre}}}, \bibinfo {author}
  {\bibfnamefont {K.~K.}\ \bibnamefont {{Schaffer}}}, \bibinfo {author}
  {\bibfnamefont {G.}~\bibnamefont {{Smecher}}}, \bibinfo {author}
  {\bibfnamefont {A.~A.}\ \bibnamefont {{Stark}}}, \bibinfo {author}
  {\bibfnamefont {C.}~\bibnamefont {{Tucker}}}, \bibinfo {author}
  {\bibfnamefont {K.}~\bibnamefont {{Vanderlinde}}}, \bibinfo {author}
  {\bibfnamefont {J.~D.}\ \bibnamefont {{Vieira}}}, \bibinfo {author}
  {\bibfnamefont {G.}~\bibnamefont {{Wang}}}, \bibinfo {author} {\bibfnamefont
  {N.}~\bibnamefont {{Whitehorn}}}, \bibinfo {author} {\bibfnamefont
  {V.}~\bibnamefont {{Yefremenko}}}, \ and\ \bibinfo {author} {\bibfnamefont
  {O.}~\bibnamefont {{Zahn}}},\ }\href {\doibase 10.1088/0004-637X/810/1/50}
  {\bibfield  {journal} {\bibinfo  {journal} {\apj}\ }\textbf {\bibinfo
  {volume} {810}},\ \bibinfo {eid} {50} (\bibinfo {year} {2015})},\ \Eprint
  {http://arxiv.org/abs/1412.4760} {arXiv:1412.4760} \BibitemShut {NoStop}%
\bibitem [{\citenamefont {{Planck Collaboration}}\ \emph
  {et~al.}(2016)\citenamefont {{Planck Collaboration}}, \citenamefont {{Ade}},
  \citenamefont {{Aghanim}}, \citenamefont {{Arnaud}}, \citenamefont
  {{Ashdown}}, \citenamefont {{Aumont}}, \citenamefont {{Baccigalupi}},
  \citenamefont {{Banday}}, \citenamefont {{Barreiro}}, \citenamefont
  {{Bartlett}},\ and\ \citenamefont {et~al.}}]{2016PlanckLensing}%
  \BibitemOpen
  \bibfield  {author} {\bibinfo {author} {\bibnamefont {{Planck
  Collaboration}}}, \bibinfo {author} {\bibfnamefont {P.~A.~R.}\ \bibnamefont
  {{Ade}}}, \bibinfo {author} {\bibfnamefont {N.}~\bibnamefont {{Aghanim}}},
  \bibinfo {author} {\bibfnamefont {M.}~\bibnamefont {{Arnaud}}}, \bibinfo
  {author} {\bibfnamefont {M.}~\bibnamefont {{Ashdown}}}, \bibinfo {author}
  {\bibfnamefont {J.}~\bibnamefont {{Aumont}}}, \bibinfo {author}
  {\bibfnamefont {C.}~\bibnamefont {{Baccigalupi}}}, \bibinfo {author}
  {\bibfnamefont {A.~J.}\ \bibnamefont {{Banday}}}, \bibinfo {author}
  {\bibfnamefont {R.~B.}\ \bibnamefont {{Barreiro}}}, \bibinfo {author}
  {\bibfnamefont {J.~G.}\ \bibnamefont {{Bartlett}}}, \ and\ \bibinfo {author}
  {\bibnamefont {et~al.}},\ }\href {\doibase 10.1051/0004-6361/201525941}
  {\bibfield  {journal} {\bibinfo  {journal} {\aap}\ }\textbf {\bibinfo
  {volume} {594}},\ \bibinfo {eid} {A15} (\bibinfo {year} {2016})},\ \Eprint
  {http://arxiv.org/abs/1502.01591} {arXiv:1502.01591} \BibitemShut {NoStop}%
\bibitem [{\citenamefont {{Sherwin}}\ \emph {et~al.}(2017)\citenamefont
  {{Sherwin}}, \citenamefont {{van Engelen}}, \citenamefont {{Sehgal}},
  \citenamefont {{Madhavacheril}}, \citenamefont {{Addison}}, \citenamefont
  {{Aiola}}, \citenamefont {{Allison}}, \citenamefont {{Battaglia}},
  \citenamefont {{Becker}}, \citenamefont {{Beall}}, \citenamefont {{Bond}},
  \citenamefont {{Calabrese}}, \citenamefont {{Datta}}, \citenamefont
  {{Devlin}}, \citenamefont {{D{\"u}nner}}, \citenamefont {{Dunkley}},
  \citenamefont {{Fox}}, \citenamefont {{Gallardo}}, \citenamefont {{Halpern}},
  \citenamefont {{Hasselfield}}, \citenamefont {{Henderson}}, \citenamefont
  {{Hill}}, \citenamefont {{Hilton}}, \citenamefont {{Hubmayr}}, \citenamefont
  {{Hughes}}, \citenamefont {{Hincks}}, \citenamefont {{Hlozek}}, \citenamefont
  {{Huffenberger}}, \citenamefont {{Koopman}}, \citenamefont {{Kosowsky}},
  \citenamefont {{Louis}}, \citenamefont {{Maurin}}, \citenamefont {{McMahon}},
  \citenamefont {{Moodley}}, \citenamefont {{Naess}}, \citenamefont {{Nati}},
  \citenamefont {{Newburgh}}, \citenamefont {{Niemack}}, \citenamefont
  {{Page}}, \citenamefont {{Sievers}}, \citenamefont {{Spergel}}, \citenamefont
  {{Staggs}}, \citenamefont {{Thornton}}, \citenamefont {{Van Lanen}},
  \citenamefont {{Vavagiakis}},\ and\ \citenamefont {{Wollack}}}]{2017ACT}%
  \BibitemOpen
  \bibfield  {author} {\bibinfo {author} {\bibfnamefont {B.~D.}\ \bibnamefont
  {{Sherwin}}}, \bibinfo {author} {\bibfnamefont {A.}~\bibnamefont {{van
  Engelen}}}, \bibinfo {author} {\bibfnamefont {N.}~\bibnamefont {{Sehgal}}},
  \bibinfo {author} {\bibfnamefont {M.}~\bibnamefont {{Madhavacheril}}},
  \bibinfo {author} {\bibfnamefont {G.~E.}\ \bibnamefont {{Addison}}}, \bibinfo
  {author} {\bibfnamefont {S.}~\bibnamefont {{Aiola}}}, \bibinfo {author}
  {\bibfnamefont {R.}~\bibnamefont {{Allison}}}, \bibinfo {author}
  {\bibfnamefont {N.}~\bibnamefont {{Battaglia}}}, \bibinfo {author}
  {\bibfnamefont {D.~T.}\ \bibnamefont {{Becker}}}, \bibinfo {author}
  {\bibfnamefont {J.~A.}\ \bibnamefont {{Beall}}}, \bibinfo {author}
  {\bibfnamefont {J.~R.}\ \bibnamefont {{Bond}}}, \bibinfo {author}
  {\bibfnamefont {E.}~\bibnamefont {{Calabrese}}}, \bibinfo {author}
  {\bibfnamefont {R.}~\bibnamefont {{Datta}}}, \bibinfo {author} {\bibfnamefont
  {M.~J.}\ \bibnamefont {{Devlin}}}, \bibinfo {author} {\bibfnamefont
  {R.}~\bibnamefont {{D{\"u}nner}}}, \bibinfo {author} {\bibfnamefont
  {J.}~\bibnamefont {{Dunkley}}}, \bibinfo {author} {\bibfnamefont {A.~E.}\
  \bibnamefont {{Fox}}}, \bibinfo {author} {\bibfnamefont {P.}~\bibnamefont
  {{Gallardo}}}, \bibinfo {author} {\bibfnamefont {M.}~\bibnamefont
  {{Halpern}}}, \bibinfo {author} {\bibfnamefont {M.}~\bibnamefont
  {{Hasselfield}}}, \bibinfo {author} {\bibfnamefont {S.}~\bibnamefont
  {{Henderson}}}, \bibinfo {author} {\bibfnamefont {J.~C.}\ \bibnamefont
  {{Hill}}}, \bibinfo {author} {\bibfnamefont {G.~C.}\ \bibnamefont
  {{Hilton}}}, \bibinfo {author} {\bibfnamefont {J.}~\bibnamefont {{Hubmayr}}},
  \bibinfo {author} {\bibfnamefont {J.~P.}\ \bibnamefont {{Hughes}}}, \bibinfo
  {author} {\bibfnamefont {A.~D.}\ \bibnamefont {{Hincks}}}, \bibinfo {author}
  {\bibfnamefont {R.}~\bibnamefont {{Hlozek}}}, \bibinfo {author}
  {\bibfnamefont {K.~M.}\ \bibnamefont {{Huffenberger}}}, \bibinfo {author}
  {\bibfnamefont {B.}~\bibnamefont {{Koopman}}}, \bibinfo {author}
  {\bibfnamefont {A.}~\bibnamefont {{Kosowsky}}}, \bibinfo {author}
  {\bibfnamefont {T.}~\bibnamefont {{Louis}}}, \bibinfo {author} {\bibfnamefont
  {L.}~\bibnamefont {{Maurin}}}, \bibinfo {author} {\bibfnamefont
  {J.}~\bibnamefont {{McMahon}}}, \bibinfo {author} {\bibfnamefont
  {K.}~\bibnamefont {{Moodley}}}, \bibinfo {author} {\bibfnamefont
  {S.}~\bibnamefont {{Naess}}}, \bibinfo {author} {\bibfnamefont
  {F.}~\bibnamefont {{Nati}}}, \bibinfo {author} {\bibfnamefont
  {L.}~\bibnamefont {{Newburgh}}}, \bibinfo {author} {\bibfnamefont {M.~D.}\
  \bibnamefont {{Niemack}}}, \bibinfo {author} {\bibfnamefont {L.~A.}\
  \bibnamefont {{Page}}}, \bibinfo {author} {\bibfnamefont {J.}~\bibnamefont
  {{Sievers}}}, \bibinfo {author} {\bibfnamefont {D.~N.}\ \bibnamefont
  {{Spergel}}}, \bibinfo {author} {\bibfnamefont {S.~T.}\ \bibnamefont
  {{Staggs}}}, \bibinfo {author} {\bibfnamefont {R.~J.}\ \bibnamefont
  {{Thornton}}}, \bibinfo {author} {\bibfnamefont {J.}~\bibnamefont {{Van
  Lanen}}}, \bibinfo {author} {\bibfnamefont {E.}~\bibnamefont {{Vavagiakis}}},
  \ and\ \bibinfo {author} {\bibfnamefont {E.~J.}\ \bibnamefont {{Wollack}}},\
  }\href {\doibase 10.1103/PhysRevD.95.123529} {\bibfield  {journal} {\bibinfo
  {journal} {\prd}\ }\textbf {\bibinfo {volume} {95}},\ \bibinfo {eid} {123529}
  (\bibinfo {year} {2017})},\ \Eprint {http://arxiv.org/abs/1611.09753}
  {arXiv:1611.09753} \BibitemShut {NoStop}%
\bibitem [{\citenamefont {{Simard}}\ \emph {et~al.}(2017)\citenamefont
  {{Simard}}, \citenamefont {{Omori}}, \citenamefont {{Aylor}}, \citenamefont
  {{Baxter}}, \citenamefont {{Benson}}, \citenamefont {{Bleem}}, \citenamefont
  {{Carlstrom}}, \citenamefont {{Chang}}, \citenamefont {{Cho}}, \citenamefont
  {{Chown}}, \citenamefont {{Crawford}}, \citenamefont {{Crites}},
  \citenamefont {{de Haan}}, \citenamefont {{Dobbs}}, \citenamefont
  {{Everett}}, \citenamefont {{George}}, \citenamefont {{Halverson}},
  \citenamefont {{Harrington}}, \citenamefont {{Henning}}, \citenamefont
  {{Holder}}, \citenamefont {{Hou}}, \citenamefont {{Holzapfel}}, \citenamefont
  {{Hrubes}}, \citenamefont {{Knox}}, \citenamefont {{Lee}}, \citenamefont
  {{Leitch}}, \citenamefont {{Luong-Van}}, \citenamefont {{Manzotti}},
  \citenamefont {{McMahon}}, \citenamefont {{Meyer}}, \citenamefont {{Mocanu}},
  \citenamefont {{Mohr}}, \citenamefont {{Natoli}}, \citenamefont {{Padin}},
  \citenamefont {{Pryke}}, \citenamefont {{Reichardt}}, \citenamefont {{Ruhl}},
  \citenamefont {{Sayre}}, \citenamefont {{Schaffer}}, \citenamefont
  {{Shirokoff}}, \citenamefont {{Staniszewski}}, \citenamefont {{Stark}},
  \citenamefont {{Story}}, \citenamefont {{Vanderlinde}}, \citenamefont
  {{Vieira}}, \citenamefont {{Williamson}},\ and\ \citenamefont
  {{Wu}}}]{2017SPTParams}%
  \BibitemOpen
  \bibfield  {author} {\bibinfo {author} {\bibfnamefont {G.}~\bibnamefont
  {{Simard}}}, \bibinfo {author} {\bibfnamefont {Y.}~\bibnamefont {{Omori}}},
  \bibinfo {author} {\bibfnamefont {K.}~\bibnamefont {{Aylor}}}, \bibinfo
  {author} {\bibfnamefont {E.~J.}\ \bibnamefont {{Baxter}}}, \bibinfo {author}
  {\bibfnamefont {B.~A.}\ \bibnamefont {{Benson}}}, \bibinfo {author}
  {\bibfnamefont {L.~E.}\ \bibnamefont {{Bleem}}}, \bibinfo {author}
  {\bibfnamefont {J.~E.}\ \bibnamefont {{Carlstrom}}}, \bibinfo {author}
  {\bibfnamefont {C.~L.}\ \bibnamefont {{Chang}}}, \bibinfo {author}
  {\bibfnamefont {H.}~\bibnamefont {{Cho}}}, \bibinfo {author} {\bibfnamefont
  {R.}~\bibnamefont {{Chown}}}, \bibinfo {author} {\bibfnamefont {T.~M.}\
  \bibnamefont {{Crawford}}}, \bibinfo {author} {\bibfnamefont {A.~T.}\
  \bibnamefont {{Crites}}}, \bibinfo {author} {\bibfnamefont {T.}~\bibnamefont
  {{de Haan}}}, \bibinfo {author} {\bibfnamefont {M.~A.}\ \bibnamefont
  {{Dobbs}}}, \bibinfo {author} {\bibfnamefont {W.~B.}\ \bibnamefont
  {{Everett}}}, \bibinfo {author} {\bibfnamefont {E.~M.}\ \bibnamefont
  {{George}}}, \bibinfo {author} {\bibfnamefont {N.~W.}\ \bibnamefont
  {{Halverson}}}, \bibinfo {author} {\bibfnamefont {N.~L.}\ \bibnamefont
  {{Harrington}}}, \bibinfo {author} {\bibfnamefont {J.~W.}\ \bibnamefont
  {{Henning}}}, \bibinfo {author} {\bibfnamefont {G.~P.}\ \bibnamefont
  {{Holder}}}, \bibinfo {author} {\bibfnamefont {Z.}~\bibnamefont {{Hou}}},
  \bibinfo {author} {\bibfnamefont {W.~L.}\ \bibnamefont {{Holzapfel}}},
  \bibinfo {author} {\bibfnamefont {J.~D.}\ \bibnamefont {{Hrubes}}}, \bibinfo
  {author} {\bibfnamefont {L.}~\bibnamefont {{Knox}}}, \bibinfo {author}
  {\bibfnamefont {A.~T.}\ \bibnamefont {{Lee}}}, \bibinfo {author}
  {\bibfnamefont {E.~M.}\ \bibnamefont {{Leitch}}}, \bibinfo {author}
  {\bibfnamefont {D.}~\bibnamefont {{Luong-Van}}}, \bibinfo {author}
  {\bibfnamefont {A.}~\bibnamefont {{Manzotti}}}, \bibinfo {author}
  {\bibfnamefont {J.~J.}\ \bibnamefont {{McMahon}}}, \bibinfo {author}
  {\bibfnamefont {S.~S.}\ \bibnamefont {{Meyer}}}, \bibinfo {author}
  {\bibfnamefont {L.~M.}\ \bibnamefont {{Mocanu}}}, \bibinfo {author}
  {\bibfnamefont {J.~J.}\ \bibnamefont {{Mohr}}}, \bibinfo {author}
  {\bibfnamefont {T.}~\bibnamefont {{Natoli}}}, \bibinfo {author}
  {\bibfnamefont {S.}~\bibnamefont {{Padin}}}, \bibinfo {author} {\bibfnamefont
  {C.}~\bibnamefont {{Pryke}}}, \bibinfo {author} {\bibfnamefont {C.~L.}\
  \bibnamefont {{Reichardt}}}, \bibinfo {author} {\bibfnamefont {J.~E.}\
  \bibnamefont {{Ruhl}}}, \bibinfo {author} {\bibfnamefont {J.~T.}\
  \bibnamefont {{Sayre}}}, \bibinfo {author} {\bibfnamefont {K.~K.}\
  \bibnamefont {{Schaffer}}}, \bibinfo {author} {\bibfnamefont
  {E.}~\bibnamefont {{Shirokoff}}}, \bibinfo {author} {\bibfnamefont
  {Z.}~\bibnamefont {{Staniszewski}}}, \bibinfo {author} {\bibfnamefont
  {A.~A.}\ \bibnamefont {{Stark}}}, \bibinfo {author} {\bibfnamefont {K.~T.}\
  \bibnamefont {{Story}}}, \bibinfo {author} {\bibfnamefont {K.}~\bibnamefont
  {{Vanderlinde}}}, \bibinfo {author} {\bibfnamefont {J.~D.}\ \bibnamefont
  {{Vieira}}}, \bibinfo {author} {\bibfnamefont {R.}~\bibnamefont
  {{Williamson}}}, \ and\ \bibinfo {author} {\bibfnamefont {W.~L.~K.}\
  \bibnamefont {{Wu}}},\ }\href@noop {} {\bibfield  {journal} {\bibinfo
  {journal} {ArXiv e-prints}\ } (\bibinfo {year} {2017})},\ \Eprint
  {http://arxiv.org/abs/1712.07541} {arXiv:1712.07541} \BibitemShut {NoStop}%
\bibitem [{\citenamefont {{Henderson}}\ \emph {et~al.}(2016)\citenamefont
  {{Henderson}}, \citenamefont {{Allison}}, \citenamefont {{Austermann}},
  \citenamefont {{Baildon}}, \citenamefont {{Battaglia}}, \citenamefont
  {{Beall}}, \citenamefont {{Becker}}, \citenamefont {{De Bernardis}},
  \citenamefont {{Bond}}, \citenamefont {{Calabrese}}, \citenamefont {{Choi}},
  \citenamefont {{Coughlin}}, \citenamefont {{Crowley}}, \citenamefont
  {{Datta}}, \citenamefont {{Devlin}}, \citenamefont {{Duff}}, \citenamefont
  {{Dunkley}}, \citenamefont {{D{\"u}nner}}, \citenamefont {{van Engelen}},
  \citenamefont {{Gallardo}}, \citenamefont {{Grace}}, \citenamefont
  {{Hasselfield}}, \citenamefont {{Hills}}, \citenamefont {{Hilton}},
  \citenamefont {{Hincks}}, \citenamefont {{Hlo{\^z}ek}}, \citenamefont {{Ho}},
  \citenamefont {{Hubmayr}}, \citenamefont {{Huffenberger}}, \citenamefont
  {{Hughes}}, \citenamefont {{Irwin}}, \citenamefont {{Koopman}}, \citenamefont
  {{Kosowsky}}, \citenamefont {{Li}}, \citenamefont {{McMahon}}, \citenamefont
  {{Munson}}, \citenamefont {{Nati}}, \citenamefont {{Newburgh}}, \citenamefont
  {{Niemack}}, \citenamefont {{Niraula}}, \citenamefont {{Page}}, \citenamefont
  {{Pappas}}, \citenamefont {{Salatino}}, \citenamefont {{Schillaci}},
  \citenamefont {{Schmitt}}, \citenamefont {{Sehgal}}, \citenamefont
  {{Sherwin}}, \citenamefont {{Sievers}}, \citenamefont {{Simon}},
  \citenamefont {{Spergel}}, \citenamefont {{Staggs}}, \citenamefont
  {{Stevens}}, \citenamefont {{Thornton}}, \citenamefont {{Van Lanen}},
  \citenamefont {{Vavagiakis}}, \citenamefont {{Ward}},\ and\ \citenamefont
  {{Wollack}}}]{AdvACT}%
  \BibitemOpen
  \bibfield  {author} {\bibinfo {author} {\bibfnamefont {S.~W.}\ \bibnamefont
  {{Henderson}}}, \bibinfo {author} {\bibfnamefont {R.}~\bibnamefont
  {{Allison}}}, \bibinfo {author} {\bibfnamefont {J.}~\bibnamefont
  {{Austermann}}}, \bibinfo {author} {\bibfnamefont {T.}~\bibnamefont
  {{Baildon}}}, \bibinfo {author} {\bibfnamefont {N.}~\bibnamefont
  {{Battaglia}}}, \bibinfo {author} {\bibfnamefont {J.~A.}\ \bibnamefont
  {{Beall}}}, \bibinfo {author} {\bibfnamefont {D.}~\bibnamefont {{Becker}}},
  \bibinfo {author} {\bibfnamefont {F.}~\bibnamefont {{De Bernardis}}},
  \bibinfo {author} {\bibfnamefont {J.~R.}\ \bibnamefont {{Bond}}}, \bibinfo
  {author} {\bibfnamefont {E.}~\bibnamefont {{Calabrese}}}, \bibinfo {author}
  {\bibfnamefont {S.~K.}\ \bibnamefont {{Choi}}}, \bibinfo {author}
  {\bibfnamefont {K.~P.}\ \bibnamefont {{Coughlin}}}, \bibinfo {author}
  {\bibfnamefont {K.~T.}\ \bibnamefont {{Crowley}}}, \bibinfo {author}
  {\bibfnamefont {R.}~\bibnamefont {{Datta}}}, \bibinfo {author} {\bibfnamefont
  {M.~J.}\ \bibnamefont {{Devlin}}}, \bibinfo {author} {\bibfnamefont {S.~M.}\
  \bibnamefont {{Duff}}}, \bibinfo {author} {\bibfnamefont {J.}~\bibnamefont
  {{Dunkley}}}, \bibinfo {author} {\bibfnamefont {R.}~\bibnamefont
  {{D{\"u}nner}}}, \bibinfo {author} {\bibfnamefont {A.}~\bibnamefont {{van
  Engelen}}}, \bibinfo {author} {\bibfnamefont {P.~A.}\ \bibnamefont
  {{Gallardo}}}, \bibinfo {author} {\bibfnamefont {E.}~\bibnamefont {{Grace}}},
  \bibinfo {author} {\bibfnamefont {M.}~\bibnamefont {{Hasselfield}}}, \bibinfo
  {author} {\bibfnamefont {F.}~\bibnamefont {{Hills}}}, \bibinfo {author}
  {\bibfnamefont {G.~C.}\ \bibnamefont {{Hilton}}}, \bibinfo {author}
  {\bibfnamefont {A.~D.}\ \bibnamefont {{Hincks}}}, \bibinfo {author}
  {\bibfnamefont {R.}~\bibnamefont {{Hlo{\^z}ek}}}, \bibinfo {author}
  {\bibfnamefont {S.~P.}\ \bibnamefont {{Ho}}}, \bibinfo {author}
  {\bibfnamefont {J.}~\bibnamefont {{Hubmayr}}}, \bibinfo {author}
  {\bibfnamefont {K.}~\bibnamefont {{Huffenberger}}}, \bibinfo {author}
  {\bibfnamefont {J.~P.}\ \bibnamefont {{Hughes}}}, \bibinfo {author}
  {\bibfnamefont {K.~D.}\ \bibnamefont {{Irwin}}}, \bibinfo {author}
  {\bibfnamefont {B.~J.}\ \bibnamefont {{Koopman}}}, \bibinfo {author}
  {\bibfnamefont {A.~B.}\ \bibnamefont {{Kosowsky}}}, \bibinfo {author}
  {\bibfnamefont {D.}~\bibnamefont {{Li}}}, \bibinfo {author} {\bibfnamefont
  {J.}~\bibnamefont {{McMahon}}}, \bibinfo {author} {\bibfnamefont
  {C.}~\bibnamefont {{Munson}}}, \bibinfo {author} {\bibfnamefont
  {F.}~\bibnamefont {{Nati}}}, \bibinfo {author} {\bibfnamefont
  {L.}~\bibnamefont {{Newburgh}}}, \bibinfo {author} {\bibfnamefont {M.~D.}\
  \bibnamefont {{Niemack}}}, \bibinfo {author} {\bibfnamefont {P.}~\bibnamefont
  {{Niraula}}}, \bibinfo {author} {\bibfnamefont {L.~A.}\ \bibnamefont
  {{Page}}}, \bibinfo {author} {\bibfnamefont {C.~G.}\ \bibnamefont
  {{Pappas}}}, \bibinfo {author} {\bibfnamefont {M.}~\bibnamefont
  {{Salatino}}}, \bibinfo {author} {\bibfnamefont {A.}~\bibnamefont
  {{Schillaci}}}, \bibinfo {author} {\bibfnamefont {B.~L.}\ \bibnamefont
  {{Schmitt}}}, \bibinfo {author} {\bibfnamefont {N.}~\bibnamefont {{Sehgal}}},
  \bibinfo {author} {\bibfnamefont {B.~D.}\ \bibnamefont {{Sherwin}}}, \bibinfo
  {author} {\bibfnamefont {J.~L.}\ \bibnamefont {{Sievers}}}, \bibinfo {author}
  {\bibfnamefont {S.~M.}\ \bibnamefont {{Simon}}}, \bibinfo {author}
  {\bibfnamefont {D.~N.}\ \bibnamefont {{Spergel}}}, \bibinfo {author}
  {\bibfnamefont {S.~T.}\ \bibnamefont {{Staggs}}}, \bibinfo {author}
  {\bibfnamefont {J.~R.}\ \bibnamefont {{Stevens}}}, \bibinfo {author}
  {\bibfnamefont {R.}~\bibnamefont {{Thornton}}}, \bibinfo {author}
  {\bibfnamefont {J.}~\bibnamefont {{Van Lanen}}}, \bibinfo {author}
  {\bibfnamefont {E.~M.}\ \bibnamefont {{Vavagiakis}}}, \bibinfo {author}
  {\bibfnamefont {J.~T.}\ \bibnamefont {{Ward}}}, \ and\ \bibinfo {author}
  {\bibfnamefont {E.~J.}\ \bibnamefont {{Wollack}}},\ }\href {\doibase
  10.1007/s10909-016-1575-z} {\bibfield  {journal} {\bibinfo  {journal}
  {Journal of Low Temperature Physics}\ }\textbf {\bibinfo {volume} {184}},\
  \bibinfo {pages} {772} (\bibinfo {year} {2016})},\ \Eprint
  {http://arxiv.org/abs/1510.02809} {arXiv:1510.02809 [astro-ph.IM]}
  \BibitemShut {NoStop}%
\bibitem [{\citenamefont {{Suzuki}}\ \emph {et~al.}(2016)\citenamefont
  {{Suzuki}}, \citenamefont {{Ade}}, \citenamefont {{Akiba}}, \citenamefont
  {{Aleman}}, \citenamefont {{Arnold}}, \citenamefont {{Baccigalupi}},
  \citenamefont {{Barch}}, \citenamefont {{Barron}}, \citenamefont {{Bender}},
  \citenamefont {{Boettger}}, \citenamefont {{Borrill}}, \citenamefont
  {{Chapman}}, \citenamefont {{Chinone}}, \citenamefont {{Cukierman}},
  \citenamefont {{Dobbs}}, \citenamefont {{Ducout}}, \citenamefont {{Dunner}},
  \citenamefont {{Elleflot}}, \citenamefont {{Errard}}, \citenamefont
  {{Fabbian}}, \citenamefont {{Feeney}}, \citenamefont {{Feng}}, \citenamefont
  {{Fujino}}, \citenamefont {{Fuller}}, \citenamefont {{Gilbert}},
  \citenamefont {{Goeckner-Wald}}, \citenamefont {{Groh}}, \citenamefont
  {{Haan}}, \citenamefont {{Hall}}, \citenamefont {{Halverson}}, \citenamefont
  {{Hamada}}, \citenamefont {{Hasegawa}}, \citenamefont {{Hattori}},
  \citenamefont {{Hazumi}}, \citenamefont {{Hill}}, \citenamefont
  {{Holzapfel}}, \citenamefont {{Hori}}, \citenamefont {{Howe}}, \citenamefont
  {{Inoue}}, \citenamefont {{Irie}}, \citenamefont {{Jaehnig}}, \citenamefont
  {{Jaffe}}, \citenamefont {{Jeong}}, \citenamefont {{Katayama}}, \citenamefont
  {{Kaufman}}, \citenamefont {{Kazemzadeh}}, \citenamefont {{Keating}},
  \citenamefont {{Kermish}}, \citenamefont {{Keskitalo}}, \citenamefont
  {{Kisner}}, \citenamefont {{Kusaka}}, \citenamefont {{Jeune}}, \citenamefont
  {{Lee}}, \citenamefont {{Leon}}, \citenamefont {{Linder}}, \citenamefont
  {{Lowry}}, \citenamefont {{Matsuda}}, \citenamefont {{Matsumura}},
  \citenamefont {{Miller}}, \citenamefont {{Mizukami}}, \citenamefont
  {{Montgomery}}, \citenamefont {{Navaroli}}, \citenamefont {{Nishino}},
  \citenamefont {{Peloton}}, \citenamefont {{Poletti}}, \citenamefont
  {{Puglisi}}, \citenamefont {{Rebeiz}}, \citenamefont {{Raum}}, \citenamefont
  {{Reichardt}}, \citenamefont {{Richards}}, \citenamefont {{Ross}},
  \citenamefont {{Rotermund}}, \citenamefont {{Segawa}}, \citenamefont
  {{Sherwin}}, \citenamefont {{Shirley}}, \citenamefont {{Siritanasak}},
  \citenamefont {{Stebor}}, \citenamefont {{Stompor}}, \citenamefont
  {{Suzuki}}, \citenamefont {{Tajima}}, \citenamefont {{Takada}}, \citenamefont
  {{Takakura}}, \citenamefont {{Takatori}}, \citenamefont {{Tikhomirov}},
  \citenamefont {{Tomaru}}, \citenamefont {{Westbrook}}, \citenamefont
  {{Whitehorn}}, \citenamefont {{Yamashita}}, \citenamefont {{Zahn}},\ and\
  \citenamefont {{Zahn}}}]{SimArr}%
  \BibitemOpen
  \bibfield  {author} {\bibinfo {author} {\bibfnamefont {A.}~\bibnamefont
  {{Suzuki}}}, \bibinfo {author} {\bibfnamefont {P.}~\bibnamefont {{Ade}}},
  \bibinfo {author} {\bibfnamefont {Y.}~\bibnamefont {{Akiba}}}, \bibinfo
  {author} {\bibfnamefont {C.}~\bibnamefont {{Aleman}}}, \bibinfo {author}
  {\bibfnamefont {K.}~\bibnamefont {{Arnold}}}, \bibinfo {author}
  {\bibfnamefont {C.}~\bibnamefont {{Baccigalupi}}}, \bibinfo {author}
  {\bibfnamefont {B.}~\bibnamefont {{Barch}}}, \bibinfo {author} {\bibfnamefont
  {D.}~\bibnamefont {{Barron}}}, \bibinfo {author} {\bibfnamefont
  {A.}~\bibnamefont {{Bender}}}, \bibinfo {author} {\bibfnamefont
  {D.}~\bibnamefont {{Boettger}}}, \bibinfo {author} {\bibfnamefont
  {J.}~\bibnamefont {{Borrill}}}, \bibinfo {author} {\bibfnamefont
  {S.}~\bibnamefont {{Chapman}}}, \bibinfo {author} {\bibfnamefont
  {Y.}~\bibnamefont {{Chinone}}}, \bibinfo {author} {\bibfnamefont
  {A.}~\bibnamefont {{Cukierman}}}, \bibinfo {author} {\bibfnamefont
  {M.}~\bibnamefont {{Dobbs}}}, \bibinfo {author} {\bibfnamefont
  {A.}~\bibnamefont {{Ducout}}}, \bibinfo {author} {\bibfnamefont
  {R.}~\bibnamefont {{Dunner}}}, \bibinfo {author} {\bibfnamefont
  {T.}~\bibnamefont {{Elleflot}}}, \bibinfo {author} {\bibfnamefont
  {J.}~\bibnamefont {{Errard}}}, \bibinfo {author} {\bibfnamefont
  {G.}~\bibnamefont {{Fabbian}}}, \bibinfo {author} {\bibfnamefont
  {S.}~\bibnamefont {{Feeney}}}, \bibinfo {author} {\bibfnamefont
  {C.}~\bibnamefont {{Feng}}}, \bibinfo {author} {\bibfnamefont
  {T.}~\bibnamefont {{Fujino}}}, \bibinfo {author} {\bibfnamefont
  {G.}~\bibnamefont {{Fuller}}}, \bibinfo {author} {\bibfnamefont
  {A.}~\bibnamefont {{Gilbert}}}, \bibinfo {author} {\bibfnamefont
  {N.}~\bibnamefont {{Goeckner-Wald}}}, \bibinfo {author} {\bibfnamefont
  {J.}~\bibnamefont {{Groh}}}, \bibinfo {author} {\bibfnamefont {T.~D.}\
  \bibnamefont {{Haan}}}, \bibinfo {author} {\bibfnamefont {G.}~\bibnamefont
  {{Hall}}}, \bibinfo {author} {\bibfnamefont {N.}~\bibnamefont {{Halverson}}},
  \bibinfo {author} {\bibfnamefont {T.}~\bibnamefont {{Hamada}}}, \bibinfo
  {author} {\bibfnamefont {M.}~\bibnamefont {{Hasegawa}}}, \bibinfo {author}
  {\bibfnamefont {K.}~\bibnamefont {{Hattori}}}, \bibinfo {author}
  {\bibfnamefont {M.}~\bibnamefont {{Hazumi}}}, \bibinfo {author}
  {\bibfnamefont {C.}~\bibnamefont {{Hill}}}, \bibinfo {author} {\bibfnamefont
  {W.}~\bibnamefont {{Holzapfel}}}, \bibinfo {author} {\bibfnamefont
  {Y.}~\bibnamefont {{Hori}}}, \bibinfo {author} {\bibfnamefont
  {L.}~\bibnamefont {{Howe}}}, \bibinfo {author} {\bibfnamefont
  {Y.}~\bibnamefont {{Inoue}}}, \bibinfo {author} {\bibfnamefont
  {F.}~\bibnamefont {{Irie}}}, \bibinfo {author} {\bibfnamefont
  {G.}~\bibnamefont {{Jaehnig}}}, \bibinfo {author} {\bibfnamefont
  {A.}~\bibnamefont {{Jaffe}}}, \bibinfo {author} {\bibfnamefont
  {O.}~\bibnamefont {{Jeong}}}, \bibinfo {author} {\bibfnamefont
  {N.}~\bibnamefont {{Katayama}}}, \bibinfo {author} {\bibfnamefont
  {J.}~\bibnamefont {{Kaufman}}}, \bibinfo {author} {\bibfnamefont
  {K.}~\bibnamefont {{Kazemzadeh}}}, \bibinfo {author} {\bibfnamefont
  {B.}~\bibnamefont {{Keating}}}, \bibinfo {author} {\bibfnamefont
  {Z.}~\bibnamefont {{Kermish}}}, \bibinfo {author} {\bibfnamefont
  {R.}~\bibnamefont {{Keskitalo}}}, \bibinfo {author} {\bibfnamefont
  {T.}~\bibnamefont {{Kisner}}}, \bibinfo {author} {\bibfnamefont
  {A.}~\bibnamefont {{Kusaka}}}, \bibinfo {author} {\bibfnamefont {M.~L.}\
  \bibnamefont {{Jeune}}}, \bibinfo {author} {\bibfnamefont {A.}~\bibnamefont
  {{Lee}}}, \bibinfo {author} {\bibfnamefont {D.}~\bibnamefont {{Leon}}},
  \bibinfo {author} {\bibfnamefont {E.}~\bibnamefont {{Linder}}}, \bibinfo
  {author} {\bibfnamefont {L.}~\bibnamefont {{Lowry}}}, \bibinfo {author}
  {\bibfnamefont {F.}~\bibnamefont {{Matsuda}}}, \bibinfo {author}
  {\bibfnamefont {T.}~\bibnamefont {{Matsumura}}}, \bibinfo {author}
  {\bibfnamefont {N.}~\bibnamefont {{Miller}}}, \bibinfo {author}
  {\bibfnamefont {K.}~\bibnamefont {{Mizukami}}}, \bibinfo {author}
  {\bibfnamefont {J.}~\bibnamefont {{Montgomery}}}, \bibinfo {author}
  {\bibfnamefont {M.}~\bibnamefont {{Navaroli}}}, \bibinfo {author}
  {\bibfnamefont {H.}~\bibnamefont {{Nishino}}}, \bibinfo {author}
  {\bibfnamefont {J.}~\bibnamefont {{Peloton}}}, \bibinfo {author}
  {\bibfnamefont {D.}~\bibnamefont {{Poletti}}}, \bibinfo {author}
  {\bibfnamefont {G.}~\bibnamefont {{Puglisi}}}, \bibinfo {author}
  {\bibfnamefont {G.}~\bibnamefont {{Rebeiz}}}, \bibinfo {author}
  {\bibfnamefont {C.}~\bibnamefont {{Raum}}}, \bibinfo {author} {\bibfnamefont
  {C.}~\bibnamefont {{Reichardt}}}, \bibinfo {author} {\bibfnamefont
  {P.}~\bibnamefont {{Richards}}}, \bibinfo {author} {\bibfnamefont
  {C.}~\bibnamefont {{Ross}}}, \bibinfo {author} {\bibfnamefont
  {K.}~\bibnamefont {{Rotermund}}}, \bibinfo {author} {\bibfnamefont
  {Y.}~\bibnamefont {{Segawa}}}, \bibinfo {author} {\bibfnamefont
  {B.}~\bibnamefont {{Sherwin}}}, \bibinfo {author} {\bibfnamefont
  {I.}~\bibnamefont {{Shirley}}}, \bibinfo {author} {\bibfnamefont
  {P.}~\bibnamefont {{Siritanasak}}}, \bibinfo {author} {\bibfnamefont
  {N.}~\bibnamefont {{Stebor}}}, \bibinfo {author} {\bibfnamefont
  {R.}~\bibnamefont {{Stompor}}}, \bibinfo {author} {\bibfnamefont
  {J.}~\bibnamefont {{Suzuki}}}, \bibinfo {author} {\bibfnamefont
  {O.}~\bibnamefont {{Tajima}}}, \bibinfo {author} {\bibfnamefont
  {S.}~\bibnamefont {{Takada}}}, \bibinfo {author} {\bibfnamefont
  {S.}~\bibnamefont {{Takakura}}}, \bibinfo {author} {\bibfnamefont
  {S.}~\bibnamefont {{Takatori}}}, \bibinfo {author} {\bibfnamefont
  {A.}~\bibnamefont {{Tikhomirov}}}, \bibinfo {author} {\bibfnamefont
  {T.}~\bibnamefont {{Tomaru}}}, \bibinfo {author} {\bibfnamefont
  {B.}~\bibnamefont {{Westbrook}}}, \bibinfo {author} {\bibfnamefont
  {N.}~\bibnamefont {{Whitehorn}}}, \bibinfo {author} {\bibfnamefont
  {T.}~\bibnamefont {{Yamashita}}}, \bibinfo {author} {\bibfnamefont
  {A.}~\bibnamefont {{Zahn}}}, \ and\ \bibinfo {author} {\bibfnamefont
  {O.}~\bibnamefont {{Zahn}}},\ }\href {\doibase 10.1007/s10909-015-1425-4}
  {\bibfield  {journal} {\bibinfo  {journal} {Journal of Low Temperature
  Physics}\ }\textbf {\bibinfo {volume} {184}},\ \bibinfo {pages} {805}
  (\bibinfo {year} {2016})},\ \Eprint {http://arxiv.org/abs/1512.07299}
  {arXiv:1512.07299 [astro-ph.IM]} \BibitemShut {NoStop}%
\bibitem [{\citenamefont {{Benson}}\ \emph {et~al.}(2014)\citenamefont
  {{Benson}}, \citenamefont {{Ade}}, \citenamefont {{Ahmed}}, \citenamefont
  {{Allen}}, \citenamefont {{Arnold}}, \citenamefont {{Austermann}},
  \citenamefont {{Bender}}, \citenamefont {{Bleem}}, \citenamefont
  {{Carlstrom}}, \citenamefont {{Chang}}, \citenamefont {{Cho}}, \citenamefont
  {{Cliche}}, \citenamefont {{Crawford}}, \citenamefont {{Cukierman}},
  \citenamefont {{de Haan}}, \citenamefont {{Dobbs}}, \citenamefont
  {{Dutcher}}, \citenamefont {{Everett}}, \citenamefont {{Gilbert}},
  \citenamefont {{Halverson}}, \citenamefont {{Hanson}}, \citenamefont
  {{Harrington}}, \citenamefont {{Hattori}}, \citenamefont {{Henning}},
  \citenamefont {{Hilton}}, \citenamefont {{Holder}}, \citenamefont
  {{Holzapfel}}, \citenamefont {{Irwin}}, \citenamefont {{Keisler}},
  \citenamefont {{Knox}}, \citenamefont {{Kubik}}, \citenamefont {{Kuo}},
  \citenamefont {{Lee}}, \citenamefont {{Leitch}}, \citenamefont {{Li}},
  \citenamefont {{McDonald}}, \citenamefont {{Meyer}}, \citenamefont
  {{Montgomery}}, \citenamefont {{Myers}}, \citenamefont {{Natoli}},
  \citenamefont {{Nguyen}}, \citenamefont {{Novosad}}, \citenamefont {{Padin}},
  \citenamefont {{Pan}}, \citenamefont {{Pearson}}, \citenamefont
  {{Reichardt}}, \citenamefont {{Ruhl}}, \citenamefont {{Saliwanchik}},
  \citenamefont {{Simard}}, \citenamefont {{Smecher}}, \citenamefont {{Sayre}},
  \citenamefont {{Shirokoff}}, \citenamefont {{Stark}}, \citenamefont
  {{Story}}, \citenamefont {{Suzuki}}, \citenamefont {{Thompson}},
  \citenamefont {{Tucker}}, \citenamefont {{Vanderlinde}}, \citenamefont
  {{Vieira}}, \citenamefont {{Vikhlinin}}, \citenamefont {{Wang}},
  \citenamefont {{Yefremenko}},\ and\ \citenamefont {{Yoon}}}]{SPT3G}%
  \BibitemOpen
  \bibfield  {author} {\bibinfo {author} {\bibfnamefont {B.~A.}\ \bibnamefont
  {{Benson}}}, \bibinfo {author} {\bibfnamefont {P.~A.~R.}\ \bibnamefont
  {{Ade}}}, \bibinfo {author} {\bibfnamefont {Z.}~\bibnamefont {{Ahmed}}},
  \bibinfo {author} {\bibfnamefont {S.~W.}\ \bibnamefont {{Allen}}}, \bibinfo
  {author} {\bibfnamefont {K.}~\bibnamefont {{Arnold}}}, \bibinfo {author}
  {\bibfnamefont {J.~E.}\ \bibnamefont {{Austermann}}}, \bibinfo {author}
  {\bibfnamefont {A.~N.}\ \bibnamefont {{Bender}}}, \bibinfo {author}
  {\bibfnamefont {L.~E.}\ \bibnamefont {{Bleem}}}, \bibinfo {author}
  {\bibfnamefont {J.~E.}\ \bibnamefont {{Carlstrom}}}, \bibinfo {author}
  {\bibfnamefont {C.~L.}\ \bibnamefont {{Chang}}}, \bibinfo {author}
  {\bibfnamefont {H.~M.}\ \bibnamefont {{Cho}}}, \bibinfo {author}
  {\bibfnamefont {J.~F.}\ \bibnamefont {{Cliche}}}, \bibinfo {author}
  {\bibfnamefont {T.~M.}\ \bibnamefont {{Crawford}}}, \bibinfo {author}
  {\bibfnamefont {A.}~\bibnamefont {{Cukierman}}}, \bibinfo {author}
  {\bibfnamefont {T.}~\bibnamefont {{de Haan}}}, \bibinfo {author}
  {\bibfnamefont {M.~A.}\ \bibnamefont {{Dobbs}}}, \bibinfo {author}
  {\bibfnamefont {D.}~\bibnamefont {{Dutcher}}}, \bibinfo {author}
  {\bibfnamefont {W.}~\bibnamefont {{Everett}}}, \bibinfo {author}
  {\bibfnamefont {A.}~\bibnamefont {{Gilbert}}}, \bibinfo {author}
  {\bibfnamefont {N.~W.}\ \bibnamefont {{Halverson}}}, \bibinfo {author}
  {\bibfnamefont {D.}~\bibnamefont {{Hanson}}}, \bibinfo {author}
  {\bibfnamefont {N.~L.}\ \bibnamefont {{Harrington}}}, \bibinfo {author}
  {\bibfnamefont {K.}~\bibnamefont {{Hattori}}}, \bibinfo {author}
  {\bibfnamefont {J.~W.}\ \bibnamefont {{Henning}}}, \bibinfo {author}
  {\bibfnamefont {G.~C.}\ \bibnamefont {{Hilton}}}, \bibinfo {author}
  {\bibfnamefont {G.~P.}\ \bibnamefont {{Holder}}}, \bibinfo {author}
  {\bibfnamefont {W.~L.}\ \bibnamefont {{Holzapfel}}}, \bibinfo {author}
  {\bibfnamefont {K.~D.}\ \bibnamefont {{Irwin}}}, \bibinfo {author}
  {\bibfnamefont {R.}~\bibnamefont {{Keisler}}}, \bibinfo {author}
  {\bibfnamefont {L.}~\bibnamefont {{Knox}}}, \bibinfo {author} {\bibfnamefont
  {D.}~\bibnamefont {{Kubik}}}, \bibinfo {author} {\bibfnamefont {C.~L.}\
  \bibnamefont {{Kuo}}}, \bibinfo {author} {\bibfnamefont {A.~T.}\ \bibnamefont
  {{Lee}}}, \bibinfo {author} {\bibfnamefont {E.~M.}\ \bibnamefont {{Leitch}}},
  \bibinfo {author} {\bibfnamefont {D.}~\bibnamefont {{Li}}}, \bibinfo {author}
  {\bibfnamefont {M.}~\bibnamefont {{McDonald}}}, \bibinfo {author}
  {\bibfnamefont {S.~S.}\ \bibnamefont {{Meyer}}}, \bibinfo {author}
  {\bibfnamefont {J.}~\bibnamefont {{Montgomery}}}, \bibinfo {author}
  {\bibfnamefont {M.}~\bibnamefont {{Myers}}}, \bibinfo {author} {\bibfnamefont
  {T.}~\bibnamefont {{Natoli}}}, \bibinfo {author} {\bibfnamefont
  {H.}~\bibnamefont {{Nguyen}}}, \bibinfo {author} {\bibfnamefont
  {V.}~\bibnamefont {{Novosad}}}, \bibinfo {author} {\bibfnamefont
  {S.}~\bibnamefont {{Padin}}}, \bibinfo {author} {\bibfnamefont
  {Z.}~\bibnamefont {{Pan}}}, \bibinfo {author} {\bibfnamefont
  {J.}~\bibnamefont {{Pearson}}}, \bibinfo {author} {\bibfnamefont
  {C.}~\bibnamefont {{Reichardt}}}, \bibinfo {author} {\bibfnamefont {J.~E.}\
  \bibnamefont {{Ruhl}}}, \bibinfo {author} {\bibfnamefont {B.~R.}\
  \bibnamefont {{Saliwanchik}}}, \bibinfo {author} {\bibfnamefont
  {G.}~\bibnamefont {{Simard}}}, \bibinfo {author} {\bibfnamefont
  {G.}~\bibnamefont {{Smecher}}}, \bibinfo {author} {\bibfnamefont {J.~T.}\
  \bibnamefont {{Sayre}}}, \bibinfo {author} {\bibfnamefont {E.}~\bibnamefont
  {{Shirokoff}}}, \bibinfo {author} {\bibfnamefont {A.~A.}\ \bibnamefont
  {{Stark}}}, \bibinfo {author} {\bibfnamefont {K.}~\bibnamefont {{Story}}},
  \bibinfo {author} {\bibfnamefont {A.}~\bibnamefont {{Suzuki}}}, \bibinfo
  {author} {\bibfnamefont {K.~L.}\ \bibnamefont {{Thompson}}}, \bibinfo
  {author} {\bibfnamefont {C.}~\bibnamefont {{Tucker}}}, \bibinfo {author}
  {\bibfnamefont {K.}~\bibnamefont {{Vanderlinde}}}, \bibinfo {author}
  {\bibfnamefont {J.~D.}\ \bibnamefont {{Vieira}}}, \bibinfo {author}
  {\bibfnamefont {A.}~\bibnamefont {{Vikhlinin}}}, \bibinfo {author}
  {\bibfnamefont {G.}~\bibnamefont {{Wang}}}, \bibinfo {author} {\bibfnamefont
  {V.}~\bibnamefont {{Yefremenko}}}, \ and\ \bibinfo {author} {\bibfnamefont
  {K.~W.}\ \bibnamefont {{Yoon}}},\ }in\ \href {\doibase 10.1117/12.2057305}
  {\emph {\bibinfo {booktitle} {Millimeter, Submillimeter, and Far-Infrared
  Detectors and Instrumentation for Astronomy VII}}},\ \bibinfo {series}
  {\procspie}, Vol.\ \bibinfo {volume} {9153}\ (\bibinfo {year} {2014})\ p.\
  \bibinfo {pages} {91531P},\ \Eprint {http://arxiv.org/abs/1407.2973}
  {arXiv:1407.2973 [astro-ph.IM]} \BibitemShut {NoStop}%
\bibitem [{\citenamefont {{Abazajian}}\ \emph {et~al.}(2016)\citenamefont
  {{Abazajian}}, \citenamefont {{Adshead}}, \citenamefont {{Ahmed}},
  \citenamefont {{Allen}}, \citenamefont {{Alonso}}, \citenamefont {{Arnold}},
  \citenamefont {{Baccigalupi}}, \citenamefont {{Bartlett}}, \citenamefont
  {{Battaglia}}, \citenamefont {{Benson}}, \citenamefont {{Bischoff}},
  \citenamefont {{Borrill}}, \citenamefont {{Buza}}, \citenamefont
  {{Calabrese}}, \citenamefont {{Caldwell}}, \citenamefont {{Carlstrom}},
  \citenamefont {{Chang}}, \citenamefont {{Crawford}}, \citenamefont
  {{Cyr-Racine}}, \citenamefont {{De Bernardis}}, \citenamefont {{de Haan}},
  \citenamefont {{di Serego Alighieri}}, \citenamefont {{Dunkley}},
  \citenamefont {{Dvorkin}}, \citenamefont {{Errard}}, \citenamefont
  {{Fabbian}}, \citenamefont {{Feeney}}, \citenamefont {{Ferraro}},
  \citenamefont {{Filippini}}, \citenamefont {{Flauger}}, \citenamefont
  {{Fuller}}, \citenamefont {{Gluscevic}}, \citenamefont {{Green}},
  \citenamefont {{Grin}}, \citenamefont {{Grohs}}, \citenamefont {{Henning}},
  \citenamefont {{Hill}}, \citenamefont {{Hlozek}}, \citenamefont {{Holder}},
  \citenamefont {{Holzapfel}}, \citenamefont {{Hu}}, \citenamefont
  {{Huffenberger}}, \citenamefont {{Keskitalo}}, \citenamefont {{Knox}},
  \citenamefont {{Kosowsky}}, \citenamefont {{Kovac}}, \citenamefont
  {{Kovetz}}, \citenamefont {{Kuo}}, \citenamefont {{Kusaka}}, \citenamefont
  {{Le Jeune}}, \citenamefont {{Lee}}, \citenamefont {{Lilley}}, \citenamefont
  {{Loverde}}, \citenamefont {{Madhavacheril}}, \citenamefont {{Mantz}},
  \citenamefont {{Marsh}}, \citenamefont {{McMahon}}, \citenamefont
  {{Meerburg}}, \citenamefont {{Meyers}}, \citenamefont {{Miller}},
  \citenamefont {{Munoz}}, \citenamefont {{Nguyen}}, \citenamefont {{Niemack}},
  \citenamefont {{Peloso}}, \citenamefont {{Peloton}}, \citenamefont
  {{Pogosian}}, \citenamefont {{Pryke}}, \citenamefont {{Raveri}},
  \citenamefont {{Reichardt}}, \citenamefont {{Rocha}}, \citenamefont
  {{Rotti}}, \citenamefont {{Schaan}}, \citenamefont {{Schmittfull}},
  \citenamefont {{Scott}}, \citenamefont {{Sehgal}}, \citenamefont
  {{Shandera}}, \citenamefont {{Sherwin}}, \citenamefont {{Smith}},
  \citenamefont {{Sorbo}}, \citenamefont {{Starkman}}, \citenamefont {{Story}},
  \citenamefont {{van Engelen}}, \citenamefont {{Vieira}}, \citenamefont
  {{Watson}}, \citenamefont {{Whitehorn}},\ and\ \citenamefont {{Kimmy
  Wu}}}]{CMB-S4}%
  \BibitemOpen
  \bibfield  {author} {\bibinfo {author} {\bibfnamefont {K.~N.}\ \bibnamefont
  {{Abazajian}}}, \bibinfo {author} {\bibfnamefont {P.}~\bibnamefont
  {{Adshead}}}, \bibinfo {author} {\bibfnamefont {Z.}~\bibnamefont {{Ahmed}}},
  \bibinfo {author} {\bibfnamefont {S.~W.}\ \bibnamefont {{Allen}}}, \bibinfo
  {author} {\bibfnamefont {D.}~\bibnamefont {{Alonso}}}, \bibinfo {author}
  {\bibfnamefont {K.~S.}\ \bibnamefont {{Arnold}}}, \bibinfo {author}
  {\bibfnamefont {C.}~\bibnamefont {{Baccigalupi}}}, \bibinfo {author}
  {\bibfnamefont {J.~G.}\ \bibnamefont {{Bartlett}}}, \bibinfo {author}
  {\bibfnamefont {N.}~\bibnamefont {{Battaglia}}}, \bibinfo {author}
  {\bibfnamefont {B.~A.}\ \bibnamefont {{Benson}}}, \bibinfo {author}
  {\bibfnamefont {C.~A.}\ \bibnamefont {{Bischoff}}}, \bibinfo {author}
  {\bibfnamefont {J.}~\bibnamefont {{Borrill}}}, \bibinfo {author}
  {\bibfnamefont {V.}~\bibnamefont {{Buza}}}, \bibinfo {author} {\bibfnamefont
  {E.}~\bibnamefont {{Calabrese}}}, \bibinfo {author} {\bibfnamefont
  {R.}~\bibnamefont {{Caldwell}}}, \bibinfo {author} {\bibfnamefont {J.~E.}\
  \bibnamefont {{Carlstrom}}}, \bibinfo {author} {\bibfnamefont {C.~L.}\
  \bibnamefont {{Chang}}}, \bibinfo {author} {\bibfnamefont {T.~M.}\
  \bibnamefont {{Crawford}}}, \bibinfo {author} {\bibfnamefont {F.-Y.}\
  \bibnamefont {{Cyr-Racine}}}, \bibinfo {author} {\bibfnamefont
  {F.}~\bibnamefont {{De Bernardis}}}, \bibinfo {author} {\bibfnamefont
  {T.}~\bibnamefont {{de Haan}}}, \bibinfo {author} {\bibfnamefont
  {S.}~\bibnamefont {{di Serego Alighieri}}}, \bibinfo {author} {\bibfnamefont
  {J.}~\bibnamefont {{Dunkley}}}, \bibinfo {author} {\bibfnamefont
  {C.}~\bibnamefont {{Dvorkin}}}, \bibinfo {author} {\bibfnamefont
  {J.}~\bibnamefont {{Errard}}}, \bibinfo {author} {\bibfnamefont
  {G.}~\bibnamefont {{Fabbian}}}, \bibinfo {author} {\bibfnamefont
  {S.}~\bibnamefont {{Feeney}}}, \bibinfo {author} {\bibfnamefont
  {S.}~\bibnamefont {{Ferraro}}}, \bibinfo {author} {\bibfnamefont {J.~P.}\
  \bibnamefont {{Filippini}}}, \bibinfo {author} {\bibfnamefont
  {R.}~\bibnamefont {{Flauger}}}, \bibinfo {author} {\bibfnamefont {G.~M.}\
  \bibnamefont {{Fuller}}}, \bibinfo {author} {\bibfnamefont {V.}~\bibnamefont
  {{Gluscevic}}}, \bibinfo {author} {\bibfnamefont {D.}~\bibnamefont
  {{Green}}}, \bibinfo {author} {\bibfnamefont {D.}~\bibnamefont {{Grin}}},
  \bibinfo {author} {\bibfnamefont {E.}~\bibnamefont {{Grohs}}}, \bibinfo
  {author} {\bibfnamefont {J.~W.}\ \bibnamefont {{Henning}}}, \bibinfo {author}
  {\bibfnamefont {J.~C.}\ \bibnamefont {{Hill}}}, \bibinfo {author}
  {\bibfnamefont {R.}~\bibnamefont {{Hlozek}}}, \bibinfo {author}
  {\bibfnamefont {G.}~\bibnamefont {{Holder}}}, \bibinfo {author}
  {\bibfnamefont {W.}~\bibnamefont {{Holzapfel}}}, \bibinfo {author}
  {\bibfnamefont {W.}~\bibnamefont {{Hu}}}, \bibinfo {author} {\bibfnamefont
  {K.~M.}\ \bibnamefont {{Huffenberger}}}, \bibinfo {author} {\bibfnamefont
  {R.}~\bibnamefont {{Keskitalo}}}, \bibinfo {author} {\bibfnamefont
  {L.}~\bibnamefont {{Knox}}}, \bibinfo {author} {\bibfnamefont
  {A.}~\bibnamefont {{Kosowsky}}}, \bibinfo {author} {\bibfnamefont
  {J.}~\bibnamefont {{Kovac}}}, \bibinfo {author} {\bibfnamefont {E.~D.}\
  \bibnamefont {{Kovetz}}}, \bibinfo {author} {\bibfnamefont {C.-L.}\
  \bibnamefont {{Kuo}}}, \bibinfo {author} {\bibfnamefont {A.}~\bibnamefont
  {{Kusaka}}}, \bibinfo {author} {\bibfnamefont {M.}~\bibnamefont {{Le
  Jeune}}}, \bibinfo {author} {\bibfnamefont {A.~T.}\ \bibnamefont {{Lee}}},
  \bibinfo {author} {\bibfnamefont {M.}~\bibnamefont {{Lilley}}}, \bibinfo
  {author} {\bibfnamefont {M.}~\bibnamefont {{Loverde}}}, \bibinfo {author}
  {\bibfnamefont {M.~S.}\ \bibnamefont {{Madhavacheril}}}, \bibinfo {author}
  {\bibfnamefont {A.}~\bibnamefont {{Mantz}}}, \bibinfo {author} {\bibfnamefont
  {D.~J.~E.}\ \bibnamefont {{Marsh}}}, \bibinfo {author} {\bibfnamefont
  {J.}~\bibnamefont {{McMahon}}}, \bibinfo {author} {\bibfnamefont {P.~D.}\
  \bibnamefont {{Meerburg}}}, \bibinfo {author} {\bibfnamefont
  {J.}~\bibnamefont {{Meyers}}}, \bibinfo {author} {\bibfnamefont {A.~D.}\
  \bibnamefont {{Miller}}}, \bibinfo {author} {\bibfnamefont {J.~B.}\
  \bibnamefont {{Munoz}}}, \bibinfo {author} {\bibfnamefont {H.~N.}\
  \bibnamefont {{Nguyen}}}, \bibinfo {author} {\bibfnamefont {M.~D.}\
  \bibnamefont {{Niemack}}}, \bibinfo {author} {\bibfnamefont {M.}~\bibnamefont
  {{Peloso}}}, \bibinfo {author} {\bibfnamefont {J.}~\bibnamefont {{Peloton}}},
  \bibinfo {author} {\bibfnamefont {L.}~\bibnamefont {{Pogosian}}}, \bibinfo
  {author} {\bibfnamefont {C.}~\bibnamefont {{Pryke}}}, \bibinfo {author}
  {\bibfnamefont {M.}~\bibnamefont {{Raveri}}}, \bibinfo {author}
  {\bibfnamefont {C.~L.}\ \bibnamefont {{Reichardt}}}, \bibinfo {author}
  {\bibfnamefont {G.}~\bibnamefont {{Rocha}}}, \bibinfo {author} {\bibfnamefont
  {A.}~\bibnamefont {{Rotti}}}, \bibinfo {author} {\bibfnamefont
  {E.}~\bibnamefont {{Schaan}}}, \bibinfo {author} {\bibfnamefont {M.~M.}\
  \bibnamefont {{Schmittfull}}}, \bibinfo {author} {\bibfnamefont
  {D.}~\bibnamefont {{Scott}}}, \bibinfo {author} {\bibfnamefont
  {N.}~\bibnamefont {{Sehgal}}}, \bibinfo {author} {\bibfnamefont
  {S.}~\bibnamefont {{Shandera}}}, \bibinfo {author} {\bibfnamefont {B.~D.}\
  \bibnamefont {{Sherwin}}}, \bibinfo {author} {\bibfnamefont {T.~L.}\
  \bibnamefont {{Smith}}}, \bibinfo {author} {\bibfnamefont {L.}~\bibnamefont
  {{Sorbo}}}, \bibinfo {author} {\bibfnamefont {G.~D.}\ \bibnamefont
  {{Starkman}}}, \bibinfo {author} {\bibfnamefont {K.~T.}\ \bibnamefont
  {{Story}}}, \bibinfo {author} {\bibfnamefont {A.}~\bibnamefont {{van
  Engelen}}}, \bibinfo {author} {\bibfnamefont {J.~D.}\ \bibnamefont
  {{Vieira}}}, \bibinfo {author} {\bibfnamefont {S.}~\bibnamefont {{Watson}}},
  \bibinfo {author} {\bibfnamefont {N.}~\bibnamefont {{Whitehorn}}}, \ and\
  \bibinfo {author} {\bibfnamefont {W.~L.}\ \bibnamefont {{Kimmy Wu}}},\
  }\href@noop {} {\bibfield  {journal} {\bibinfo  {journal} {ArXiv e-prints}\ }
  (\bibinfo {year} {2016})},\ \Eprint {http://arxiv.org/abs/1610.02743}
  {arXiv:1610.02743} \BibitemShut {NoStop}%
\bibitem [{\citenamefont {{Hu}}(2001)}]{HuQuadEsti}%
  \BibitemOpen
  \bibfield  {author} {\bibinfo {author} {\bibfnamefont {W.}~\bibnamefont
  {{Hu}}},\ }\href {\doibase 10.1086/323253} {\bibfield  {journal} {\bibinfo
  {journal} {\apjl}\ }\textbf {\bibinfo {volume} {557}},\ \bibinfo {pages}
  {L79} (\bibinfo {year} {2001})},\ \Eprint
  {http://arxiv.org/abs/astro-ph/0105424} {astro-ph/0105424} \BibitemShut
  {NoStop}%
\bibitem [{\citenamefont {{Hu}}\ and\ \citenamefont
  {{Okamoto}}(2002)}]{HuOkamPol}%
  \BibitemOpen
  \bibfield  {author} {\bibinfo {author} {\bibfnamefont {W.}~\bibnamefont
  {{Hu}}}\ and\ \bibinfo {author} {\bibfnamefont {T.}~\bibnamefont
  {{Okamoto}}},\ }\href {\doibase 10.1086/341110} {\bibfield  {journal}
  {\bibinfo  {journal} {\apj}\ }\textbf {\bibinfo {volume} {574}},\ \bibinfo
  {pages} {566} (\bibinfo {year} {2002})},\ \Eprint
  {http://arxiv.org/abs/astro-ph/0111606} {astro-ph/0111606} \BibitemShut
  {NoStop}%
\bibitem [{\citenamefont {Kesden}\ \emph {et~al.}(2003)\citenamefont {Kesden},
  \citenamefont {Cooray},\ and\ \citenamefont {Kamionkowski}}]{Kesden2003}%
  \BibitemOpen
  \bibfield  {author} {\bibinfo {author} {\bibfnamefont {M.}~\bibnamefont
  {Kesden}}, \bibinfo {author} {\bibfnamefont {A.}~\bibnamefont {Cooray}}, \
  and\ \bibinfo {author} {\bibfnamefont {M.}~\bibnamefont {Kamionkowski}},\
  }\href {\doibase 10.1103/PhysRevD.67.123507} {\bibfield  {journal} {\bibinfo
  {journal} {Phys. Rev. D}\ }\textbf {\bibinfo {volume} {67}},\ \bibinfo
  {pages} {123507} (\bibinfo {year} {2003})}\BibitemShut {NoStop}%
\bibitem [{\citenamefont {{Hanson}}\ \emph {et~al.}(2011)\citenamefont
  {{Hanson}}, \citenamefont {{Challinor}}, \citenamefont {{Efstathiou}},\ and\
  \citenamefont {{Bielewicz}}}]{HansonN2}%
  \BibitemOpen
  \bibfield  {author} {\bibinfo {author} {\bibfnamefont {D.}~\bibnamefont
  {{Hanson}}}, \bibinfo {author} {\bibfnamefont {A.}~\bibnamefont
  {{Challinor}}}, \bibinfo {author} {\bibfnamefont {G.}~\bibnamefont
  {{Efstathiou}}}, \ and\ \bibinfo {author} {\bibfnamefont {P.}~\bibnamefont
  {{Bielewicz}}},\ }\href {\doibase 10.1103/PhysRevD.83.043005} {\bibfield
  {journal} {\bibinfo  {journal} {\prd}\ }\textbf {\bibinfo {volume} {83}},\
  \bibinfo {eid} {043005} (\bibinfo {year} {2011})},\ \Eprint
  {http://arxiv.org/abs/1008.4403} {arXiv:1008.4403 [astro-ph.CO]} \BibitemShut
  {NoStop}%
\bibitem [{\citenamefont {{Hanson}}\ \emph {et~al.}(2009)\citenamefont
  {{Hanson}}, \citenamefont {{Rocha}},\ and\ \citenamefont
  {{G{\'o}rski}}}]{2009Hanson}%
  \BibitemOpen
  \bibfield  {author} {\bibinfo {author} {\bibfnamefont {D.}~\bibnamefont
  {{Hanson}}}, \bibinfo {author} {\bibfnamefont {G.}~\bibnamefont {{Rocha}}}, \
  and\ \bibinfo {author} {\bibfnamefont {K.}~\bibnamefont {{G{\'o}rski}}},\
  }\href {\doibase 10.1111/j.1365-2966.2009.15614.x} {\bibfield  {journal}
  {\bibinfo  {journal} {\mnras}\ }\textbf {\bibinfo {volume} {400}},\ \bibinfo
  {pages} {2169} (\bibinfo {year} {2009})},\ \Eprint
  {http://arxiv.org/abs/0907.1927} {arXiv:0907.1927 [astro-ph.CO]} \BibitemShut
  {NoStop}%
\bibitem [{\citenamefont {{Hanson}}\ \emph
  {et~al.}(2010{\natexlab{b}})\citenamefont {{Hanson}}, \citenamefont
  {{Lewis}},\ and\ \citenamefont {{Challinor}}}]{HansonBeam}%
  \BibitemOpen
  \bibfield  {author} {\bibinfo {author} {\bibfnamefont {D.}~\bibnamefont
  {{Hanson}}}, \bibinfo {author} {\bibfnamefont {A.}~\bibnamefont {{Lewis}}}, \
  and\ \bibinfo {author} {\bibfnamefont {A.}~\bibnamefont {{Challinor}}},\
  }\href {\doibase 10.1103/PhysRevD.81.103003} {\bibfield  {journal} {\bibinfo
  {journal} {\prd}\ }\textbf {\bibinfo {volume} {81}},\ \bibinfo {eid} {103003}
  (\bibinfo {year} {2010}{\natexlab{b}})},\ \Eprint
  {http://arxiv.org/abs/1003.0198} {arXiv:1003.0198 [astro-ph.CO]} \BibitemShut
  {NoStop}%
\bibitem [{\citenamefont {{Fantaye}}\ \emph {et~al.}(2012)\citenamefont
  {{Fantaye}}, \citenamefont {{Baccigalupi}}, \citenamefont {{Leach}},\ and\
  \citenamefont {{Yadav}}}]{2012JCAPFantaye}%
  \BibitemOpen
  \bibfield  {author} {\bibinfo {author} {\bibfnamefont {Y.}~\bibnamefont
  {{Fantaye}}}, \bibinfo {author} {\bibfnamefont {C.}~\bibnamefont
  {{Baccigalupi}}}, \bibinfo {author} {\bibfnamefont {S.~M.}\ \bibnamefont
  {{Leach}}}, \ and\ \bibinfo {author} {\bibfnamefont {A.~P.~S.}\ \bibnamefont
  {{Yadav}}},\ }\href {\doibase 10.1088/1475-7516/2012/12/017} {\bibfield
  {journal} {\bibinfo  {journal} {\jcap}\ }\textbf {\bibinfo {volume} {12}},\
  \bibinfo {eid} {017} (\bibinfo {year} {2012})},\ \Eprint
  {http://arxiv.org/abs/1207.0508} {arXiv:1207.0508 [astro-ph.CO]} \BibitemShut
  {NoStop}%
\bibitem [{\citenamefont {{van Engelen}}\ \emph {et~al.}(2014)\citenamefont
  {{van Engelen}}, \citenamefont {{Bhattacharya}}, \citenamefont {{Sehgal}},
  \citenamefont {{Holder}}, \citenamefont {{Zahn}},\ and\ \citenamefont
  {{Nagai}}}]{2014vanEngelen}%
  \BibitemOpen
  \bibfield  {author} {\bibinfo {author} {\bibfnamefont {A.}~\bibnamefont {{van
  Engelen}}}, \bibinfo {author} {\bibfnamefont {S.}~\bibnamefont
  {{Bhattacharya}}}, \bibinfo {author} {\bibfnamefont {N.}~\bibnamefont
  {{Sehgal}}}, \bibinfo {author} {\bibfnamefont {G.~P.}\ \bibnamefont
  {{Holder}}}, \bibinfo {author} {\bibfnamefont {O.}~\bibnamefont {{Zahn}}}, \
  and\ \bibinfo {author} {\bibfnamefont {D.}~\bibnamefont {{Nagai}}},\ }\href
  {\doibase 10.1088/0004-637X/786/1/13} {\bibfield  {journal} {\bibinfo
  {journal} {\apj}\ }\textbf {\bibinfo {volume} {786}},\ \bibinfo {eid} {13}
  (\bibinfo {year} {2014})},\ \Eprint {http://arxiv.org/abs/1310.7023}
  {arXiv:1310.7023} \BibitemShut {NoStop}%
\bibitem [{\citenamefont {{Osborne}}\ \emph {et~al.}(2014)\citenamefont
  {{Osborne}}, \citenamefont {{Hanson}},\ and\ \citenamefont
  {{Dor{\'e}}}}]{2014Osborne}%
  \BibitemOpen
  \bibfield  {author} {\bibinfo {author} {\bibfnamefont {S.~J.}\ \bibnamefont
  {{Osborne}}}, \bibinfo {author} {\bibfnamefont {D.}~\bibnamefont {{Hanson}}},
  \ and\ \bibinfo {author} {\bibfnamefont {O.}~\bibnamefont {{Dor{\'e}}}},\
  }\href {\doibase 10.1088/1475-7516/2014/03/024} {\bibfield  {journal}
  {\bibinfo  {journal} {\jcap}\ }\textbf {\bibinfo {volume} {3}},\ \bibinfo
  {eid} {024} (\bibinfo {year} {2014})},\ \Eprint
  {http://arxiv.org/abs/1310.7547} {arXiv:1310.7547} \BibitemShut {NoStop}%
\bibitem [{\citenamefont {{Ferraro}}\ and\ \citenamefont
  {{Hill}}(2018)}]{2018FerrHill}%
  \BibitemOpen
  \bibfield  {author} {\bibinfo {author} {\bibfnamefont {S.}~\bibnamefont
  {{Ferraro}}}\ and\ \bibinfo {author} {\bibfnamefont {J.~C.}\ \bibnamefont
  {{Hill}}},\ }\href {\doibase 10.1103/PhysRevD.97.023512} {\bibfield
  {journal} {\bibinfo  {journal} {\prd}\ }\textbf {\bibinfo {volume} {97}},\
  \bibinfo {eid} {023512} (\bibinfo {year} {2018})},\ \Eprint
  {http://arxiv.org/abs/1705.06751} {arXiv:1705.06751} \BibitemShut {NoStop}%
\bibitem [{\citenamefont {{Lewis}}\ \emph {et~al.}(2011)\citenamefont
  {{Lewis}}, \citenamefont {{Challinor}},\ and\ \citenamefont
  {{Hanson}}}]{Lewis2011}%
  \BibitemOpen
  \bibfield  {author} {\bibinfo {author} {\bibfnamefont {A.}~\bibnamefont
  {{Lewis}}}, \bibinfo {author} {\bibfnamefont {A.}~\bibnamefont
  {{Challinor}}}, \ and\ \bibinfo {author} {\bibfnamefont {D.}~\bibnamefont
  {{Hanson}}},\ }\href {\doibase 10.1088/1475-7516/2011/03/018} {\bibfield
  {journal} {\bibinfo  {journal} {\jcap}\ }\textbf {\bibinfo {volume} {3}},\
  \bibinfo {eid} {018} (\bibinfo {year} {2011})},\ \Eprint
  {http://arxiv.org/abs/1101.2234} {arXiv:1101.2234} \BibitemShut {NoStop}%
\bibitem [{\citenamefont {{Namikawa}}\ \emph {et~al.}(2013)\citenamefont
  {{Namikawa}}, \citenamefont {{Hanson}},\ and\ \citenamefont
  {{Takahashi}}}]{Namikawa2013}%
  \BibitemOpen
  \bibfield  {author} {\bibinfo {author} {\bibfnamefont {T.}~\bibnamefont
  {{Namikawa}}}, \bibinfo {author} {\bibfnamefont {D.}~\bibnamefont
  {{Hanson}}}, \ and\ \bibinfo {author} {\bibfnamefont {R.}~\bibnamefont
  {{Takahashi}}},\ }\href {\doibase 10.1093/mnras/stt195} {\bibfield  {journal}
  {\bibinfo  {journal} {\mnras}\ }\textbf {\bibinfo {volume} {431}},\ \bibinfo
  {pages} {609} (\bibinfo {year} {2013})},\ \Eprint
  {http://arxiv.org/abs/1209.0091} {arXiv:1209.0091 [astro-ph.CO]} \BibitemShut
  {NoStop}%
\bibitem [{\citenamefont {{Namikawa}}\ and\ \citenamefont
  {{Takahashi}}(2014)}]{Namikawa2014}%
  \BibitemOpen
  \bibfield  {author} {\bibinfo {author} {\bibfnamefont {T.}~\bibnamefont
  {{Namikawa}}}\ and\ \bibinfo {author} {\bibfnamefont {R.}~\bibnamefont
  {{Takahashi}}},\ }\href {\doibase 10.1093/mnras/stt2290} {\bibfield
  {journal} {\bibinfo  {journal} {\mnras}\ }\textbf {\bibinfo {volume} {438}},\
  \bibinfo {pages} {1507} (\bibinfo {year} {2014})},\ \Eprint
  {http://arxiv.org/abs/1310.2372} {arXiv:1310.2372} \BibitemShut {NoStop}%
\bibitem [{\citenamefont {{Madhavacheril}}\ and\ \citenamefont
  {{Hill}}(2018)}]{2018MadHill}%
  \BibitemOpen
  \bibfield  {author} {\bibinfo {author} {\bibfnamefont {M.~S.}\ \bibnamefont
  {{Madhavacheril}}}\ and\ \bibinfo {author} {\bibfnamefont {J.~C.}\
  \bibnamefont {{Hill}}},\ }\href@noop {} {\bibfield  {journal} {\bibinfo
  {journal} {ArXiv e-prints}\ } (\bibinfo {year} {2018})},\ \Eprint
  {http://arxiv.org/abs/1802.08230} {arXiv:1802.08230} \BibitemShut {NoStop}%
\bibitem [{\citenamefont {{B{\"o}hm}}\ \emph {et~al.}(2016)\citenamefont
  {{B{\"o}hm}}, \citenamefont {{Schmittfull}},\ and\ \citenamefont
  {{Sherwin}}}]{N32}%
  \BibitemOpen
  \bibfield  {author} {\bibinfo {author} {\bibfnamefont {V.}~\bibnamefont
  {{B{\"o}hm}}}, \bibinfo {author} {\bibfnamefont {M.}~\bibnamefont
  {{Schmittfull}}}, \ and\ \bibinfo {author} {\bibfnamefont {B.~D.}\
  \bibnamefont {{Sherwin}}},\ }\href {\doibase 10.1103/PhysRevD.94.043519}
  {\bibfield  {journal} {\bibinfo  {journal} {\prd}\ }\textbf {\bibinfo
  {volume} {94}},\ \bibinfo {eid} {043519} (\bibinfo {year} {2016})},\ \Eprint
  {http://arxiv.org/abs/1605.01392} {arXiv:1605.01392} \BibitemShut {NoStop}%
\bibitem [{\citenamefont {{Pratten}}\ and\ \citenamefont
  {{Lewis}}(2016)}]{postBornPratten}%
  \BibitemOpen
  \bibfield  {author} {\bibinfo {author} {\bibfnamefont {G.}~\bibnamefont
  {{Pratten}}}\ and\ \bibinfo {author} {\bibfnamefont {A.}~\bibnamefont
  {{Lewis}}},\ }\href {\doibase 10.1088/1475-7516/2016/08/047} {\bibfield
  {journal} {\bibinfo  {journal} {\jcap}\ }\textbf {\bibinfo {volume} {8}},\
  \bibinfo {eid} {047} (\bibinfo {year} {2016})},\ \Eprint
  {http://arxiv.org/abs/1605.05662} {arXiv:1605.05662} \BibitemShut {NoStop}%
\bibitem [{\citenamefont {{Petri}}\ \emph {et~al.}(2017)\citenamefont
  {{Petri}}, \citenamefont {{Haiman}},\ and\ \citenamefont
  {{May}}}]{2017PhRvDPetri}%
  \BibitemOpen
  \bibfield  {author} {\bibinfo {author} {\bibfnamefont {A.}~\bibnamefont
  {{Petri}}}, \bibinfo {author} {\bibfnamefont {Z.}~\bibnamefont {{Haiman}}}, \
  and\ \bibinfo {author} {\bibfnamefont {M.}~\bibnamefont {{May}}},\ }\href
  {\doibase 10.1103/PhysRevD.95.123503} {\bibfield  {journal} {\bibinfo
  {journal} {\prd}\ }\textbf {\bibinfo {volume} {95}},\ \bibinfo {eid} {123503}
  (\bibinfo {year} {2017})},\ \Eprint {http://arxiv.org/abs/1612.00852}
  {arXiv:1612.00852} \BibitemShut {NoStop}%
\bibitem [{\citenamefont {{Fabbian}}\ \emph {et~al.}(2018)\citenamefont
  {{Fabbian}}, \citenamefont {{Calabrese}},\ and\ \citenamefont
  {{Carbone}}}]{2018JCAPFabbian}%
  \BibitemOpen
  \bibfield  {author} {\bibinfo {author} {\bibfnamefont {G.}~\bibnamefont
  {{Fabbian}}}, \bibinfo {author} {\bibfnamefont {M.}~\bibnamefont
  {{Calabrese}}}, \ and\ \bibinfo {author} {\bibfnamefont {C.}~\bibnamefont
  {{Carbone}}},\ }\href {\doibase 10.1088/1475-7516/2018/02/050} {\bibfield
  {journal} {\bibinfo  {journal} {\jcap}\ }\textbf {\bibinfo {volume} {2}},\
  \bibinfo {eid} {050} (\bibinfo {year} {2018})},\ \Eprint
  {http://arxiv.org/abs/1702.03317} {arXiv:1702.03317} \BibitemShut {NoStop}%
\bibitem [{\citenamefont {{Zaldarriaga}}\ and\ \citenamefont
  {{Seljak}}(1999)}]{1999PhRvDZal}%
  \BibitemOpen
  \bibfield  {author} {\bibinfo {author} {\bibfnamefont {M.}~\bibnamefont
  {{Zaldarriaga}}}\ and\ \bibinfo {author} {\bibfnamefont {U.}~\bibnamefont
  {{Seljak}}},\ }\href {\doibase 10.1103/PhysRevD.59.123507} {\bibfield
  {journal} {\bibinfo  {journal} {\prd}\ }\textbf {\bibinfo {volume} {59}},\
  \bibinfo {eid} {123507} (\bibinfo {year} {1999})},\ \Eprint
  {http://arxiv.org/abs/astro-ph/9810257} {astro-ph/9810257} \BibitemShut
  {NoStop}%
\bibitem [{\citenamefont {{Anderes}}(2013)}]{Anderes2013}%
  \BibitemOpen
  \bibfield  {author} {\bibinfo {author} {\bibfnamefont {E.}~\bibnamefont
  {{Anderes}}},\ }\href {\doibase 10.1103/PhysRevD.88.083517} {\bibfield
  {journal} {\bibinfo  {journal} {\prd}\ }\textbf {\bibinfo {volume} {88}},\
  \bibinfo {eid} {083517} (\bibinfo {year} {2013})},\ \Eprint
  {http://arxiv.org/abs/1301.2576} {arXiv:1301.2576 [astro-ph.IM]} \BibitemShut
  {NoStop}%
\bibitem [{\citenamefont {{Hirata}}\ and\ \citenamefont
  {{Seljak}}(2003{\natexlab{a}})}]{2003Hirata}%
  \BibitemOpen
  \bibfield  {author} {\bibinfo {author} {\bibfnamefont {C.~M.}\ \bibnamefont
  {{Hirata}}}\ and\ \bibinfo {author} {\bibfnamefont {U.}~\bibnamefont
  {{Seljak}}},\ }\href {\doibase 10.1103/PhysRevD.67.043001} {\bibfield
  {journal} {\bibinfo  {journal} {\prd}\ }\textbf {\bibinfo {volume} {67}}
  (\bibinfo {year} {2003}{\natexlab{a}}),\
  10.1103/PhysRevD.67.043001}\BibitemShut {NoStop}%
\bibitem [{\citenamefont {{Carron}}\ and\ \citenamefont
  {{Lewis}}(2017)}]{2017Carron}%
  \BibitemOpen
  \bibfield  {author} {\bibinfo {author} {\bibfnamefont {J.}~\bibnamefont
  {{Carron}}}\ and\ \bibinfo {author} {\bibfnamefont {A.}~\bibnamefont
  {{Lewis}}},\ }\href {\doibase 10.1103/PhysRevD.96.063510} {\bibfield
  {journal} {\bibinfo  {journal} {\prd}\ }\textbf {\bibinfo {volume} {96}}
  (\bibinfo {year} {2017}),\ 10.1103/PhysRevD.96.063510}\BibitemShut {NoStop}%
\bibitem [{\citenamefont {{Millea}}\ \emph {et~al.}(2017)\citenamefont
  {{Millea}}, \citenamefont {{Anderes}},\ and\ \citenamefont
  {{Wandelt}}}]{2017Millea}%
  \BibitemOpen
  \bibfield  {author} {\bibinfo {author} {\bibfnamefont {M.}~\bibnamefont
  {{Millea}}}, \bibinfo {author} {\bibfnamefont {E.}~\bibnamefont {{Anderes}}},
  \ and\ \bibinfo {author} {\bibfnamefont {B.~D.}\ \bibnamefont {{Wandelt}}},\
  }\href@noop {} {\bibfield  {journal} {\bibinfo  {journal} {ArXiv e-prints}\ }
  (\bibinfo {year} {2017})},\ \Eprint {http://arxiv.org/abs/1708.06753}
  {arXiv:1708.06753} \BibitemShut {NoStop}%
\bibitem [{\citenamefont {{Bucher}}\ \emph {et~al.}(2012)\citenamefont
  {{Bucher}}, \citenamefont {{Carvalho}}, \citenamefont {{Moodley}},\ and\
  \citenamefont {{Remazeilles}}}]{Bucher2012}%
  \BibitemOpen
  \bibfield  {author} {\bibinfo {author} {\bibfnamefont {M.}~\bibnamefont
  {{Bucher}}}, \bibinfo {author} {\bibfnamefont {C.~S.}\ \bibnamefont
  {{Carvalho}}}, \bibinfo {author} {\bibfnamefont {K.}~\bibnamefont
  {{Moodley}}}, \ and\ \bibinfo {author} {\bibfnamefont {M.}~\bibnamefont
  {{Remazeilles}}},\ }\href {\doibase 10.1103/PhysRevD.85.043016} {\bibfield
  {journal} {\bibinfo  {journal} {\prd}\ }\textbf {\bibinfo {volume} {85}},\
  \bibinfo {eid} {043016} (\bibinfo {year} {2012})},\ \Eprint
  {http://arxiv.org/abs/1004.3285} {arXiv:1004.3285 [astro-ph.CO]} \BibitemShut
  {NoStop}%
\bibitem [{\citenamefont {{Prince}}\ \emph {et~al.}(2018)\citenamefont
  {{Prince}}, \citenamefont {{Moodley}}, \citenamefont {{Ridl}},\ and\
  \citenamefont {{Bucher}}}]{Prince2018}%
  \BibitemOpen
  \bibfield  {author} {\bibinfo {author} {\bibfnamefont {H.}~\bibnamefont
  {{Prince}}}, \bibinfo {author} {\bibfnamefont {K.}~\bibnamefont {{Moodley}}},
  \bibinfo {author} {\bibfnamefont {J.}~\bibnamefont {{Ridl}}}, \ and\ \bibinfo
  {author} {\bibfnamefont {M.}~\bibnamefont {{Bucher}}},\ }\href {\doibase
  10.1088/1475-7516/2018/01/034} {\bibfield  {journal} {\bibinfo  {journal}
  {\jcap}\ }\textbf {\bibinfo {volume} {1}},\ \bibinfo {eid} {034} (\bibinfo
  {year} {2018})},\ \Eprint {http://arxiv.org/abs/1709.02227}
  {arXiv:1709.02227} \BibitemShut {NoStop}%
\bibitem [{\citenamefont {{Schaan}}\ and\ \citenamefont
  {{Ferraro}}(2018)}]{SchaanFerr2018_2}%
  \BibitemOpen
  \bibfield  {author} {\bibinfo {author} {\bibfnamefont {E.}~\bibnamefont
  {{Schaan}}}\ and\ \bibinfo {author} {\bibfnamefont {S.}~\bibnamefont
  {{Ferraro}}},\ }\href@noop {} {\bibfield  {journal} {\bibinfo  {journal}
  {ArXiv e-prints}\ } (\bibinfo {year} {2018})},\ \Eprint
  {http://arxiv.org/abs/1804.06403} {arXiv:1804.06403} \BibitemShut {NoStop}%
\bibitem [{\citenamefont {{Namikawa}}(2016)}]{BkkkNamikawa}%
  \BibitemOpen
  \bibfield  {author} {\bibinfo {author} {\bibfnamefont {T.}~\bibnamefont
  {{Namikawa}}},\ }\href {\doibase 10.1103/PhysRevD.93.121301} {\bibfield
  {journal} {\bibinfo  {journal} {\prd}\ }\textbf {\bibinfo {volume} {93}},\
  \bibinfo {eid} {121301} (\bibinfo {year} {2016})},\ \Eprint
  {http://arxiv.org/abs/1604.08578} {arXiv:1604.08578} \BibitemShut {NoStop}%
\bibitem [{\citenamefont {{Marozzi}}\ \emph {et~al.}(2016)\citenamefont
  {{Marozzi}}, \citenamefont {{Fanizza}}, \citenamefont {{Di Dio}},\ and\
  \citenamefont {{Durrer}}}]{postBornMarozzi}%
  \BibitemOpen
  \bibfield  {author} {\bibinfo {author} {\bibfnamefont {G.}~\bibnamefont
  {{Marozzi}}}, \bibinfo {author} {\bibfnamefont {G.}~\bibnamefont
  {{Fanizza}}}, \bibinfo {author} {\bibfnamefont {E.}~\bibnamefont {{Di Dio}}},
  \ and\ \bibinfo {author} {\bibfnamefont {R.}~\bibnamefont {{Durrer}}},\
  }\href {\doibase 10.1088/1475-7516/2016/09/028} {\bibfield  {journal}
  {\bibinfo  {journal} {\jcap}\ }\textbf {\bibinfo {volume} {9}},\ \bibinfo
  {eid} {028} (\bibinfo {year} {2016})},\ \Eprint
  {http://arxiv.org/abs/1605.08761} {arXiv:1605.08761} \BibitemShut {NoStop}%
\bibitem [{\citenamefont {{Liu}}\ \emph {et~al.}(2016)\citenamefont {{Liu}},
  \citenamefont {{Hill}}, \citenamefont {{Sherwin}}, \citenamefont {{Petri}},
  \citenamefont {{B{\"o}hm}},\ and\ \citenamefont {{Haiman}}}]{JiaPeaks2016}%
  \BibitemOpen
  \bibfield  {author} {\bibinfo {author} {\bibfnamefont {J.}~\bibnamefont
  {{Liu}}}, \bibinfo {author} {\bibfnamefont {J.~C.}\ \bibnamefont {{Hill}}},
  \bibinfo {author} {\bibfnamefont {B.~D.}\ \bibnamefont {{Sherwin}}}, \bibinfo
  {author} {\bibfnamefont {A.}~\bibnamefont {{Petri}}}, \bibinfo {author}
  {\bibfnamefont {V.}~\bibnamefont {{B{\"o}hm}}}, \ and\ \bibinfo {author}
  {\bibfnamefont {Z.}~\bibnamefont {{Haiman}}},\ }\href {\doibase
  10.1103/PhysRevD.94.103501} {\bibfield  {journal} {\bibinfo  {journal}
  {\prd}\ }\textbf {\bibinfo {volume} {94}},\ \bibinfo {eid} {103501} (\bibinfo
  {year} {2016})},\ \Eprint {http://arxiv.org/abs/1608.03169}
  {arXiv:1608.03169} \BibitemShut {NoStop}%
\bibitem [{\citenamefont {{Springel}}(2005)}]{Gadget-2}%
  \BibitemOpen
  \bibfield  {author} {\bibinfo {author} {\bibfnamefont {V.}~\bibnamefont
  {{Springel}}},\ }\href {\doibase 10.1111/j.1365-2966.2005.09655.x} {\bibfield
   {journal} {\bibinfo  {journal} {\mnras}\ }\textbf {\bibinfo {volume}
  {364}},\ \bibinfo {pages} {1105} (\bibinfo {year} {2005})},\ \Eprint
  {http://arxiv.org/abs/astro-ph/0505010} {astro-ph/0505010} \BibitemShut
  {NoStop}%
\bibitem [{\citenamefont {Lewis}\ \emph {et~al.}(2000)\citenamefont {Lewis},
  \citenamefont {Challinor},\ and\ \citenamefont {Lasenby}}]{CAMB}%
  \BibitemOpen
  \bibfield  {author} {\bibinfo {author} {\bibfnamefont {A.}~\bibnamefont
  {Lewis}}, \bibinfo {author} {\bibfnamefont {A.}~\bibnamefont {Challinor}}, \
  and\ \bibinfo {author} {\bibfnamefont {A.}~\bibnamefont {Lasenby}},\ }\href
  {\doibase 10.1086/309179} {\bibfield  {journal} {\bibinfo  {journal}
  {Astrophys. J.}\ }\textbf {\bibinfo {volume} {538}},\ \bibinfo {pages} {473}
  (\bibinfo {year} {2000})},\ \Eprint {http://arxiv.org/abs/astro-ph/9911177}
  {arXiv:astro-ph/9911177 [astro-ph]} \BibitemShut {NoStop}%
\bibitem [{\citenamefont {{Petri}}(2016)}]{LensTools}%
  \BibitemOpen
  \bibfield  {author} {\bibinfo {author} {\bibfnamefont {A.}~\bibnamefont
  {{Petri}}},\ }\href {\doibase 10.1016/j.ascom.2016.06.001} {\bibfield
  {journal} {\bibinfo  {journal} {Astronomy and Computing}\ }\textbf {\bibinfo
  {volume} {17}},\ \bibinfo {pages} {73} (\bibinfo {year} {2016})},\ \Eprint
  {http://arxiv.org/abs/1606.01903} {arXiv:1606.01903} \BibitemShut {NoStop}%
\bibitem [{\citenamefont {{Petri}}\ \emph {et~al.}(2016)\citenamefont
  {{Petri}}, \citenamefont {{Haiman}},\ and\ \citenamefont
  {{May}}}]{Petri1026sampleVar}%
  \BibitemOpen
  \bibfield  {author} {\bibinfo {author} {\bibfnamefont {A.}~\bibnamefont
  {{Petri}}}, \bibinfo {author} {\bibfnamefont {Z.}~\bibnamefont {{Haiman}}}, \
  and\ \bibinfo {author} {\bibfnamefont {M.}~\bibnamefont {{May}}},\ }\href
  {\doibase 10.1103/PhysRevD.93.063524} {\bibfield  {journal} {\bibinfo
  {journal} {\prd}\ }\textbf {\bibinfo {volume} {93}},\ \bibinfo {eid} {063524}
  (\bibinfo {year} {2016})},\ \Eprint {http://arxiv.org/abs/1601.06792}
  {arXiv:1601.06792} \BibitemShut {NoStop}%
\bibitem [{\citenamefont {{Blas}}\ \emph {et~al.}(2011)\citenamefont {{Blas}},
  \citenamefont {{Lesgourgues}},\ and\ \citenamefont {{Tram}}}]{CLASSII}%
  \BibitemOpen
  \bibfield  {author} {\bibinfo {author} {\bibfnamefont {D.}~\bibnamefont
  {{Blas}}}, \bibinfo {author} {\bibfnamefont {J.}~\bibnamefont
  {{Lesgourgues}}}, \ and\ \bibinfo {author} {\bibfnamefont {T.}~\bibnamefont
  {{Tram}}},\ }\href {\doibase 10.1088/1475-7516/2011/07/034} {\bibfield
  {journal} {\bibinfo  {journal} {\jcap}\ }\textbf {\bibinfo {volume} {7}},\
  \bibinfo {eid} {034} (\bibinfo {year} {2011})},\ \Eprint
  {http://arxiv.org/abs/1104.2933} {arXiv:1104.2933} \BibitemShut {NoStop}%
\bibitem [{\citenamefont {{Takahashi}}\ \emph {et~al.}(2012)\citenamefont
  {{Takahashi}}, \citenamefont {{Sato}}, \citenamefont {{Nishimichi}},
  \citenamefont {{Taruya}},\ and\ \citenamefont {{Oguri}}}]{HALOFITT}%
  \BibitemOpen
  \bibfield  {author} {\bibinfo {author} {\bibfnamefont {R.}~\bibnamefont
  {{Takahashi}}}, \bibinfo {author} {\bibfnamefont {M.}~\bibnamefont {{Sato}}},
  \bibinfo {author} {\bibfnamefont {T.}~\bibnamefont {{Nishimichi}}}, \bibinfo
  {author} {\bibfnamefont {A.}~\bibnamefont {{Taruya}}}, \ and\ \bibinfo
  {author} {\bibfnamefont {M.}~\bibnamefont {{Oguri}}},\ }\href {\doibase
  10.1088/0004-637X/761/2/152} {\bibfield  {journal} {\bibinfo  {journal}
  {\apj}\ }\textbf {\bibinfo {volume} {761}},\ \bibinfo {eid} {152} (\bibinfo
  {year} {2012})},\ \Eprint {http://arxiv.org/abs/1208.2701} {arXiv:1208.2701}
  \BibitemShut {NoStop}%
\bibitem [{\citenamefont {{Gil-Mar{\'{\i}}n}}\ \emph
  {et~al.}(2012)\citenamefont {{Gil-Mar{\'{\i}}n}}, \citenamefont {{Wagner}},
  \citenamefont {{Fragkoudi}}, \citenamefont {{Jimenez}},\ and\ \citenamefont
  {{Verde}}}]{GilMarin2011}%
  \BibitemOpen
  \bibfield  {author} {\bibinfo {author} {\bibfnamefont {H.}~\bibnamefont
  {{Gil-Mar{\'{\i}}n}}}, \bibinfo {author} {\bibfnamefont {C.}~\bibnamefont
  {{Wagner}}}, \bibinfo {author} {\bibfnamefont {F.}~\bibnamefont
  {{Fragkoudi}}}, \bibinfo {author} {\bibfnamefont {R.}~\bibnamefont
  {{Jimenez}}}, \ and\ \bibinfo {author} {\bibfnamefont {L.}~\bibnamefont
  {{Verde}}},\ }\href {\doibase 10.1088/1475-7516/2012/02/047} {\bibfield
  {journal} {\bibinfo  {journal} {\jcap}\ }\textbf {\bibinfo {volume} {2}},\
  \bibinfo {eid} {047} (\bibinfo {year} {2012})},\ \Eprint
  {http://arxiv.org/abs/1111.4477} {arXiv:1111.4477 [astro-ph.CO]} \BibitemShut
  {NoStop}%
\bibitem [{\citenamefont {{Louis}}\ \emph {et~al.}(2013)\citenamefont
  {{Louis}}, \citenamefont {{N{\ae}ss}}, \citenamefont {{Das}}, \citenamefont
  {{Dunkley}},\ and\ \citenamefont {{Sherwin}}}]{LouisLens2013}%
  \BibitemOpen
  \bibfield  {author} {\bibinfo {author} {\bibfnamefont {T.}~\bibnamefont
  {{Louis}}}, \bibinfo {author} {\bibfnamefont {S.}~\bibnamefont {{N{\ae}ss}}},
  \bibinfo {author} {\bibfnamefont {S.}~\bibnamefont {{Das}}}, \bibinfo
  {author} {\bibfnamefont {J.}~\bibnamefont {{Dunkley}}}, \ and\ \bibinfo
  {author} {\bibfnamefont {B.}~\bibnamefont {{Sherwin}}},\ }\href {\doibase
  10.1093/mnras/stt1421} {\bibfield  {journal} {\bibinfo  {journal} {\mnras}\
  }\textbf {\bibinfo {volume} {435}},\ \bibinfo {pages} {2040} (\bibinfo {year}
  {2013})},\ \Eprint {http://arxiv.org/abs/1306.6692} {arXiv:1306.6692
  [astro-ph.CO]} \BibitemShut {NoStop}%
\bibitem [{\citenamefont {{Lewis}}\ and\ \citenamefont
  {{Pratten}}(2016)}]{BiSpecCMBPow}%
  \BibitemOpen
  \bibfield  {author} {\bibinfo {author} {\bibfnamefont {A.}~\bibnamefont
  {{Lewis}}}\ and\ \bibinfo {author} {\bibfnamefont {G.}~\bibnamefont
  {{Pratten}}},\ }\href {\doibase 10.1088/1475-7516/2016/12/003} {\bibfield
  {journal} {\bibinfo  {journal} {\jcap}\ }\textbf {\bibinfo {volume} {12}},\
  \bibinfo {eid} {003} (\bibinfo {year} {2016})},\ \Eprint
  {http://arxiv.org/abs/1608.01263} {arXiv:1608.01263} \BibitemShut {NoStop}%
\bibitem [{\citenamefont {{Schmittfull}}\ \emph {et~al.}(2013)\citenamefont
  {{Schmittfull}}, \citenamefont {{Challinor}}, \citenamefont {{Hanson}},\ and\
  \citenamefont {{Lewis}}}]{2013Schmittfull}%
  \BibitemOpen
  \bibfield  {author} {\bibinfo {author} {\bibfnamefont {M.~M.}\ \bibnamefont
  {{Schmittfull}}}, \bibinfo {author} {\bibfnamefont {A.}~\bibnamefont
  {{Challinor}}}, \bibinfo {author} {\bibfnamefont {D.}~\bibnamefont
  {{Hanson}}}, \ and\ \bibinfo {author} {\bibfnamefont {A.}~\bibnamefont
  {{Lewis}}},\ }\href {\doibase 10.1103/PhysRevD.88.063012} {\bibfield
  {journal} {\bibinfo  {journal} {\prd}\ }\textbf {\bibinfo {volume} {88}},\
  \bibinfo {eid} {063012} (\bibinfo {year} {2013})},\ \Eprint
  {http://arxiv.org/abs/1308.0286} {arXiv:1308.0286 [astro-ph.CO]} \BibitemShut
  {NoStop}%
\bibitem [{\citenamefont {{Peloton}}\ \emph {et~al.}(2017)\citenamefont
  {{Peloton}}, \citenamefont {{Schmittfull}}, \citenamefont {{Lewis}},
  \citenamefont {{Carron}},\ and\ \citenamefont {{Zahn}}}]{2017Peloton}%
  \BibitemOpen
  \bibfield  {author} {\bibinfo {author} {\bibfnamefont {J.}~\bibnamefont
  {{Peloton}}}, \bibinfo {author} {\bibfnamefont {M.}~\bibnamefont
  {{Schmittfull}}}, \bibinfo {author} {\bibfnamefont {A.}~\bibnamefont
  {{Lewis}}}, \bibinfo {author} {\bibfnamefont {J.}~\bibnamefont {{Carron}}}, \
  and\ \bibinfo {author} {\bibfnamefont {O.}~\bibnamefont {{Zahn}}},\ }\href
  {\doibase 10.1103/PhysRevD.95.043508} {\bibfield  {journal} {\bibinfo
  {journal} {\prd}\ }\textbf {\bibinfo {volume} {95}},\ \bibinfo {eid} {043508}
  (\bibinfo {year} {2017})},\ \Eprint {http://arxiv.org/abs/1611.01446}
  {arXiv:1611.01446} \BibitemShut {NoStop}%
\bibitem [{\citenamefont {{Suzuki}}\ \emph {et~al.}(2018)\citenamefont
  {{Suzuki}}, \citenamefont {{Ade}}, \citenamefont {{Akiba}}, \citenamefont
  {{Alonso}}, \citenamefont {{Arnold}}, \citenamefont {{Aumont}}, \citenamefont
  {{Baccigalupi}}, \citenamefont {{Barron}}, \citenamefont {{Basak}},
  \citenamefont {{Beckman}}, \citenamefont {{Borrill}}, \citenamefont
  {{Boulanger}}, \citenamefont {{Bucher}}, \citenamefont {{Calabrese}},
  \citenamefont {{Chinone}}, \citenamefont {{Cho}}, \citenamefont
  {{Cukierman}}, \citenamefont {{Curtis}}, \citenamefont {{de Haan}},
  \citenamefont {{Dobbs}}, \citenamefont {{Dominjon}}, \citenamefont
  {{Dotani}}, \citenamefont {{Duband}}, \citenamefont {{Ducout}}, \citenamefont
  {{Dunkley}}, \citenamefont {{Duval}}, \citenamefont {{Elleflot}},
  \citenamefont {{Eriksen}}, \citenamefont {{Errard}}, \citenamefont
  {{Fischer}}, \citenamefont {{Fujino}}, \citenamefont {{Funaki}},
  \citenamefont {{Fuskeland}}, \citenamefont {{Ganga}}, \citenamefont
  {{Goeckner-Wald}}, \citenamefont {{Grain}}, \citenamefont {{Halverson}},
  \citenamefont {{Hamada}}, \citenamefont {{Hasebe}}, \citenamefont
  {{Hasegawa}}, \citenamefont {{Hattori}}, \citenamefont {{Hattori}},
  \citenamefont {{Hayes}}, \citenamefont {{Hazumi}}, \citenamefont
  {{Hidehira}}, \citenamefont {{Hill}}, \citenamefont {{Hilton}}, \citenamefont
  {{Hubmayr}}, \citenamefont {{Ichiki}}, \citenamefont {{Iida}}, \citenamefont
  {{Imada}}, \citenamefont {{Inoue}}, \citenamefont {{Inoue}}, \citenamefont
  {{D.}}, \citenamefont {{Ishino}}, \citenamefont {{Jeong}}, \citenamefont
  {{Kanai}}, \citenamefont {{Kaneko}}, \citenamefont {{Kashima}}, \citenamefont
  {{Katayama}}, \citenamefont {{Kawasaki}}, \citenamefont {{Kernasovskiy}},
  \citenamefont {{Keskitalo}}, \citenamefont {{Kibayashi}}, \citenamefont
  {{Kida}}, \citenamefont {{Kimura}}, \citenamefont {{Kisner}}, \citenamefont
  {{Kohri}}, \citenamefont {{Komatsu}}, \citenamefont {{Komatsu}},
  \citenamefont {{Kuo}}, \citenamefont {{Kurinsky}}, \citenamefont {{Kusaka}},
  \citenamefont {{Lazarian}}, \citenamefont {{Lee}}, \citenamefont {{Li}},
  \citenamefont {{Linder}}, \citenamefont {{Maffei}}, \citenamefont
  {{Mangilli}}, \citenamefont {{Maki}}, \citenamefont {{Matsumura}},
  \citenamefont {{Matsuura}}, \citenamefont {{Meilhan}}, \citenamefont
  {{Mima}}, \citenamefont {{Minami}}, \citenamefont {{Mitsuda}}, \citenamefont
  {{Montier}}, \citenamefont {{Nagai}}, \citenamefont {{Nagasaki}},
  \citenamefont {{Nagata}}, \citenamefont {{Nakajima}}, \citenamefont
  {{Nakamura}}, \citenamefont {{Namikawa}}, \citenamefont {{Naruse}},
  \citenamefont {{Nishino}}, \citenamefont {{Nitta}}, \citenamefont
  {{Noguchi}}, \citenamefont {{Ogawa}}, \citenamefont {{Oguri}}, \citenamefont
  {{Okada}}, \citenamefont {{Okamoto}}, \citenamefont {{Okamura}},
  \citenamefont {{Otani}}, \citenamefont {{Patanchon}}, \citenamefont
  {{Pisano}}, \citenamefont {{Rebeiz}}, \citenamefont {{Remazeilles}},
  \citenamefont {{Richards}}, \citenamefont {{Sakai}}, \citenamefont
  {{Sakurai}}, \citenamefont {{Sato}}, \citenamefont {{Sato}}, \citenamefont
  {{Sawada}}, \citenamefont {{Segawa}}, \citenamefont {{Sekimoto}},
  \citenamefont {{Seljak}}, \citenamefont {{Sherwin}}, \citenamefont
  {{Shimizu}}, \citenamefont {{Shinozaki}}, \citenamefont {{Stompor}},
  \citenamefont {{Sugai}}, \citenamefont {{Sugita}}, \citenamefont {{Suzuki}},
  \citenamefont {{Tajima}}, \citenamefont {{Takada}}, \citenamefont {{Takaku}},
  \citenamefont {{Takakura}}, \citenamefont {{Takatori}}, \citenamefont
  {{Tanabe}}, \citenamefont {{Taylor}}, \citenamefont {{Thompson}},
  \citenamefont {{Thorne}}, \citenamefont {{Tomaru}}, \citenamefont {{Tomida}},
  \citenamefont {{Tomita}}, \citenamefont {{Tristram}}, \citenamefont
  {{Tucker}}, \citenamefont {{Turin}}, \citenamefont {{Tsujimoto}},
  \citenamefont {{Uozumi}}, \citenamefont {{Utsunomiya}}, \citenamefont
  {{Uzawa}}, \citenamefont {{Vansyngel}}, \citenamefont {{Wehus}},
  \citenamefont {{Westbrook}}, \citenamefont {{Willer}}, \citenamefont
  {{Whitehorn}}, \citenamefont {{Yamada}}, \citenamefont {{Yamamoto}},
  \citenamefont {{Yamasaki}}, \citenamefont {{Yamashita}},\ and\ \citenamefont
  {{Yoshida}}}]{2018LiteBird}%
  \BibitemOpen
  \bibfield  {author} {\bibinfo {author} {\bibfnamefont {A.}~\bibnamefont
  {{Suzuki}}}, \bibinfo {author} {\bibfnamefont {P.~A.~R.}\ \bibnamefont
  {{Ade}}}, \bibinfo {author} {\bibfnamefont {Y.}~\bibnamefont {{Akiba}}},
  \bibinfo {author} {\bibfnamefont {D.}~\bibnamefont {{Alonso}}}, \bibinfo
  {author} {\bibfnamefont {K.}~\bibnamefont {{Arnold}}}, \bibinfo {author}
  {\bibfnamefont {J.}~\bibnamefont {{Aumont}}}, \bibinfo {author}
  {\bibfnamefont {C.}~\bibnamefont {{Baccigalupi}}}, \bibinfo {author}
  {\bibfnamefont {D.}~\bibnamefont {{Barron}}}, \bibinfo {author}
  {\bibfnamefont {S.}~\bibnamefont {{Basak}}}, \bibinfo {author} {\bibfnamefont
  {S.}~\bibnamefont {{Beckman}}}, \bibinfo {author} {\bibfnamefont
  {J.}~\bibnamefont {{Borrill}}}, \bibinfo {author} {\bibfnamefont
  {F.}~\bibnamefont {{Boulanger}}}, \bibinfo {author} {\bibfnamefont
  {M.}~\bibnamefont {{Bucher}}}, \bibinfo {author} {\bibfnamefont
  {E.}~\bibnamefont {{Calabrese}}}, \bibinfo {author} {\bibfnamefont
  {Y.}~\bibnamefont {{Chinone}}}, \bibinfo {author} {\bibfnamefont
  {H.}~\bibnamefont {{Cho}}}, \bibinfo {author} {\bibfnamefont
  {A.}~\bibnamefont {{Cukierman}}}, \bibinfo {author} {\bibfnamefont {D.~W.}\
  \bibnamefont {{Curtis}}}, \bibinfo {author} {\bibfnamefont {T.}~\bibnamefont
  {{de Haan}}}, \bibinfo {author} {\bibfnamefont {M.}~\bibnamefont {{Dobbs}}},
  \bibinfo {author} {\bibfnamefont {A.}~\bibnamefont {{Dominjon}}}, \bibinfo
  {author} {\bibfnamefont {T.}~\bibnamefont {{Dotani}}}, \bibinfo {author}
  {\bibfnamefont {L.}~\bibnamefont {{Duband}}}, \bibinfo {author}
  {\bibfnamefont {A.}~\bibnamefont {{Ducout}}}, \bibinfo {author}
  {\bibfnamefont {J.}~\bibnamefont {{Dunkley}}}, \bibinfo {author}
  {\bibfnamefont {J.~M.}\ \bibnamefont {{Duval}}}, \bibinfo {author}
  {\bibfnamefont {T.}~\bibnamefont {{Elleflot}}}, \bibinfo {author}
  {\bibfnamefont {H.~K.}\ \bibnamefont {{Eriksen}}}, \bibinfo {author}
  {\bibfnamefont {J.}~\bibnamefont {{Errard}}}, \bibinfo {author}
  {\bibfnamefont {J.}~\bibnamefont {{Fischer}}}, \bibinfo {author}
  {\bibfnamefont {T.}~\bibnamefont {{Fujino}}}, \bibinfo {author}
  {\bibfnamefont {T.}~\bibnamefont {{Funaki}}}, \bibinfo {author}
  {\bibfnamefont {U.}~\bibnamefont {{Fuskeland}}}, \bibinfo {author}
  {\bibfnamefont {K.}~\bibnamefont {{Ganga}}}, \bibinfo {author} {\bibfnamefont
  {N.}~\bibnamefont {{Goeckner-Wald}}}, \bibinfo {author} {\bibfnamefont
  {J.}~\bibnamefont {{Grain}}}, \bibinfo {author} {\bibfnamefont {N.~W.}\
  \bibnamefont {{Halverson}}}, \bibinfo {author} {\bibfnamefont
  {T.}~\bibnamefont {{Hamada}}}, \bibinfo {author} {\bibfnamefont
  {T.}~\bibnamefont {{Hasebe}}}, \bibinfo {author} {\bibfnamefont
  {M.}~\bibnamefont {{Hasegawa}}}, \bibinfo {author} {\bibfnamefont
  {K.}~\bibnamefont {{Hattori}}}, \bibinfo {author} {\bibfnamefont
  {M.}~\bibnamefont {{Hattori}}}, \bibinfo {author} {\bibfnamefont
  {L.}~\bibnamefont {{Hayes}}}, \bibinfo {author} {\bibfnamefont
  {M.}~\bibnamefont {{Hazumi}}}, \bibinfo {author} {\bibfnamefont
  {N.}~\bibnamefont {{Hidehira}}}, \bibinfo {author} {\bibfnamefont {C.~A.}\
  \bibnamefont {{Hill}}}, \bibinfo {author} {\bibfnamefont {G.}~\bibnamefont
  {{Hilton}}}, \bibinfo {author} {\bibfnamefont {J.}~\bibnamefont {{Hubmayr}}},
  \bibinfo {author} {\bibfnamefont {K.}~\bibnamefont {{Ichiki}}}, \bibinfo
  {author} {\bibfnamefont {T.}~\bibnamefont {{Iida}}}, \bibinfo {author}
  {\bibfnamefont {H.}~\bibnamefont {{Imada}}}, \bibinfo {author} {\bibfnamefont
  {M.}~\bibnamefont {{Inoue}}}, \bibinfo {author} {\bibfnamefont
  {Y.}~\bibnamefont {{Inoue}}}, \bibinfo {author} {\bibfnamefont
  {K.}~\bibnamefont {{D.}}}, \bibinfo {author} {\bibfnamefont {H.}~\bibnamefont
  {{Ishino}}}, \bibinfo {author} {\bibfnamefont {O.}~\bibnamefont {{Jeong}}},
  \bibinfo {author} {\bibfnamefont {H.}~\bibnamefont {{Kanai}}}, \bibinfo
  {author} {\bibfnamefont {D.}~\bibnamefont {{Kaneko}}}, \bibinfo {author}
  {\bibfnamefont {S.}~\bibnamefont {{Kashima}}}, \bibinfo {author}
  {\bibfnamefont {N.}~\bibnamefont {{Katayama}}}, \bibinfo {author}
  {\bibfnamefont {T.}~\bibnamefont {{Kawasaki}}}, \bibinfo {author}
  {\bibfnamefont {S.~A.}\ \bibnamefont {{Kernasovskiy}}}, \bibinfo {author}
  {\bibfnamefont {R.}~\bibnamefont {{Keskitalo}}}, \bibinfo {author}
  {\bibfnamefont {A.}~\bibnamefont {{Kibayashi}}}, \bibinfo {author}
  {\bibfnamefont {Y.}~\bibnamefont {{Kida}}}, \bibinfo {author} {\bibfnamefont
  {K.}~\bibnamefont {{Kimura}}}, \bibinfo {author} {\bibfnamefont
  {T.}~\bibnamefont {{Kisner}}}, \bibinfo {author} {\bibfnamefont
  {K.}~\bibnamefont {{Kohri}}}, \bibinfo {author} {\bibfnamefont
  {E.}~\bibnamefont {{Komatsu}}}, \bibinfo {author} {\bibfnamefont
  {K.}~\bibnamefont {{Komatsu}}}, \bibinfo {author} {\bibfnamefont {C.~L.}\
  \bibnamefont {{Kuo}}}, \bibinfo {author} {\bibfnamefont {N.~A.}\ \bibnamefont
  {{Kurinsky}}}, \bibinfo {author} {\bibfnamefont {A.}~\bibnamefont
  {{Kusaka}}}, \bibinfo {author} {\bibfnamefont {A.}~\bibnamefont
  {{Lazarian}}}, \bibinfo {author} {\bibfnamefont {A.~T.}\ \bibnamefont
  {{Lee}}}, \bibinfo {author} {\bibfnamefont {D.}~\bibnamefont {{Li}}},
  \bibinfo {author} {\bibfnamefont {E.}~\bibnamefont {{Linder}}}, \bibinfo
  {author} {\bibfnamefont {B.}~\bibnamefont {{Maffei}}}, \bibinfo {author}
  {\bibfnamefont {A.}~\bibnamefont {{Mangilli}}}, \bibinfo {author}
  {\bibfnamefont {M.}~\bibnamefont {{Maki}}}, \bibinfo {author} {\bibfnamefont
  {T.}~\bibnamefont {{Matsumura}}}, \bibinfo {author} {\bibfnamefont
  {S.}~\bibnamefont {{Matsuura}}}, \bibinfo {author} {\bibfnamefont
  {D.}~\bibnamefont {{Meilhan}}}, \bibinfo {author} {\bibfnamefont
  {S.}~\bibnamefont {{Mima}}}, \bibinfo {author} {\bibfnamefont
  {Y.}~\bibnamefont {{Minami}}}, \bibinfo {author} {\bibfnamefont
  {K.}~\bibnamefont {{Mitsuda}}}, \bibinfo {author} {\bibfnamefont
  {L.}~\bibnamefont {{Montier}}}, \bibinfo {author} {\bibfnamefont
  {M.}~\bibnamefont {{Nagai}}}, \bibinfo {author} {\bibfnamefont
  {T.}~\bibnamefont {{Nagasaki}}}, \bibinfo {author} {\bibfnamefont
  {R.}~\bibnamefont {{Nagata}}}, \bibinfo {author} {\bibfnamefont
  {M.}~\bibnamefont {{Nakajima}}}, \bibinfo {author} {\bibfnamefont
  {S.}~\bibnamefont {{Nakamura}}}, \bibinfo {author} {\bibfnamefont
  {T.}~\bibnamefont {{Namikawa}}}, \bibinfo {author} {\bibfnamefont
  {M.}~\bibnamefont {{Naruse}}}, \bibinfo {author} {\bibfnamefont
  {H.}~\bibnamefont {{Nishino}}}, \bibinfo {author} {\bibfnamefont
  {T.}~\bibnamefont {{Nitta}}}, \bibinfo {author} {\bibfnamefont
  {T.}~\bibnamefont {{Noguchi}}}, \bibinfo {author} {\bibfnamefont
  {H.}~\bibnamefont {{Ogawa}}}, \bibinfo {author} {\bibfnamefont
  {S.}~\bibnamefont {{Oguri}}}, \bibinfo {author} {\bibfnamefont
  {N.}~\bibnamefont {{Okada}}}, \bibinfo {author} {\bibfnamefont
  {A.}~\bibnamefont {{Okamoto}}}, \bibinfo {author} {\bibfnamefont
  {T.}~\bibnamefont {{Okamura}}}, \bibinfo {author} {\bibfnamefont
  {C.}~\bibnamefont {{Otani}}}, \bibinfo {author} {\bibfnamefont
  {G.}~\bibnamefont {{Patanchon}}}, \bibinfo {author} {\bibfnamefont
  {G.}~\bibnamefont {{Pisano}}}, \bibinfo {author} {\bibfnamefont
  {G.}~\bibnamefont {{Rebeiz}}}, \bibinfo {author} {\bibfnamefont
  {M.}~\bibnamefont {{Remazeilles}}}, \bibinfo {author} {\bibfnamefont {P.~L.}\
  \bibnamefont {{Richards}}}, \bibinfo {author} {\bibfnamefont
  {S.}~\bibnamefont {{Sakai}}}, \bibinfo {author} {\bibfnamefont
  {Y.}~\bibnamefont {{Sakurai}}}, \bibinfo {author} {\bibfnamefont
  {Y.}~\bibnamefont {{Sato}}}, \bibinfo {author} {\bibfnamefont
  {N.}~\bibnamefont {{Sato}}}, \bibinfo {author} {\bibfnamefont
  {M.}~\bibnamefont {{Sawada}}}, \bibinfo {author} {\bibfnamefont
  {Y.}~\bibnamefont {{Segawa}}}, \bibinfo {author} {\bibfnamefont
  {Y.}~\bibnamefont {{Sekimoto}}}, \bibinfo {author} {\bibfnamefont
  {U.}~\bibnamefont {{Seljak}}}, \bibinfo {author} {\bibfnamefont {B.~D.}\
  \bibnamefont {{Sherwin}}}, \bibinfo {author} {\bibfnamefont {T.}~\bibnamefont
  {{Shimizu}}}, \bibinfo {author} {\bibfnamefont {K.}~\bibnamefont
  {{Shinozaki}}}, \bibinfo {author} {\bibfnamefont {R.}~\bibnamefont
  {{Stompor}}}, \bibinfo {author} {\bibfnamefont {H.}~\bibnamefont {{Sugai}}},
  \bibinfo {author} {\bibfnamefont {H.}~\bibnamefont {{Sugita}}}, \bibinfo
  {author} {\bibfnamefont {J.}~\bibnamefont {{Suzuki}}}, \bibinfo {author}
  {\bibfnamefont {O.}~\bibnamefont {{Tajima}}}, \bibinfo {author}
  {\bibfnamefont {S.}~\bibnamefont {{Takada}}}, \bibinfo {author}
  {\bibfnamefont {R.}~\bibnamefont {{Takaku}}}, \bibinfo {author}
  {\bibfnamefont {S.}~\bibnamefont {{Takakura}}}, \bibinfo {author}
  {\bibfnamefont {S.}~\bibnamefont {{Takatori}}}, \bibinfo {author}
  {\bibfnamefont {D.}~\bibnamefont {{Tanabe}}}, \bibinfo {author}
  {\bibfnamefont {E.}~\bibnamefont {{Taylor}}}, \bibinfo {author}
  {\bibfnamefont {K.~L.}\ \bibnamefont {{Thompson}}}, \bibinfo {author}
  {\bibfnamefont {B.}~\bibnamefont {{Thorne}}}, \bibinfo {author}
  {\bibfnamefont {T.}~\bibnamefont {{Tomaru}}}, \bibinfo {author}
  {\bibfnamefont {T.}~\bibnamefont {{Tomida}}}, \bibinfo {author}
  {\bibfnamefont {N.}~\bibnamefont {{Tomita}}}, \bibinfo {author}
  {\bibfnamefont {M.}~\bibnamefont {{Tristram}}}, \bibinfo {author}
  {\bibfnamefont {C.}~\bibnamefont {{Tucker}}}, \bibinfo {author}
  {\bibfnamefont {P.}~\bibnamefont {{Turin}}}, \bibinfo {author} {\bibfnamefont
  {M.}~\bibnamefont {{Tsujimoto}}}, \bibinfo {author} {\bibfnamefont
  {S.}~\bibnamefont {{Uozumi}}}, \bibinfo {author} {\bibfnamefont
  {S.}~\bibnamefont {{Utsunomiya}}}, \bibinfo {author} {\bibfnamefont
  {Y.}~\bibnamefont {{Uzawa}}}, \bibinfo {author} {\bibfnamefont
  {F.}~\bibnamefont {{Vansyngel}}}, \bibinfo {author} {\bibfnamefont {I.~K.}\
  \bibnamefont {{Wehus}}}, \bibinfo {author} {\bibfnamefont {B.}~\bibnamefont
  {{Westbrook}}}, \bibinfo {author} {\bibfnamefont {M.}~\bibnamefont
  {{Willer}}}, \bibinfo {author} {\bibfnamefont {N.}~\bibnamefont
  {{Whitehorn}}}, \bibinfo {author} {\bibfnamefont {Y.}~\bibnamefont
  {{Yamada}}}, \bibinfo {author} {\bibfnamefont {R.}~\bibnamefont
  {{Yamamoto}}}, \bibinfo {author} {\bibfnamefont {N.}~\bibnamefont
  {{Yamasaki}}}, \bibinfo {author} {\bibfnamefont {T.}~\bibnamefont
  {{Yamashita}}}, \ and\ \bibinfo {author} {\bibfnamefont {M.}~\bibnamefont
  {{Yoshida}}},\ }\href@noop {} {\bibfield  {journal} {\bibinfo  {journal}
  {ArXiv e-prints}\ } (\bibinfo {year} {2018})},\ \Eprint
  {http://arxiv.org/abs/1801.06987} {arXiv:1801.06987 [astro-ph.IM]}
  \BibitemShut {NoStop}%
\bibitem [{\citenamefont {{Hanany}}\ and\ \citenamefont {{Inflation Probe
  Mission Study Team}}(2018)}]{2018Pico}%
  \BibitemOpen
  \bibfield  {author} {\bibinfo {author} {\bibfnamefont {S.}~\bibnamefont
  {{Hanany}}}\ and\ \bibinfo {author} {\bibnamefont {{Inflation Probe Mission
  Study Team}}},\ }in\ \href@noop {} {\emph {\bibinfo {booktitle} {American
  Astronomical Society Meeting Abstracts \#231}}},\ \bibinfo {series} {American
  Astronomical Society Meeting Abstracts}, Vol.\ \bibinfo {volume} {231}\
  (\bibinfo {year} {2018})\ p.\ \bibinfo {pages} {121.01}\BibitemShut {NoStop}%
\bibitem [{\citenamefont {{Merkel}}\ and\ \citenamefont
  {{Sch{\"a}fer}}(2017)}]{2017Merkel}%
  \BibitemOpen
  \bibfield  {author} {\bibinfo {author} {\bibfnamefont {P.~M.}\ \bibnamefont
  {{Merkel}}}\ and\ \bibinfo {author} {\bibfnamefont {B.~M.}\ \bibnamefont
  {{Sch{\"a}fer}}},\ }\href {\doibase 10.1093/mnras/stx1664} {\bibfield
  {journal} {\bibinfo  {journal} {\mnras}\ }\textbf {\bibinfo {volume} {471}},\
  \bibinfo {pages} {2431} (\bibinfo {year} {2017})}\BibitemShut {NoStop}%
\bibitem [{\citenamefont {{Stebbins}}(1996)}]{1996Stebbins}%
  \BibitemOpen
  \bibfield  {author} {\bibinfo {author} {\bibfnamefont {A.}~\bibnamefont
  {{Stebbins}}},\ }\href@noop {} {\bibfield  {journal} {\bibinfo  {journal}
  {ArXiv Astrophysics e-prints}\ } (\bibinfo {year} {1996})},\ \Eprint
  {http://arxiv.org/abs/astro-ph/9609149} {astro-ph/9609149} \BibitemShut
  {NoStop}%
\bibitem [{\citenamefont {{Hirata}}\ and\ \citenamefont
  {{Seljak}}(2003{\natexlab{b}})}]{2003PhRvDHirata}%
  \BibitemOpen
  \bibfield  {author} {\bibinfo {author} {\bibfnamefont {C.~M.}\ \bibnamefont
  {{Hirata}}}\ and\ \bibinfo {author} {\bibfnamefont {U.}~\bibnamefont
  {{Seljak}}},\ }\href {\doibase 10.1103/PhysRevD.68.083002} {\bibfield
  {journal} {\bibinfo  {journal} {\prd}\ }\textbf {\bibinfo {volume} {68}},\
  \bibinfo {eid} {083002} (\bibinfo {year} {2003}{\natexlab{b}})},\ \Eprint
  {http://arxiv.org/abs/astro-ph/0306354} {astro-ph/0306354} \BibitemShut
  {NoStop}%
\bibitem [{\citenamefont {{Cooray}}\ \emph {et~al.}(2005)\citenamefont
  {{Cooray}}, \citenamefont {{Kamionkowski}},\ and\ \citenamefont
  {{Caldwell}}}]{2005PhRvDCooray}%
  \BibitemOpen
  \bibfield  {author} {\bibinfo {author} {\bibfnamefont {A.}~\bibnamefont
  {{Cooray}}}, \bibinfo {author} {\bibfnamefont {M.}~\bibnamefont
  {{Kamionkowski}}}, \ and\ \bibinfo {author} {\bibfnamefont {R.~R.}\
  \bibnamefont {{Caldwell}}},\ }\href {\doibase 10.1103/PhysRevD.71.123527}
  {\bibfield  {journal} {\bibinfo  {journal} {\prd}\ }\textbf {\bibinfo
  {volume} {71}},\ \bibinfo {eid} {123527} (\bibinfo {year} {2005})},\ \Eprint
  {http://arxiv.org/abs/astro-ph/0503002} {astro-ph/0503002} \BibitemShut
  {NoStop}%
\end{thebibliography}%
\end{document}